\documentclass[prb, twocolumn, superscriptaddress, floatfix, nofootinbib]{revtex4-2}
\usepackage{graphicx}
\usepackage{float}
\usepackage{amsmath,amssymb,amstext,dsfont,tikz,graphicx,physics,mathtools,bm,
simpler-wick,color}
\usepackage[colorlinks=true, allcolors=purple]{hyperref}

\newcommand{\vm}{\vb{m}}
\newcommand{\vk}{\vb{k}}

\DeclareMathOperator{\gs}{g.s.}
\DeclareMathOperator{\nc}{n.c.}
\DeclareMathOperator{\avg}{avg}
\DeclareMathOperator{\eff}{eff}

\makeatletter 
    
\renewcommand\onecolumngrid{
\do@columngrid{one}{\@ne}%
\def\set@footnotewidth{\onecolumngrid}
\def\footnoterule{\kern-6pt\hrule width 1.5in\kern6pt}%
}

\renewcommand\twocolumngrid{
        \def\footnoterule{
        \dimen@\skip\footins\divide\dimen@\thr@@
        \kern-\dimen@\hrule width.5in\kern\dimen@}
        \do@columngrid{mlt}{\tw@}
}%

\makeatother

\begin{document}

\title{Nonlocality and entanglement in measured critical quantum Ising chains}

\author{Zack Weinstein}
\thanks{Z.W. and R.S. contributed equally to this work.}
\affiliation{Department of Physics, University of California, Berkeley, California 94720, USA}
\author{Rohith Sajith}
\thanks{Z.W. and R.S. contributed equally to this work.}
\affiliation{Department of Physics, University of California, Berkeley, California 94720, USA}
\author{Ehud Altman}
\affiliation{Department of Physics, University of California, Berkeley, California 94720, USA}
\affiliation{Materials Sciences Division, Lawrence Berkeley National Laboratory, Berkeley, CA 94720, USA}
\author{Samuel J. Garratt}
\affiliation{Department of Physics, University of California, Berkeley, California 94720, USA}

\begin{abstract}
We study the effects of measurements, performed with a finite density in space, on the ground state of the one-dimensional transverse-field Ising model at criticality. Local degrees of freedom in critical states exhibit long-range entanglement, and as a result, local measurements can have highly nonlocal effects. Our analytical investigation of correlations and entanglement in the ensemble of measured states is based on properties of the Ising conformal field theory (CFT), where measurements appear as (1+0)-dimensional defects in the (1+1)-dimensional Euclidean spacetime. So that we can verify our predictions using large-scale free-fermion numerics, we restrict ourselves to parity-symmetric measurements. To describe their averaged effects analytically we use a replica approach, and we show that the defect arising in the replica theory is an irrelevant perturbation to the Ising CFT. Strikingly, the asymptotic scalings of averaged correlations and entanglement entropy are therefore unchanged relative to the ground state. In contrast, the defect generated by postselecting on the most likely measurement outcomes is exactly marginal. We then find that the exponent governing postmeasurement order parameter correlations, as well as the ``effective central charge'' governing the scaling of entanglement entropy, vary continuously with the density of measurements in space. Our work establishes new connections between the effects of measurements on many-body quantum states and of physical defects on low-energy equilibrium properties.
\end{abstract}

\maketitle

\section{Introduction}

Measuring one of the qubits in a Bell pair nonlocally alters the state of the unmeasured qubit. In many-body systems, where the entanglement of a state can be highly complex, the nonlocal effects of measurements can give rise to a remarkable variety of different structures \cite{briegel_2009_measurement,li_2021_conformal,piroli_quantum_2021,bao_2021_teleportation,tantivasadakarn_long_2021,verresen_efficiently_2021,lin_2022_probing,lu_measurement_2022,garratt_measurements_2022,zhu_nishimori_2022,lee_2022_decoding,zhu_nishimori_2022,tantivasadakarn_shortest_2022,tantivasadakarn_hierarchy_2022,antonini_2022_holographic,antonini2022holographic}. Strikingly, even when starting from a state with short-ranged entanglement, one can use measurements to create topological order and other long-range entangled states \cite{piroli_quantum_2021,tantivasadakarn_long_2021,verresen_efficiently_2021,lu_measurement_2022,tantivasadakarn_hierarchy_2022,tantivasadakarn_shortest_2022,lee_2022_decoding,zhu_nishimori_2022}. In the context of the measurement-induced phase transition in quantum circuits \cite{skinner_measurement-induced_2019,li_quantum_2018,li_measurement-driven_2019,gullans_dynamical_2020,choi_quantum_2020,bao_theory_2020,jian_measurement-induced_2020,potter_entanglement_2022,fisher_random_2022}, the nonlocal effects of measurements are known to be crucial for the emergence of conformal symmetry at the critical point \cite{li_2021_conformal}.

Given such a diverse range of phenomena, it is important to seek unifying principles underlying the effects of measurements on many entangled degrees of freedom. Critical ground states \cite{sachdev_quantum_2011} in one spatial dimension here offer a high degree of theoretical control because universal structures are described at long distances by (1+1)-dimensional conformal field theories (CFTs) \cite{belavin_infinite_1984, di_francesco_conformal_1997,henkel1999conformal}. Moreover, as shown in Ref.~\cite{garratt_measurements_2022}, studies of the effects of measurements on these states are closely related to problems arising in the theory of surface critical phenomena \cite{cardy_conformal_1984,cardy1996scaling}. This connection has more recently appeared in studies of the effects of local decoherence on topological \cite{lee_2022_symmetry,bao_2023_mixed,lee_2023_criticality} and critical states \cite{lee_2023_criticality,zou2023channeling}. 

In this work we set out to understand the effects of measurements on the ground state of the transverse-field Ising model (TFIM) at criticality, which is described at long distances by the Ising CFT \cite{sachdev_quantum_2011}, and to study the entanglement entropy of the postmeasurement quantum states. The structure of these states can be understood by considering the introduction of (1+0)-dimensional defects to the Ising CFT, a problem which has been studied extensively both in and out of equilibrum \cite{bariev_effect_1979,mccoy_two-spin_1980,igloi_1993_inhomogeneous,oshikawa_boundary_1997,igloi_entanglement_2009,eisler_entanglement_2010,igloi_entanglement_2009}.
\begin{figure}[t]
	\includegraphics[width = 0.9\columnwidth]{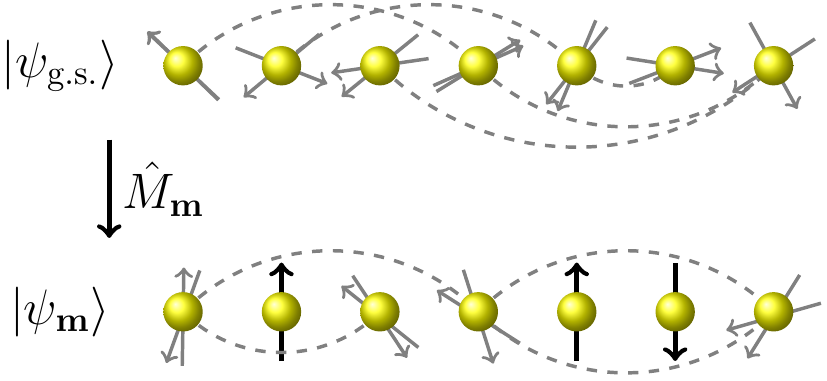}
	\caption{Schematic depiction of the measurement protocol considered in this work. The ground-state $\ket{\psi_{\gs}}$ of the critical transverse-field Ising model (\ref{eq:tfi_ham}) is measured using the measurement operator $\hat{M}_{\vm}$, which is a product of an extensively large set of local projectors. The remaining state $\ket{\psi_{\vm}} \propto \hat{M}_{\vm} \ket{\psi_{\gs}}$ retains nontrivial long-range correlations and entanglement scaling.}
	\label{fig:cartoon}	
\end{figure}

Our study of the TFIM is motivated in part by the aim of observing measurement-induced collective phenomena in experimental quantum simulators. As discussed in Refs.~\cite{gullans_2020_scalable,li_2022_cross,garratt_measurements_2022,lee_2022_decoding,feng2022measurement,li_21_robust}, the effects of large numbers of measurements can, quite generally, be observed when experimental data are complemented by results from a simulation, thereby avoiding the infamous `postselection problem' \cite{skinner_measurement-induced_2019,li_measurement-driven_2019,gullans_2020_scalable,ippoliti_2021_postselection}. This raises the question of which phenomena can be observed in exact simulations. The TFIM is a natural setting to explore this because, within certain measurement schemes, the many-body state can be represented exactly with polynomial computational resources. With this hybrid of quantum and classical simulation in mind, throughout this work we will emphasize the connection between the effects of projective measurements on a lattice model, which bears a close relation to the situation in experiment, and that of defects in a CFT.

We focus on parity-preserving local measurements performed with a finite density in space (see Fig.~\ref{fig:cartoon}). We first consider physical quantities averaged over the ensemble of measurement outcomes, weighting the contributions from different outcomes according to the Born rule. To characterize a postmeasurement state one can compute the expectation value of an observable, but since this object is linear in the postmeasurement density matrix, a naive average over runs of the experiment converts our local measurements into a dephasing channel with strictly local effects \cite{nielsen_quantum_2010}. In order to diagnose the nonlocal effects of measurement on average, it is instead necessary to consider quantities postmeasurement that are nonlinear in the density matrix, such as connected correlation functions and the entanglement entropy of a subregion. Following Ref.~\cite{garratt_measurements_2022} we use a replica approach to study averages of these nonlinear objects. These observables can then be studied at long distances using a replicated Ising CFT, where measurements give rise to an interreplica coupling at a fixed imaginary time (a `spacelike' defect). For averages over parity-preserving measurements this defect is irrelevant under the renormalization group (RG), and consequently long-distance properties of the ensemble of postmeasurement states are not significantly modified relative to the ground state. Remarkably, critical correlations are therefore robust to measurements: the exponents governing power-law correlations between unmeasured qubits, and the prefactor of the logarithmically scaling entanglement entropy, are unchanged.

Following this, we consider the effects of `forced' measurements, which in practice would correspond to postselecting for a particular set of outcomes. We focus our attention on the single most likely measurement outcome for a given set of measurement locations; these generate a marginal defect which, in the Ising CFT, appears as a finite density of energy operators inserted along a line of fixed imaginary time. This type of defect has been analyzed in the context of classical statistical mechanics \cite{mccoy_two-spin_1980,bariev_effect_1979,igloi_1993_inhomogeneous,oshikawa_boundary_1997}, and is known to result in order parameter correlations with a power-law exponent which continuously varies with the strength of the defect. Correspondingly, we numerically observe order parameter correlations in the measured states with a power-law exponent which continuously varies with the density of measurements in space. A defect of this kind at a fixed point in space (i.e. a `timelike' defect) has also been shown to result in a logarithmically scaling half-system entanglement entropy with a continuously varying effective central charge \cite{peschel_entanglement_2005,igloi_entanglement_2009,eisler_entanglement_2010,peschel_exact_2012,brehm_entanglement_2015,roy_entanglement_2022}. We derive a general relation between the effects of forced measurements and of physical defects on the entanglement entropy in CFTs, which suggests a similar logarithmic entanglement entropy of arbitrary subregions of the measured state with the same varying effective central charge. We confirm this relation numerically in TFIMs.

The effects of measurements on the entanglement of critical ground states has previously been investigated in Refs.~\cite{rajabpour2015post,rajabpour2016entanglement,lin_2022_probing}. In Refs.~\cite{rajabpour2015post,rajabpour2016entanglement}, the authors used CFT techniques to compute the entanglement entropy between two unmeasured subsystems after completely disentangling a finite region of space via measurements. Conversely, Ref.~\cite{lin_2022_probing} considers the remaining entanglement after measuring nearly all degrees of freedom of a critical state, such as the ground state of the critical TFIM, as a diagnostic of the state's ``sign structure." In contrast to these previous works, we focus here on measurements performed with a finite density throughout all of space. Separately, we note that in the dynamics of continuously monitored TFIMs and free-fermionic systems ~\cite{chen_emergent_2020,alberton_entanglement_2021,jian_criticality_2022,bao_symmetry_2021,turkeshi_measurement-induced_2021,buchhold_effective_2021,botzung_engineered_2021,boorman_diagnostics_2022,turkeshi_entanglement_2022,piccitto_entanglement_2022,kells_topological_2023,turkeshi_entanglement_2023}, a continuously varying effective central charge has previously been identified in the steady-state entanglement entropy. However, the physical mechanism behind this entanglement scaling is entirely different from that of the present work; in particular, it does not share the same connection to surface critical phenomena employed here.

This paper is organized as follows. First, in Sec.~\ref{sec:model} we describe the lattice model and standard properties of the Ising CFT. In Sec.~\ref{sec:Born}, we study the full ensemble of measurement outcomes by calculating postmeasurement correlation functions weighted according to their Born probabilities. There we also show explicitly the connection between projective measurements on the lattice and a defect in the CFT. In Sec.~\ref{sec:postselection}, we then consider postselecting on the most likely measurement outcomes. Here we derive the connection between the entanglement entropies of subregions in (i) the postmeasurement states and (ii) the ground state of a Hamiltonian featuring defects at fixed points in space. We discuss our results and suggest future directions in Sec.~\ref{sec:discussion}.

\section{Model}\label{sec:model}
Here we review basic definitions and properties of the transverse-field Ising model (TFIM). We consider a one-dimensional system of $N$ qubits with periodic boundary conditions, with the Hamiltonian
\begin{equation}
\label{eq:tfi_ham}
	H = -J \sum_{j = 1}^N \Big\{ Z_j Z_{j+1} + g X_j \Big\} ,
\end{equation}
where $Z_j$ and $X_j$ are the standard Pauli matrices acting on the $j$th qubit, and $Z_{N+1} = Z_1$. We fix $J = 1$ throughout. The TFIM exhibits a quantum phase transition between paramagnetic and ferromagnetic ground states for $g > 1$ and $g < 1$ respectively \cite{sachdev_quantum_2011,fradkin_field_2013}; the critical point at $g = 1$, which is our primary focus, features a gapless spectrum and is described at long distances by the Ising conformal field theory (CFT) \cite{di_francesco_conformal_1997}. 

For both analytical convenience and numerical tractability, it is useful to rewrite $H$ as a model of Majorana fermions using the Jordan-Wigner transformation \cite{sachdev_quantum_2011}. Defining the Majorana fermion operators $\gamma_{2j-1} = \qty[ \prod_{i < j} X_i ] Z_j$ and $\gamma_{2j} = \qty[ \prod_{i < j} X_i ] Y_j$, which satisfy $\acomm{\gamma_i}{\gamma_j} = 2\delta_{ij}$, $H$ takes the form
\begin{equation}
\label{eq:ham_JW}
	H = -iJ \sum_{j = 1}^N \Big\{ g \gamma_{2j-1} \gamma_{2j} + \gamma_{2j} \gamma_{2j+1} \Big\} ,
\end{equation}
where we define $\gamma_{2N+1} = - \gamma_1 \Pi$, with $\Pi = \prod_j X_j$ the total parity. The ground state $\ket{\psi_{\gs}}$ of $H$ lies in the even-parity sector, $\Pi = +1$ \cite{schultz_two-dimensional_1964}; so long as we restrict our attention to parity-even observables, we may set $\Pi = +1$ throughout. The model $H$ is then a quadratic free-fermion Hamiltonian and is amenable to efficient numerical simulation using covariance matrix techniques \cite{terhal_classical_2002,knill_fermionic_2001,bravyi_lagrangian_2004}, as discussed in Appendix \ref{app:numerics}.

We will be interested in the effects of measurements on the critical $g = 1$ ground state $\ket{\psi_{\gs}}$ of $H$. Since our focus is on universal long-distance properties of the postmeasurement states, it will often be useful to work with the Ising CFT \cite{di_francesco_conformal_1997}, which can be obtained from the continuum limit of $H$. In the scaling limit $g \to 1$, long-distance correlation functions can be obtained from the Euclidean action
\begin{equation}
\label{eq:ising_CFT}
	\mathcal{S}_0[\psi] = \frac{1}{2} \int \dd{\tau} \dd{x} \psi^T (\partial_{\tau} - i \sigma^x \partial_x + m \sigma^y)\psi ,
\end{equation}
where $x$ is the spatial coordinate and $\tau$ is the imaginary time. Here $\psi = [\psi_1, \psi_2]^T$ is a two-component Grassmann field, with $\psi_1(x)$ and $\psi_2(x)$ reproducing correlations of $\gamma_{2j-1}$ and $\gamma_{2j}$ respectively, and $\sigma^x$ and $\sigma^y$ are the Pauli matrices acting on the two components of $\psi$. The Ising CFT is obtained upon setting $m \propto g - 1$ to zero. It is useful to note that $\psi$ has scaling dimension $[\psi] = [x^{-1/2}] = 1/2$. We provide more details in Appendix \ref{app:ising_CFT} on the correspondence between the lattice model and the continuum field theory.

Throughout this work, we primarily focus on three observables: namely, the order parameter correlation function $C(r)$, the connected energy density correlation function $G(r)$, and the entanglement entropy $S(r)$ of a contiguous subregion $A = [0:r)$ of $r$ sites. In the ground state, these are  \cite{di_francesco_conformal_1997,calabrese_entanglement_2009}
\begin{equation}
	\begin{split}
		& C_{\gs}(r) = \expval{Z_0 Z_r}_{\gs} \sim r^{-1/4}, \\
		& G_{\gs}(r) = \expval{X_0 X_r}_{\gs} - \expval{X_0}_{\gs} \expval{X_r}_{\gs} \sim r^{-2}, \\
		& S_{\gs}(r) = -\tr \rho^A_{\gs} \log \rho^A_{\gs} \sim \frac{1}{6} \log r  + b_0,
	\end{split}
 \label{eq:obs}
\end{equation}
where $\expval{\cdot}_{\gs} = \bra{\psi_{\gs}} \cdot \ket{\psi_{\gs}}$ denotes ground-state correlations at the critical point, $\rho^A_{\gs} = \tr_{A^c} \dyad{\psi_{\gs}}$ is the reduced density matrix of subsystem $A$, and $\sim$ indicates the asymptotic scaling behavior of these three observables for large $r$. In the first two equations we have omitted a nonuniversal constant prefactor; in the last equation, the coefficient $1/6$ corresponds to a central charge $c = 1/2$, while $b_0$ is a nonuniversal constant \cite{calabrese_entanglement_2009}.

\section{Born Ensemble Projective Measurements}
\label{sec:Born}
Due to its algebraic correlations and long-range entanglement, local measurements performed on the ground state $\ket{\psi_{\gs}}$ of the critical TFIM can potentially exhibit highly nonlocal effects \cite{garratt_measurements_2022}. To determine the effects of projective measurements on the ground state, we randomly perform a projective measurement of $X_j$ at each site $j$ with probability $p$, with measurement outcomes sampled according to the Born rule. The postmeasurement states remain nontrivial on the $\sim (1-p)N$ unmeasured qubits, and in this section we aim to characterize the average long-distance behavior of correlations and entanglement in the ensemble of such measured states.

Our protocol is conveniently described using a measurement operator $\hat{M}_{\vm} = \prod_{j = 1}^N \hat{M}_{m_j, j}$, which is a product of the local measurement operators
\begin{equation}
	\hat{M}_{0, j} = \sqrt{1-p}, \quad \hat{M}_{\pm 1, j} = \sqrt{p} \frac{1 \pm X_j}{2} .
\end{equation}
Here $m_j = 0$ corresponds to not performing a measurement on site $j$, while $m_j = \pm 1$ corresponds to a projective measurement of $X_j$ with result $\pm 1$. Naturally, the full set of measurement operators satisfies the probability-conserving condition $\sum_{\vm} \hat{M}_{\vm}^2 = 1$ (in other words, the set of all $\hat{M}^2_{\vm}$ constitute a positive operator-valued measure \cite{nielsen_quantum_2010}). The measurement outcome $\vm$ occurs with Born probability $p_{\vm}$ and results in the postmeasurement state $\ket{\psi_{\vm}}$, where 
\begin{equation}
	\ket{\psi_{\vm}} = \frac{\hat{M}_{\vm} \ket{\psi_{\gs}}}{\sqrt{\expval*{\hat{M}_{\vm}^2}_{\gs}}}, \quad p_{\vm} = \expval*{\hat{M}_{\vm}^2}_{\gs} .
\end{equation}
In Appendix \ref{app:ZZ} we additionally discuss projective measurements of $Z_j Z_{j+1}$ for each bond. The results are qualitatively similar to the case of $X_j$ measurements discussed here; in the continuum limit, both operators are given to leading order by the Ising CFT energy operator \cite{zou_conformal_2020}.

We would like to determine the typical behavior of long-range correlations in the states $\ket{\psi_{\vm}}$. As has been elaborated elsewhere \cite{garratt_measurements_2022}, although a given set of measurements can have nonlocal effects on the ground state, the averaged behavior of linear observables $\overline{\expval{O}_{\vm}} = \sum_{\vm} p_{\vm} \bra{\psi_{\vm}} O \ket{\psi_{\vm}}$ is identical to the behavior of observables following a series of local quantum channels. Since local quantum channels can exhibit only local effects on the ground state, the nonlocality of measurements is hidden from these averages.

Instead, we focus on the measurement-averaged behavior of observables which are nonlinear in the density matrix $\rho_{\vm} = \dyad{\psi_{\vm}}$: namely, the \textit{squared} order parameter correlation function $C^2_{\vm}(r)$, as well as the connected energy density correlation function $G_{\vm}(r)$ and the entanglement entropy $S_{\vm}(r)$, the latter two of which are already nonlinear observables. Here, the subscripts $\vm$ indicate that these observables are computed with respect to the postmeasurement state $\ket{\psi_{\vm}}$, rather than $\ket{\psi_{\gs}}$ as in Eq.~(\ref{eq:obs}). Explicitly,
\begin{equation}
\label{eq:obsm}
	\begin{split}
		& C^2_{\vm}(r) = \expval{Z_0 Z_r}^2_{\vm}, \\
		& G_{\vm}(r) = \expval{X_0 X_r}_{\vm} - \expval{X_0}_{\vm} \expval{X_r}_{\vm}, \\
		& S_{\vm}(r) = -\tr \rho^A_{\vm} \log \rho^A_{\vm}.
	\end{split}
\end{equation}
We now describe how averages of these objects, with weights given by the Born probabilities $p_{\vm}$, can be studied analytically.

\subsection{Replica field theory}
\label{subsec:ensemble_field_theory}
To analyze the average effects of measurements on these nonlinear observables, we develop a replica approach analogous to the one employed in Ref.~\cite{garratt_measurements_2022}. The resulting replica observables are described by a replicated Ising CFT in the continuum limit, and we show that the average effect of projective measurements on the ground state is to couple the replicas together along the $\tau = 0$ axis in Euclidean spacetime. A simple scaling analysis will then suggest that this coupling is irrelevant.

For purposes of illustration, consider the average of $G_{\vm}(r)$. We will comment on $C_{\vm}^2(r)$ and $S_{
\vm}(r)$ at the end of this section. Starting with just the disconnected piece, the average is given by
\begin{equation}
	\begin{split}
		\overline{ \expval{X_0}_{\vm} \expval{X_r}_{\vm} } &= \sum_{\vm} p_{\vm} \expval{X_0}_{\vm} \expval{X_r}_{\vm} \\
		&= \sum_{\vm} \frac{\expval*{X_0 \hat{M}^2_{\vm}}_{\gs}\expval*{X_r \hat{M}^2_{\vm}}_{\gs}}{\expval*{\hat{M}_{\vm}^2}_{\gs}} ,
	\end{split}
\end{equation}
where we have used $\comm*{X_j}{\hat{M}_{\vm}} = 0$. The difficulty in averaging this quantity directly lies in the nontrivial denominator arising from the normalization of $\ket{\psi_{\vm}}$. In order to compute observables of this form, we employ the following replica scheme:
\begin{equation}
\label{eq:replica}
	\begin{split}
		&\overline{ \expval{X_0}_{\vm} \expval{X_r}_{\vm} } = \lim_{n \to 1} \frac{\sum_{\vm} p_{\vm}^n \expval{X_0}_{\vm} \expval{X_r}_{\vm}}{\sum_{\vm} p_{\vm}^n} \\
		& \ \quad = \lim_{n \to 1} \frac{\sum_{\vm} \expval*{\hat{M}_{\vm}^2}^{n-2}_{\gs} \expval*{X_0 \hat{M}^2_{\vm}}_{\gs} \expval*{X_r \hat{M}_{\vm}^2}_{\gs}}{\sum_{\vm} \expval*{\hat{M}_{\vm}^2}^n_{\gs}} .
	\end{split}
\end{equation}
In this scheme, we effectively weight each set of measurement outcomes $\vm$ by the alternative probability distribution $p_{\vm}^n / \sum_{\vm'} p_{\vm'}^n$, thereby biasing the distribution towards the most likely measurement outcomes. By writing the product of expectation values as a single expectation value over an $n$-fold replicated Hilbert space, we obtain $\overline{G_{\vm}(r)}$ as the $n \to 1$ replica limit of
\begin{equation}
\label{eq:G_n}
	\overline{G^{(n)}_{\vm}(r)} = \frac{\bra{\psi_{\gs}^{\otimes n}} (X_0^{(0)} X^{(0)}_r - X^{(0)}_0 X^{(1)}_r) \hat{M}_{\avg} \ket{\psi_{\gs}^{\otimes n}}}{\bra{\psi_{\gs}^{\otimes n}} \hat{M}_{\avg} \ket{\psi_{\gs}^{\otimes n}}} .
\end{equation}
Here $X^{(\alpha)}_j$ denotes the $X_j$ operator in replica $\alpha$, while $\hat{M}_{\avg}$ is given by
\begin{align}
		\hat{M}_{\avg} &= \sum_{\vm} [\hat{M}_{\vm}^2]^{\otimes n} \label{eq:M_avg}\\
		&= \prod_{j = 1}^N \qty{ (1 - p)^n + p^n \sum_{m_j = \pm 1} \qty( \frac{1 + m_j X_j}{2} )^{\otimes n}  } \notag \\
		&\propto \prod_{j = 1}^N \qty{ 1 + \mu \sum_{r = 1}^{\lfloor n / 2 \rfloor} \sum_{1 \leq \alpha_1 < \ldots < \alpha_{2r} \leq n} \! \! \! \! \! \! \! \! X^{(\alpha_1)}_j \ldots X^{(\alpha_{2r})}_j } \notag,
\end{align}
where $\mu = [1 + 2^{n-1}(p^{-1} -1)^n]^{-1}$ is a monotonic function of $p$, and we have neglected an overall constant which cancels between the numerator and denominator. Acting on $\ket{\psi_{\gs}^{\otimes n}}$, $\hat{M}_{\avg}$ has the effect of weakly locking the multiple replicas together by favoring spin configurations in which $X^{(1)}_j = \ldots = X^{(n)}_j$.

As in Ref.~\cite{garratt_measurements_2022}, we now interpret the insertion of $\hat{M}_{\avg}$ as a spacelike defect in Euclidean spacetime. Towards this end, we rewrite both the numerator and denominator of Eq.~(\ref{eq:G_n}) using an imaginary-time path integral of Majorana fermions, and we take a continuum limit; technical details are contained in Appendix~\ref{app:cont_limit}. The denominator of Eq.~(\ref{eq:G_n}) is then given by the partition function $\mathcal{Z}^{(n)}_M$ of a multi-replica Ising field theory with an inter-replica coupling along the $\tau = 0$ line, defined by:
\begin{equation}
\label{eq:M_avg_partition_fn}
	\begin{split}
	&\mathcal{Z}^{(n)}_M = \int \prod_{\alpha = 1}^n D\psi^{(\alpha)} \, e^{-\sum_{\alpha = 1}^n \mathcal{S}_0[\psi^{(\alpha)}] - \mathcal{S}^{(n)}_M[\{\psi^{(\alpha)} \}]} ,
	\end{split}
\end{equation}
where $\mathcal{S}^{(n)}_M[\{\psi^{(\alpha)}\}]$ gives the coupling between replicas due to measurements:
\begin{equation}
	\label{eq:S_M}
	\begin{split}
		\mathcal{S}_M^{(n)} = - \mu \sum_{\alpha < \beta}  \int \dd{x} (\psi^T \sigma^y \psi)^{(\alpha)} (\psi^T \sigma^y \psi)^{(\beta)} + \ldots ,
	\end{split}
\end{equation}
where the ellipsis denotes four-replica terms and higher, which are less relevant than the two-replica term written explicitly.  The numerator of Eq.~(\ref{eq:G_n}) is given by a multi-replica correlation function having the same action. Note that the fields in $\mathcal{S}_M^{(n)}$ are evaluated strictly at $\tau = 0$. By dimensional analysis, one immediately finds that $\mu$ has dimension $-1$, and is therefore irrelevant. Furthermore, we show in Appendix \ref{app:cont_limit} that higher-order corrections in the perturbative RG cannot generate relevant or marginal terms; more precisely, we show that any possible marginal terms generated by the perturbative RG are inconsequential to observables in the $n \to 1$ replica limit.

\subsection{Correlation Functions}
Having developed a field-theoretical framework for analyzing the effects of measurements on the TFIM ground state, we now discuss the consequences for the nonlinear observables of Eq.~(\ref{eq:obsm}). In the previous section we showed that the average effect of $X_j$ measurements on the correlation function $\overline{G_{\vm}(r)}$, with outcomes sampled according to the Born rule, is to contribute an irrelevant defect-like perturbation to the replicated Ising CFT. We therefore expect $\overline{G_{\vm}(r)}$ to exhibit the same asymptotic scaling as in the unmeasured ground state. Specifically, we expect
\begin{equation}
    \overline{G_{\vm}(r)} \sim r^{-2} \quad (r \gg 1).
\end{equation}

On the other hand, the preceding analysis does not immediately apply to $\overline{C_{\vm}^2(r)}$, since $Z_j$ does not commute with the measurement operator $\hat{M}_{\vm}$ whenever site $j$ is measured. Instead, it is useful to note that both $G_{\vm}(r)$ and $C_{\vm}(r)$ vanish for every measurement realization in which either site $0$ or site $r$ is measured. We can therefore freely replace our measurement averages in both quantities with a restricted  ensemble in which sites $0$ and $r$ are unmeasured. The resulting average measurement operator in this case then commutes with both $X$ and $Z$ observables, and the above mapping follows identically for both cases. We elaborate this discussion in more detail in Appendix~\ref{app:noncommuting}, where we show explicitly that $\overline{C_{\vm}^2(r)}$ is given at long distances by
\begin{equation}
\label{eq:C2_n}
	\overline{C^2_{\vm}(r)} \sim \lim_{n \to 1} \frac{\bra{\psi_{\gs}^{\otimes n}} Z^{(0)}_0 Z^{(1)}_0 Z^{(0)}_r Z^{(1)}_r \hat{M}_{\avg} \ket{\psi_{\gs}^{\otimes n}} }{\bra{\psi_{\gs}^{\otimes n}} \hat{M}_{\avg} \ket{\psi_{\gs}^{\otimes n}} } .
\end{equation}
Whereas (\ref{eq:G_n}) is exact, Eq.~(\ref{eq:C2_n}) is expected to hold asymptotically at long distances. We may now immediately apply the analysis of the preceding section: since the contribution (\ref{eq:S_M}) to the action due to measurements is irrelevant, we again expect $\overline{C_{\vm}^2(r)}$ to asymptotically recover its ground-state scaling at long distances:
\begin{equation}
    \overline{C_{\vm}^2(r)} \sim r^{-1/2} \quad (r \gg 1).
\end{equation}

\begin{figure}[t]
    \includegraphics[width = \columnwidth]{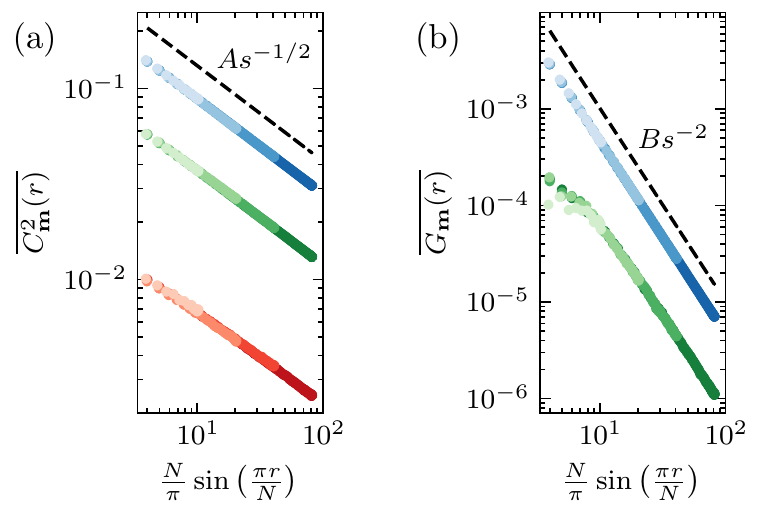}
    \caption{Ensemble-averaged correlation functions (a) $\overline{C^2_{\vm}(r)}$ and (b) $\overline{G_{\vm}(r)}$ as defined in Eq.~(\ref{eq:obsm}), for measurement probabilities $p = 0.2$ (blue), $0.5$ (green), and $0.8$ (red), and for system sizes $N = 32$, $64$, $128$, and $256$ (light to dark). Data are plotted as a function of $s = \frac{N}{\pi} \sin ( \frac{\pi r}{N} )$ to achieve scaling collapse of the various system sizes. Dotted lines depict the behavior in the unmeasured system. Both correlation functions exhibit excellent scaling collapses with the power law exponents of the unmeasured system at sufficiently large distances.}
    \label{fig:born_corr}
\end{figure}

\begin{figure}[t]
    \includegraphics[width = \columnwidth]{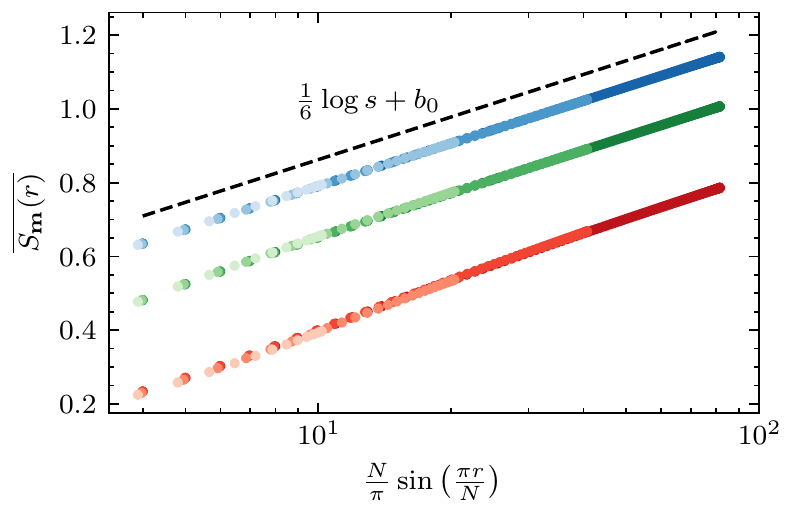}
    \caption{Ensemble-averaged entanglement entropy $\overline{S_{\vm}(r)}$ for a contiguous subregion $[0:r)$ of $r$ sites, for measurement probabilities $p = 0.2$ (blue), $0.5$ (green), and $0.8$ (red), and for system sizes $N = 32$, $64$, $128$, and $256$ (light to dark). Data are plotted as a function of $s = \frac{N}{\pi} \sin (\frac{\pi r}{N})$ to achieve scaling collapse of the various system sizes. The dotted line depicts the behavior of the unmeasured system. For each measurement probability, the entanglement entropy exhibits an excellent scaling collapse with logarithmic scaling corresponding to the central charge $c = 1/2$ of the unmeasured system at sufficiently large distances.}
    \label{fig:born_EE}
\end{figure}

We now numerically verify these analytical predictions. A crucial benefit of focusing on parity-preserving projective measurements of the TFIM is that our analytical predictions can be confirmed using large-scale free-fermion numerics \cite{terhal_classical_2002,knill_fermionic_2001,bravyi_lagrangian_2004}. We provide explicit details of our numerical approach in Appendix \ref{app:numerics}; in short, since arbitrary $k$-point correlations of the quadratic Hamiltonian (\ref{eq:ham_JW}) can be obtained using Wick's theorem, the full physical content of the state $\ket{\psi_{\gs}}$ is contained in the $N(2N-1)$ entries of the covariance matrix $G_{ij} = \expval{i \gamma_i \gamma_j}_{\gs} - i\delta_{ij}$, rather than in $2^N$ complex amplitudes (as would be the case in a generic nonintegrable system).

Figure~\ref{fig:born_corr} depicts the ensemble-averaged correlation functions $\overline{G_{\vm}(r)}$ and $\overline{C^2_{\vm}(r)}$ for several measurement probabilities and system sizes, computed numerically using Monte-Carlo sampling of both the measurement locations and outcomes. As demonstrated in Appendix~\ref{app:FSS}, conformal invariance predicts that both of these correlation functions depend on the single parameter
\begin{equation}
\label{eq:s}
    s = \frac{N}{\pi} \sin \qty( \frac{\pi r}{N} ) .
\end{equation}
From the analysis of Sec.~\ref{subsec:ensemble_field_theory}, we expect ${\overline{C^2_{\vm}(r)} \sim s^{-2}}$ and ${\overline{G_{\vm}(r)} \sim s^{-1/2}}$ at sufficiently large values of $s$. Figure ~\ref{fig:born_corr} supports this conclusion, with excellent finite-size scaling collapses of both correlation functions. 

Interestingly, $\overline{C^2_{\vm}(r)}$ does not exhibit any pronounced crossover behavior at any measurement probability: even at $p = 0.8$, measurements reduce the power-law prefactor without altering the $s^{-1/2}$ scaling. In contrast, $\overline{G_{\vm}(r)}$ exhibits stronger crossover behavior at short distances, and it would be interesting to understand the origin of this effect.

Note that we have omitted the $p = 0.8$ curve in $\overline{G_{\vm}(r)}$ since this exhibits strong finite-size effects. We nevertheless expect that the $s^{-2}$ decay observed at smaller measurement probabilities will be recovered at sufficiently large values of $s$.

\subsection{Entanglement Entropy}
\label{subsec:born_EE}
Finally, we address the average behavior of the entanglement entropy by noting that it can be obtained via the replica limit \cite{bao_theory_2020}
\begin{equation}
	\overline{S_{\vm}(r)} = \lim_{n \to 1} \frac{1}{1-n} \log \qty{ \frac{\sum_{\vm} p^n_{\vm} \tr [(\rho^A_{\vm})^n]}{\sum_{\vm} p_{\vm}^n } } ,
\end{equation}
where $\rho^A_{\vm} = \tr_{A^c} [ \hat{M}_{\vm} \dyad{\psi_{\gs}} \hat{M}_{\vm} ] / p_{\vm}$. Following Refs.~\cite{holzhey_geometric_1994,calabrese_entanglement_2009}, the numerator within the logarithm can be understood as the partition function of the same model defined on an $n$-sheeted Riemann surface with a branch cut running from $(\tau,x) = (0,0)$ to $(\tau, x) = (0,r)$. The impurity (\ref{eq:S_M}) due to measurements, which couples fields between sheets of the Riemann surface, can be taken to lie at $\tau = 0^-$ just below the branch cut. Given that the impurity is irrelevant, we expect that the asymptotic logarithmic scaling $\frac{1}{6} \log r$ will be recovered at sufficiently large $r$, up to a renormalization of the nonuniversal constant $b_0$.

Figure~\ref{fig:born_EE} depicts the ensemble-averaged entanglement entropy $\overline{S_{\vm}(r)}$ for several measurement probabilities and system sizes, again plotted as a function of $s$. In the unmeasured system, $S(r) \sim \frac{1}{6} \log s + b_1(p)$ for a $r$-independent constant $b_1(p)$ which decreases with $p$. Remarkably, we see from Fig.~\ref{fig:born_EE} that the logarithmic scaling of the entanglement entropy, and the prefactor $1/6$, are unaffected by measurements, even at large measurement strengths. 

Here we have shown that, on average, the correlations characteristic of the critical TFIM are robust to parity-preserving measurements. However, as we show in the next section, for the most likely measurement outcomes correlations are altered radically relative to the ground state.

\section{Forced Projective Measurements}
\label{sec:postselection}
In the previous section, we found that parity-preserving projective measurements sampled according to the Born rule fail to alter the asymptotic scaling of correlations or entanglement of the critical TFIM ground state $\ket{\psi_{\gs}}$. It is natural to ask whether an alternative measurement scheme can exhibit larger effects on these observables. In previous work \cite{garratt_measurements_2022} we found that postselected ``no-click'' density measurements performed uniformly throughout a Luttinger liquid are relevant (irrelevant) for Luttinger parameters $K < 1$ ($K > 1$). Motivated by this result, we now consider postselecting on a particular set of measurement outcomes in the TFIM. Since the $K = 1$ Luttinger liquid is related to two copies of the Ising CFT via bosonization \cite{shankar_quantum_2017}, it is particularly interesting to consider postselected measurements in the TFIM: if a finite density of postselected projective measurements are believed to behave qualitatively similarly to a uniform strength of weak measurements, then the postselected measurements are expected to contribute marginally.

Our measurement scheme is as follows: we again perform $X_j$ measurements on each site with probability $p$, but we now force the outcome $X_j = +1$ for each measured site. This outcome corresponds to qubits aligned along the local transverse fields, and is the single most likely outcome given the chosen measurement locations. It is convenient to describe such a measurement protocol with a measurement operator $\hat{K}_{\vk} = \prod_{j = 1}^N\hat{K}_{k_j, j}$ given by a product of local measurement operators $\hat{K}_{k_j, j}$, defined here as
\begin{equation}
\label{eq:postselect_K}
 	\hat{K}_{0,j} = 1, \quad \hat{K}_{1,j} = \frac{1+X_j}{2} .
\end{equation}
Unlike the previous measurement scheme, where $m_j = 0,\pm 1$ is sampled according to the Born rule, in this scheme we simply choose to measure site $j$ ($k_j = 1$) or leave site $j$ unmeasured ($k_j = 0$) with probabilities $p$ and $1-p$, respectively. The state $\ket{\psi_{\vk}}$ is obtained with probability $p_{\vk}$, where
\begin{equation}
\label{eq:postselect_psi}
	\ket{\psi_{\vk}} = \frac{\hat{K}_{\vk} \ket{\psi_{\gs}}}{\sqrt{\expval*{\hat{K}_{\vk}}_{\gs}}}, \quad p_{\vk} = p^{\abs*{\vk}}(1-p)^{N - \abs*{\vk}} ,
\end{equation}
where $\abs{\vk} = \sum_{j=1}^N k_j$ is the number of measurements performed. Our focus here will be on correlation functions $G_{\vk}(r)$ and $C_{\vk}(r)$, as well as the entanglement entropy $S_{\vk}(r)$, in the postmeasurement states $\ket{\psi_{\vk}}$. These correlation functions are defined in analogy with $G_{\vm}(r)$, $C_{\vm}(r)$, and $S_{\vm}(r)$ [see Eq.~\eqref{eq:obsm}], respectively, differing only in the fact that they are evaluated for states $\ket{\psi_{\vk}}$ rather than $\ket{\psi_{\vm}}$. We now show how averages of these objects with respect to $p_{\vk}$ can be studied analytically.

\subsection{Replica Field Theory}
\label{sec:postselect_replicas}

Since measurement outcomes are not sampled according to the Born rule, even observables linear in the postmeasurement density matrix $\rho_{\vk} = \ket{\psi_{\vk}}\bra{\psi_{\vk}}$ can be sensitive to the nonlocal effects of measurements. Due to the nontrivial denominator appearing in correlation functions arising from the normalization of $\ket{\psi_{\vk}}$, we nevertheless require a replica approach to average over measurement locations. In the following we show how the forced projective measurements appear in the field theory [see Eq.~\eqref{eq:S_K}]. Taking $G_{\vk}(r)$ again as an example, the average over disorder realizations is given by
\begin{align}
		&\overline{G_{\vk}(r)} = \sum_{\vk} p_{\vk} \qty[ \expval{X_0 X_r}_{\vk} - \expval{X_0}_{\vk} \expval{X_r}_{\vk} ] \\
		&\ = \sum_{\vk} p_{\vk} \qty[ \frac{\expval*{X_0 X_r \hat{K}_{\vk}}_{\gs}}{\expval*{\hat{K}_{\vk}}_{\gs}} - \frac{ \expval*{X_0 \hat{K}_{\vk}}_{\gs} \expval*{X_r \hat{K}_{\vk}}_{\gs} }{\expval*{\hat{K}_{\vk}}^2_{\gs}} ] \notag,
\end{align}
where we have used $\comm*{X_j}{\hat{K}_{\vk}} = 0$. Due to the absence of Born factors in the sampling probabilities $p_{\vk}$, we can employ a replica approach directly analogous to those used in the classical statistical mechanics of disordered systems \cite{nishimori_statistical_2001}. We obtain $\overline{G_{\vk}(r)}$ from the $n \to 0$ limit of the replica quantity $\overline{G^{(n)}_{\vk}(r)}$, defined as
\begin{equation}
\label{eq:G_n_postselect}
	\begin{split}
		\overline{G^{(n)}_{\vk}(r)} &= \frac{1}{\sum_{\vk} p_{\vk} \expval*{\hat{K}_{\vk}}_{\gs}^n} \sum_{\vk} p_{\vk} \Big[ \expval*{\hat{K}_{\vk}}_{\gs}^{n-1} \expval*{X_0 X_r \hat{K}_{\vk}}_{\gs} \\
		& \quad \quad \quad \quad - \expval*{\hat{K}_{\vk}}^{n-2}_{\gs} \expval*{X_0 \hat{K}_{\vk}}_{\gs} \expval*{X_r \hat{K}_{\vk}}_{\gs} \Big] \\
		&= \frac{\bra{\psi_{\gs}^{\otimes n}} (X^{(0)}_0 X^{(0)}_r - X^{(0)}_0 X^{(1)}_r) \hat{K}_{\avg} \ket{\psi_{\gs}^{\otimes n}} }{\bra{\psi_{\gs}^{\otimes n}} \hat{K}_{\avg} \ket{\psi_{\gs}^{\otimes n}}} .
	\end{split}
\end{equation}
As in the Born ensemble case, we have written the product of expectation values as an expectation value over an $n$-fold replicated ground state $\ket{\psi_{\gs}^{\otimes n}}$. The average measurement operator $\hat{K}_{\avg}$ is given by
\begin{align}
	\hat{K}_{\avg} &= \sum_{\vk} p_{\vk} [\hat{K}_{\vk}]^{\otimes n} \label{eq:K_avg} \\
		&= \prod_{j = 1}^N \qty{ (1-p) + p \qty(\frac{1 + X_j}{2})^{\otimes n} } \notag \\
		&\propto \prod_{j = 1}^N \qty{ 1 + \nu \sum_{r = 1}^n \sum_{1 \leq \alpha_1 < \ldots < \alpha_r \leq n} X^{(\alpha_1)}_j \ldots X^{(\alpha_r)}_j } ,\notag
\end{align}
where $\nu = [1 + 2^n(p^{-1} - 1)]^{-1}$ is a monotonic function of $p$. The average effect of forced measurements on the multi-replica ground-state $\ket{\psi_{\gs}^{\otimes n}}$ is once again to weakly lock the replicas together. However, unlike $\hat{M}_{\avg}$, $\hat{K}_{\avg}$ contains terms with an odd number of $X^{(\alpha)}_j$ replicas. These terms bias towards amplitudes for which $X^{(\alpha)}_j = +1$, as expected from the measurement scheme. 

We can again interpret the insertion of $\hat{K}_{\avg}$ as a defect along the $\tau = 0$ line in Euclidean spacetime. The denominator of (\ref{eq:G_n_postselect}) is given by a partition function $\mathcal{Z}^{(n)}_K$, analogous to that of Eq.~(\ref{eq:M_avg_partition_fn}):
\begin{equation}
	\begin{split}
	\mathcal{Z}^{(n)}_K = \int \prod_{\alpha = 1}^n D\psi^{(\alpha)} \, e^{-\sum_{\alpha = 1}^n \mathcal{S}_0[\psi^{(\alpha)}] - \mathcal{S}^{(n)}_K[\{\psi^{(\alpha)} \}]} ,
	\end{split}
\end{equation}
where $\mathcal{S}^{(n)}_K[\{\psi^{(\alpha)}\}]$ is given by
\begin{equation}
	\mathcal{S}^{(n)}_K= \nu \sum_{\alpha = 1}^n \int \dd{x} (\psi^T \sigma^y \psi)^{(\alpha)} + \ldots ,
\label{eq:S_K}
\end{equation}
and the ellipsis denotes irrelevant terms, including those listed explicitly in Eq.~(\ref{eq:S_M}). The translation-invariant perturbation in Eq.~\eqref{eq:S_K} represents the dominant averaged effect of the forced projective measurements. In fact, this perturbation also arises from a forced \textit{weak} measurement scheme that is manifestly translation invariant \cite{garratt_measurements_2022}, as we discuss in Appendix~\ref{app_noclick}.

Notably, this perturbation is \textit{exactly} marginal, and as we show below it has interesting consequences for the behavior of both correlation functions and the entanglement entropy. Since the leading term decouples across replicas, it will in fact be sufficient to focus on the single replica theory in discussions of long-distance properties. We then arrive at a continuum theory identical to one arising in studies of lines of weakened bonds in two-dimensional classical Ising models \cite{igloi_1993_inhomogeneous}, and therefore of local defects in the Hamiltonians of TFIMs.

\subsection{Correlation Functions}
First we discuss the effects of the perturbation \eqref{eq:S_K} on the few-body correlation functions $C_{\vk}$ and $G_{\vk}$. Exact calculations in two-dimensional classical Ising models and based on field-theoretic techniques \cite{igloi_1993_inhomogeneous} have shown that energy density correlators along the defect line retain the same scaling form as in the homogeneous Ising CFT. This can be understood simply by noting that the quadratic perturbation (\ref{eq:S_K}) does not modify the scaling dimension of the fermion operators $\psi(\tau,x)$, and therefore cannot modify the scaling form of observables which are local in the fermion representation. We therefore once again expect at sufficiently long distances
\begin{equation}
\overline{G_{\vk}(r)} \sim r^{-2} \quad (r \gg 1) ,
\end{equation}
as in the unmeasured case.

The order parameter correlations $C_{\vk}(r)$, on the other hand, are nonlocal in the fermionic representation, and can be strongly modified by the defect (\ref{eq:S_K}). In particular, Refs.~\cite{bariev_effect_1979,mccoy_two-spin_1980} demonstrated that order parameter correlations along the defect line of a classical Ising model exhibit nonuniversal scaling with a continuously varying exponent. We therefore similarly expect $\overline{C_{\vk}(r)}$ to exhibit a continuously varying power law:
\begin{equation}
    \overline{C_{\vk}(r)} \sim r^{-2\Delta(p)} \quad (r \gg 1) ,
\end{equation}
where $\Delta(p)$ defines the power-law scaling of $\overline{C_{\vk}(r)}$, with $\Delta(0) = 1/8$. Heuristically, the asymptotic limit of $\Delta(p)$ as $p \to 1$ can be inferred by writing $C_{\vk}(r)$ as
\begin{equation}
    \expval{Z_j Z_{j+r}}_{\vk} = \expval{\gamma_{2j-1} \qty[ \prod_{i = j}^{j+r-1} X_i ] \gamma_{2j+2r-1} }_{\vk} .
\end{equation}
At large values of $p$, $X_j \ket{\psi_{\vk}} = + \ket{\psi_{\vk}}$ for a large fraction of sites $j$. The leading contribution to $C_{\vk}(r)$ then comes from $\expval{i \gamma_{2j-1} \gamma_{2j+2r-1}}_{\gs} \sim r^{-1}$, and we therefore expect $\lim_{p \to 1} \Delta(p) = 1/2$.

\begin{figure}[t]
    \includegraphics[width = \columnwidth]{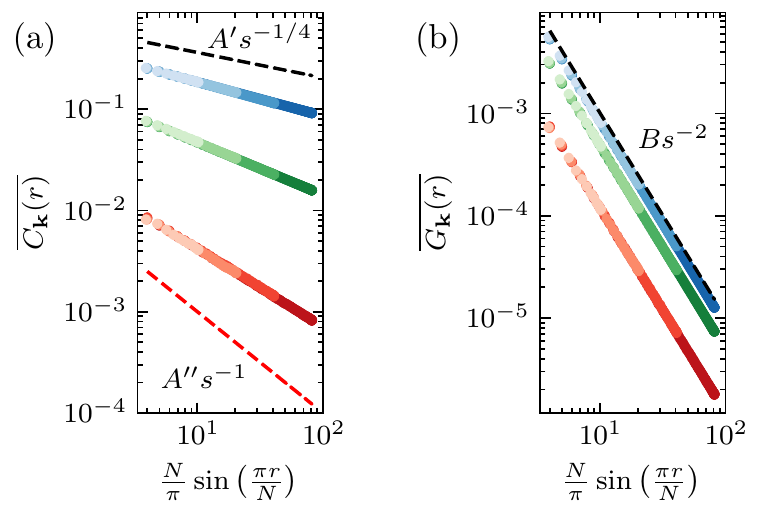}
    \caption{Ensemble-averaged correlation functions (a) $\overline{C_{\vk}(r)}$ and (b) $\overline{G_{\vk}(r)}$ in the forced-measurement ensemble, for measurement probabilities $p = 0.2$ (blue), $0.5$ (green), and $0.8$ (red), and for system sizes $N = 32$, $64$, $128$, and $256$ (light to dark). Data are plotted as a function of $s = \frac{N}{\pi} \sin ( \frac{\pi r}{N} )$ to achieve scaling collapse of the various system sizes. Black dotted lines depict the behavior in the unmeasured system, while the red dotted line depicts the asymptotic power law of $\overline{C_{\vk}(r)}$ as $p \to 1$. While $\overline{G_{\vk}(r)} \sim s^{-2}$ as in the unmeasured system, $\overline{C_{\vk}(r)} \sim s^{-2 \Delta(p)}$ exhibits power-law scaling with a continuously varying exponent $\Delta(p)$. For the measurement probabilities shown, we have $\Delta(0.2) \simeq 0.167$, $\Delta(0.5) \simeq 0.252$, and $\Delta(0.8) \simeq 0.367$.}
    \label{fig:force_corr}
\end{figure}

The analytically predicted behavior of these two correlation functions can again be verified numerically. Figure~\ref{fig:force_corr} depicts the averaged correlation functions $\overline{G_{\vk}(r)}$ and $\overline{C_{\vk}(r)}$ for various measurement probabilities and system sizes, once again plotted as a function of the single parameter $s$ [see Eq.~(\ref{eq:s})]. As in the case of measurements sampled from the Born ensemble, we observe an excellent finite-size scaling collapse of both correlation functions. We observe as predicted that $\overline{G_{\vk}(r)}$ retains its $s^{-2}$ scaling for each measurement probability, while $\overline{C_{\vk}(r)} \sim s^{-2\Delta(p)}$ obtains a continuously varying critical exponent $\Delta(p)$. With increasing $p$, $\Delta(p)$ increases monotonically towards an asymptotic value of $1/2$.

\subsection{Entanglement Entropy}
\label{subsec:force_EE}
\begin{figure*}[t]
    \includegraphics[width = \textwidth]{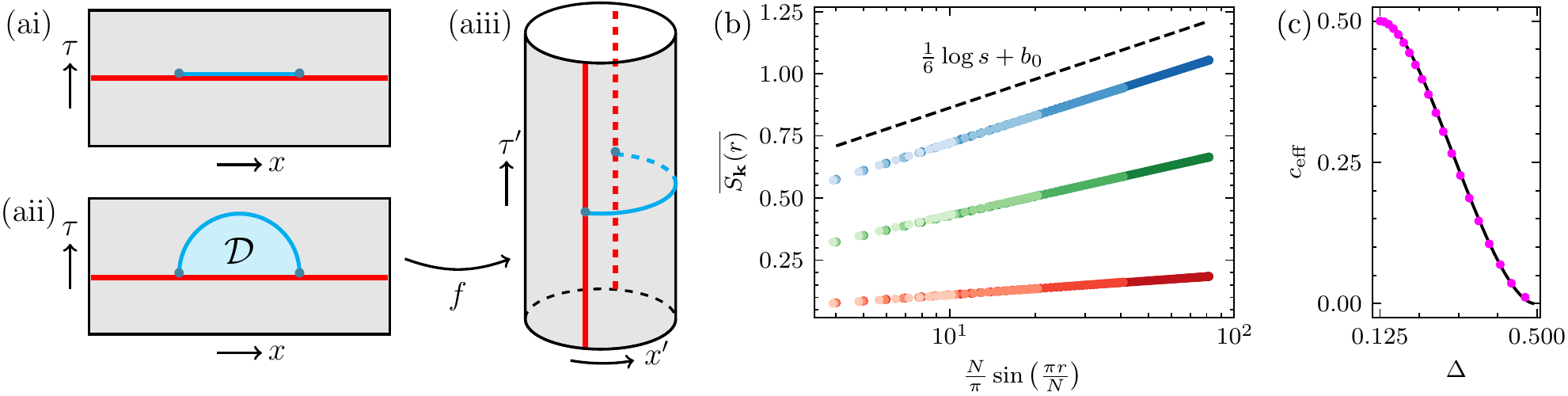}
    \caption{(a) Spacetime diagrams depicting the relation between the average entanglement entropy $\overline{S_{\vk}(r)}$ of the TFIM following forced projective measurements and the entanglement entropy $S_d(N/2)$ of a dual impurity problem. (ai): One sheet of the Riemann surface used to compute the partition function $\mathcal{Z}_n$ in Eq.~(\ref{eq:EE_CFT}). The red line denotes the measurement defect along the $\tau = 0$ line, while the blue line denotes the entanglement branch cut. (aii): By redefining the replica fields $\psi^{(\alpha)}(\tau,x)$ in region $\mathcal{D}$ as in Eq.~(\ref{eq:deform}), the branch cut is continuously deformed from the real axis onto the semicircle. (aiii): using the conformal mapping of Eq.~(\ref{eq:conf_mapping}), the infinite plane is mapped to a cylinder of circumference $L$. The spacelike measurement defect is mapped to two timelike impurities at $x' = 0$ and $x' = L/2$, while the deformed entanglement cut is mapped to a spacelike cut along the $\tau' = 0$ axis. (b) Ensemble-averaged entanglement entropy $\overline{S_{\vk}(r)}$ of a contiguous subregion of $r$ sites in the forced-measurement ensemble, for measurement probabilities $p = 0.2$ (blue), $0.5$ (green), and $0.8$ (red), and for system sizes $N = 32$, $64$, $128$, and $256$ (light to dark). Data are plotted as a function of $s = \frac{N}{\pi} \sin ( \frac{\pi r}{N} )$ to achieve scaling collapse of the various system sizes. The dotted line depicts the behavior in the unmeasured system. We observe $\overline{S_{\vk}(r)} \sim \frac{c_{\eff}(p)}{3} \log s + b_2(p)$ exhibits logarithmic scaling at all measurement probabilities with a continuously decreasing effective central charge $c_{\eff}(p)$. For the measurement probabilities shown, $c_{\eff}(0.2) \simeq 0.478$, $c_{\eff}(0.5) \simeq 0.339$, and $c_{\eff}(0.8) \simeq 0.105$. (c) Numerical comparison between the effective central charge $c_{\eff}(p)$ in the average entanglement entropy $\overline{S_{\vk}(r)}$ in the forced-measurement ensemble, and the effective central charge $c_{\eff,d}(g_d)$ of the half-system entanglement entropy $S_d(N/2)$ of a dual TFIM with defects described by the Hamiltonian (\ref{eq:defect_ham}). Purple dots: effective central charge $c_{\eff}(p)$ for several measurement probabilities $p$ between 0 and 0.95 in increments of 0.05, as a function of the effective scaling dimension $\Delta(p)$ governing the decay of $\overline{C_{\vk}(r)}$. Black curve: effective central charge $c_{\eff,d}(g_d)$ as a function of the scaling dimension $\Delta_d(g_d)$ governing the decay of order parameter correlations $\bra{\psi_d} Z_1 Z_{N/2} \ket{\psi_d}$ between the two impurities.} 
    \label{fig:force_EE}
\end{figure*}

Whereas the correlation functions $\overline{G_{\vk}(r)}$ and $\overline{C_{\vk}(r)}$ have natural interpretations in terms of analogous observables in either classical Ising models with defect lines or the TFIM with an ordinary timelike defect, the entanglement entropy $\overline{S_{\vk}(r)}$ has no immediately obvious analog in either of these models. In this section, we will utilize conformal invariance of the Ising CFT to demonstrate a nontrivial connection between $\overline{S_{\vk}(r)}$ and the entanglement entropy of a model with ordinary timelike defects. In particular, we will show that the average entanglement entropy following forced projective measurements retains its logarithmic scaling, but with an \textit{effective} central charge $c_{\eff}(p)$ which continuously decreases with increasing measurement probability:
\begin{equation}
\label{eq:EE_postselect_ceff}
	\overline{S_{\vk}(r)} \sim \frac{c_{\eff}(p)}{3} \log r + b_2(p)  \quad (r \gg 1).
\end{equation}
Here $c_{\eff}(p)$ is a monotonically decreasing function, with $c_{\eff}(0) = 1/2$ and $c_{\eff}(1) = 0$, and $b_2(p)$ is an $r$-independent contribution that is generically different from $b_1(p)$ in Sec.~\ref{subsec:born_EE}, which we also expect to continuously decrease with increasing measurement probability. We present the basic qualitative argument here, and leave certain technical details for Appendix~\ref{app:forced_entanglement}.

The irrelevance of inter-replica couplings in Eq.~\eqref{eq:S_K} indicates that it is sufficient to work directly at the fixed point [see also Appendix~\ref{app_noclick}]. We therefore consider the entanglement entropy of a contiguous subregion $A$ of length $r$ of the Ising CFT with a measurement defect along the $\tau = 0$ line, with the action
\begin{equation}
\label{eq:defect_FP}
    \mathcal{S}^*[\psi] = \mathcal{S}_0[\psi] + \nu \int \dd{x} \psi^T \sigma^y \psi ,
\end{equation}
where $\mathcal{S}_0$ is given by Eq.~(\ref{eq:ising_CFT}) with $m=0$, and in the latter term $\psi(\tau,x)$ is taken along the line $\tau = 0$. Following Refs.~\cite{holzhey_geometric_1994,calabrese_entanglement_2009}, the entanglement entropy can be computed from the $n \to 1$ limit of the ratio of two partition functions:
\begin{equation}
\label{eq:EE_CFT}
    S^*(r) = \lim_{n \to 1} \frac{1}{1-n} \log \qty{ \frac{\mathcal{Z}_n}{\mathcal{Z}_1^n} } .
\end{equation}
Here $S^*(r)$ denotes the entanglement entropy in the fixed-point model (\ref{eq:defect_FP}), $\mathcal{Z}_1 = \int D\psi \, e^{-\mathcal{S}^*[\psi]}$ is the single-replica partition function, and $\mathcal{Z}_n$ is the partition function of an $n$-fold replicated theory subjected to the boundary conditions
\begin{equation}
\label{eq:replica_bc}
	\psi^{(\alpha)}(\tau = 0^-, x) = \begin{cases}
		\psi^{(\alpha)}(\tau=0^+,x), & x \not\in A \\
		\psi^{(\alpha+1)}(\tau = 0^+,x), & x \in A .
	\end{cases}
\end{equation}
Alternatively, one can consider the $n$ fields $\psi^{(\alpha)}$ as a single field defined on an $n$-sheeted Riemann surface. The Riemann surface has a branch cut along the $\tau = 0^+$ axis, just above the measurement defect, running from $x = -r/2$ to $x = r/2$.

Utilizing conformal invariance, we are free to perform a scaling transformation so as to set $r = 2$. The entanglement branch cut then lies along the $x$-axis with branch points located at $x = \pm 1$, as depicted in Fig.~\ref{fig:force_EE}(ai). We can now continuously deform the branch cut from the real line to the unit semicircle in the upper-half plane, as shown in Fig.~\ref{fig:force_EE}(aii). As a theory defined on an $n$-sheeted Riemann surface, the precise location of the branch cut is unphysical and can be freely deformed, so long as the branch points at $x = \pm 1$ are left unmodified. Equivalently, as a theory of $n$ replicated fields subjected to the boundary conditions (\ref{eq:replica_bc}), the deformation of the branch cut amounts to defining a new set of fields $\tilde{\psi}^{(\alpha)}(\tau,x)$ via
\begin{equation}
\label{eq:deform}
	\tilde{\psi}^{(\alpha)}(\tau,x) = \begin{cases}
		\psi^{(\alpha)}(\tau,x), & (\tau,x) \not\in \mathcal{D} \\
		\psi^{(\alpha - 1)}(\tau,x), & (\tau,x) \in \mathcal{D} ,
	\end{cases}
\end{equation}
where $\mathcal{D}$ is the filled semicircle in the upper-half plane, shown in Fig.~\ref{fig:force_EE}(aii).

By deforming the entanglement cut onto the unit semicircle, we can relate the entanglement entropy in our measurement problem to that of a problem with ordinary timelike defects. Letting $z = x + i \tau$ and $z' = x' + i \tau'$, we use the conformal mapping
\begin{equation}
\label{eq:conf_mapping}
    z \mapsto z' = f(z) = -i \frac{L}{2\pi} \log z
\end{equation}
to map the infinite plane to a cylinder of circumference $L$. The measurement defect maps to \textit{two} timelike defects at locations $x' = 0$ and $x' = L/2$, while the deformed entanglement cut maps to a spacelike entanglement cut from $x' = 0$ to $x' = L/2$, as shown in Fig.~\ref{fig:force_EE}(aiii). We therefore obtain a relation between the average entanglement entropy $S^*(r)$ of the Ising CFT in the presence of forced measurements, and the half-system entanglement entropy $S^*_d(L/2)$ of the Ising CFT on a cylinder with ordinary defects at the entangling boundaries.

Having established this connection, we can now make contact with previous studies of the effects of physical defects on the entanglement entropy \cite{igloi_entanglement_2009,eisler_entanglement_2010,peschel_exact_2012,brehm_entanglement_2015,roy_entanglement_2022}. For the case presented here with exactly marginal defect lines, these works suggest that the entanglement entropy should maintain its logarithmic growth in $L$, but with an effective central charge $c^*_{\eff}(\nu)$ which continuously decreases with increasing defect strength:
\begin{equation}
    S^*_d(L/2) = \frac{c^*_{\eff}(\nu)}{3} \log L + b_d(\nu) .
\end{equation}
Using the transformation properties of correlation functions under conformal transformations, the results of the above sequence of mappings suggest the form (\ref{eq:EE_postselect_ceff}) for the entanglement entropy of a subregion of length $r$ in the original problem with forced measurements. If the microscopic measurement probability $p$ results in a fixed-point defect strength $\nu(p)$, then $c_{\eff}(p) = c^*_{\eff}(\nu(p))$.

We can once again numerically verify the predicted behavior (\ref{eq:EE_postselect_ceff}) for the entanglement entropy following forced projective measurements. Figure ~\ref{fig:force_EE}(b) depicts the average entanglement entropy $\overline{S_{\vk}(r)}$ for various measurement probabilities and system sizes; we again obtain an excellent finite-size scaling collapse by plotting as a function of the parameter $s$ [see Eq.~(\ref{eq:s})]. As predicted, $\overline{S_{\vk}(r)}$ retains its logarithmic scaling at all observed measurement probabilities $p$, with a continuously decreasing effective central charge $c_{\eff}(p)$. 

To verify the proposed connection to the Ising CFT with ordinary timelike defects, we additionally numerically simulate the ground state $\ket{\psi_d}$ of the critical TFIM with two defect transverse fields, for several system sizes $N$. The Hamiltonian is
\begin{equation}
\label{eq:defect_ham}
    H_d = H - g_d \qty[ X_1 + X_{N/2} ] ,
\end{equation}
where $H$ is the TFIM Hamiltonian in Eq.~(\ref{eq:tfi_ham}) with $g = 1$, and $g_d$ gives an enhancement of the transverse field at the defect sites $j = 1$ and $j = N/2$. Using free-fermion numerics [see Appendix~\ref{app:numerics}], we compute order parameter correlations $\bra{\psi_d} Z_1 Z_{N/2} \ket{\psi_d}$ between the two defect sites and the entanglement entropy $S_d(N/2) = -\tr \rho^{A_d}_d \log \rho^{A_d}_d$  of the subregion $A_d = [1:N/2]$ containing $N/2$ sites, including both defect sites. As expected from previous works on the TFIM with defects \cite{igloi_entanglement_2009,eisler_entanglement_2010,peschel_exact_2012,brehm_entanglement_2015,roy_entanglement_2022}, we find
\begin{equation}
    \begin{split}
        &\bra{\psi_d} Z_1 Z_{N/2} \ket{\psi_d} \sim N^{-2\Delta_d(g_d)}, \\
        &S_d(N/2) \sim \frac{c_{\eff,d}(g_d)}{3} \log N + b_3(g_d) .
    \end{split}
\end{equation}
\textit{A priori}, it is difficult to directly compare the effective central charge $c_{\eff,d}(g_d)$ in the defect model (\ref{eq:defect_ham}) with the effective central charge $c_{\eff}(p)$ following forced measurements; although both models are described at long distances by the same Ising CFT with a defect line, there is no simple relation between the microscopic parameters $p$ and $g_d$ and the defect strength $\nu$ at the fixed point. Instead, noting that both the order parameter scaling dimension $\Delta(p)$ and the effective central charge $c_{\eff}(p)$ are controlled by the fixed-point defect strength $\nu$ (and similarly for $\Delta_d(g_d)$ and $c_{\eff,d}(g_d)$), we eliminate $\nu$ altogether by plotting $c_{\eff}(p)$ as a function of $\Delta(p)$ and $c_{\eff,d}(g_d)$ as a function of $\Delta_d(g_d)$. The result is shown in Fig.~\ref{fig:force_EE}(c), with the black line denoting data obtained from the defect model~(\ref{eq:defect_ham}), and with purple dots depicting data obtained from the large-$s$ behavior of $\overline{S_{\vk}(r)}$ and $\overline{C_{\vk}(r)}$ for measurement probabilities $p$ between 0 and $0.95$ in increments of $0.05$. We find a remarkable agreement between the data of the two models, providing strong numerical support for the analytical mapping discussed in this section.

\section{Discussion}
\label{sec:discussion}
Measuring part of a many-body quantum state can give rise to surprising new correlations. In this work we have studied the effects of local measurements on the critical one-dimensional TFIM, a highly entangled system for which exact numerical calculations are possible. Our focus has been on the partial collapse of the ground state that arises from parity-preserving measurements of a finite fraction $\sim p$ of the degrees of freedom. We have shown that, although measuring all degrees of freedom ($p=1$) certainly destroys quantum correlations, if a finite fraction $\sim (1-p)$ remain unmeasured, then the original critical correlations survive on average at long distances. The origin of this robustness can be understood from properties of the Ising CFT. We have developed the replica framework of Ref.~\cite{garratt_measurements_2022} to include the physically realistic case of projective measurements, and in this way we have established a direct link between a microscopic lattice description of measurements of the TFIM and of defects in the Ising CFT. In particular, parity-preserving measurements with outcomes sampled according to the Born rule fail to alter long-distance correlations (for $p<1$) because they correspond to an irrelevant perturbation in the replica theory: While the unperturbed Ising theory in $(1+1)$ dimensions is quadratic in the fermion field $\psi(\tau,x)$, the perturbation is quartic in $\psi(\tau,x)$ and acts only on a $(1+0)$-dimensional surface of fixed imaginary time [see Eq.~(\ref{eq:S_M})].

However, postselecting on certain outcomes of parity-preserving measurements does lead to interesting new correlations. Measuring $X_j$ (or $Z_jZ_{j+1}$, see Appendix~\ref{app:ZZ}) and forcing the most likely outcome $X_j=+1$ generates a marginal perturbation in the field theory (quadratic in fermionic fields rather than quartic). We have shown in Sec.~\ref{sec:postselection} that the exponents governing the postmeasurement power laws vary continuously with the fraction of measured sites. Continuously varying power laws of this kind were identified some time ago in studies of the statistical mechanics of two-dimensional classical Ising models with modified couplings along a line \cite{bariev_effect_1979,mccoy_two-spin_1980,igloi_1993_inhomogeneous}, and the continuum description of these systems is essentially the same as for our measured ground state with forced measurements.

A quantity that is meaningful in the problem we have considered, but which does not arise naturally in classical statistical mechanics, is the entanglement entropy. In addition to modifying correlation functions, we have shown in Fig.~\ref{fig:force_EE}(b) that forced measurements lead to a variation of the effective central charge. A key contribution of this work is to show that the entanglement entropy of a finite subregion in this measurement problem can be mapped, through a conformal transformation, to an entanglement entropy of a system with two physical defects; the latter problem having been the subject of a number of previous studies \cite{peschel_entanglement_2005,igloi_entanglement_2009,eisler_entanglement_2010,peschel_exact_2012,brehm_entanglement_2015,roy_entanglement_2022}. By comparing long-distance properties of lattice models corresponding to the two sides of this duality transformation, we have confirmed numerically that the effective central charges coincide.

It is interesting to compare the entanglement scaling in the present work to that of \textit{dynamically} monitored TFIMs or free fermion models \cite{chen_emergent_2020,alberton_entanglement_2021,jian_criticality_2022,bao_symmetry_2021,turkeshi_measurement-induced_2021}. As noted earlier, previous works have observed logarithically scaling subsystem entanglement entropy in the steady states of continuously monitored free fermion dynamics, with effective central charges tuned by the rate of either Born-averaged or forced measurements. More recently, field-theoretical studies \cite{fava_nonlinear_2023,poboiko_theory_2023} have demonstrated that the monitored dynamics of free fermion chains can be described by a family of nonlinear sigma models; in contrast to prior numerical studies, these works suggest that the late-time subsystem entanglement entropy exhibits area-law scaling or $(\log r)^2$ scaling in complex fermions or Majorana fermions respectively. We emphasize that the physical effects of measurements in these dynamical problems differ strongly from the effects of measurements in the static setting of the present work: whereas a single round of commuting measurements on a critical ground state appears as a defect-like perturbation in Euclidean spacetime, measurements in the monitored dynamics of free fermions appear as a ``bulk" perturbation to a field theory.

An advantage when working with the integrable TFIM is that its ground state, and the effects of parity-preserving measurements, can be described exactly with polynomial computational resources. This has allowed us to verify the above predictions numerically.  While our numerical method relies on the fact the that system is integrable, aspects of our field-theoretic analysis do not. For example, if we introduce to the Hamiltonian an irrelevant integrability-breaking perturbation then we expect that correlations will be modified at short but not at long distances. The change in short-distance correlations could lead to a renormalization of the effective measurement probability, but we nevertheless expect long-distance postmeasurement correlations to decay with the same exponents as in the ground state. We also remark that our results can be applied to certain measurement protocols of tight-binding models (or equivalently XX models), whose Hamiltonians can be expressed as two independent critical Ising models \cite{sachdev_quantum_2011}.

If one moves away from free fermion simulations, it is natural to consider the effects of measurements which do not preserve the parity of the state. In particular, one can perform local measurements of the order parameter, i.e. of the $Z_j$ operators. Within the replica description of the Born ensemble in Sec.~\ref{sec:Born} one immediately finds that, since the scaling dimension of the order parameter is $1/8$ in the Ising CFT, measuring these operators generates a relevant perturbation. This suggests that measuring $Z_j$ typically causes the postmeasurement field theory to flow to a `strong measurement' fixed point, where the long-distance properties of correlation functions are modified relative to the ground state. 

The possibility for simulating the effects of measurements has important implications for experiments. This is because the effects of many measurements can be observed without postselection provided one has access to an appropriate simulation on a classical computer~\cite{gullans_2020_scalable,li_2022_cross,garratt_measurements_2022,lee_2022_decoding}. Only averages of quantities nonlinear in the postmeasurement density matrix, such as $\langle Z_0 Z_r \rangle_{\vm}^2$, are sensitive to the effects of measurement as distinct from dephasing, but these cannot be determined directly since each outcome $\vm$ occurs at most once (see the discussion in, e.g., Ref.~\cite{garratt_measurements_2022}). Instead of trying to determine averages such as $\overline{\langle Z_0 Z_r\rangle_{\vm}^2}$, which suffer from a postselection problem, one can weight the results of measurements of the operator $Z_0 Z_r$ by estimates for its expectation value coming from a simulation on a classical computer $\langle Z_0 Z_r\rangle^{\text{cl.}}_{\vm}$. In this way one can obtain the `quantum-classical estimator' \cite{garratt_measurements_2022} $\overline{\langle Z_0 Z_r
\rangle^{\text{cl.}}_{\vm} \langle Z_0 Z_r\rangle_{\vm}}$ (or `computationally assisted observable' \cite{lee_2022_decoding}) which is the cross-correlation between the experiment and our prediction, and the `classical-classical estimator' $\overline{(\langle Z_0 Z_r\rangle_{\vm}^{\text{cl.}})^2}$, which is simply the prediction. Coincidence between these two objects provides a necessary condition that the quantum system studied in experiment has exhibited the same behavior as the classical simulation. 

With regard to experimental platforms, Rydberg quantum simulators have proved to be a highly controllable setting for the study of quantum Ising models \cite{labuhn_2016_tunable,browaeys_2020_many}, with the important caveat that the long-range van der Waals interactions render the effective Ising models nonintegrable. However, since these interactions decay as the sixth power of the separation between qubits, they are an irrelevant perturbation to the Ising CFT, and certain coarse-grained features of a classical simulation of the integrable TFIM should match those of a quantum simulation using Rydberg atoms. It is natural to ask whether, by cross-correlating results from a Rydberg quantum simulator with the results of exact free-fermion numerics, the effects of measurements can be observed without postselection. One can also address this kind of question numerically: given two different lattice simulations of the same critical theory, to what extent are coarse-grained correlations postmeasurement sensitive to differences on short length scales?

It is also worth noting that postmeasurement correlation functions are essentially unaffected by local decoherence. Although the statistics of operators such as $Z_0 Z_r$ are in principle modified by measurements at all sites, the fact that local quantum channels have strictly local effects means that these statistics are sensitive only to decoherence involving sites $0$ and/or $r$. Using classical simulations and a noisy quantum simulator, it should therefore be possible to observe the robustness of critical Ising correlations to parity-preserving measurements, as well as their modification by measurements of the order parameter. The entanglement of a large subregion is, however, vulnerable to decoherence within it. While both measurements and decoherence tend to disentangle system degrees of freedom, they are distinguished by the nonlocal effects of the former. Building on Refs.~\cite{lee_2023_criticality,zou2023channeling}, it would be interesting to understand how this distinction manifests in the ensemble of postmeasurement states. 

\vspace{1mm}

\textit{Note Added.} We would like to draw the readers' attention to two parallel works by Yang \textit{et al.} \cite{yang2023entanglement} and Murciano \textit{et al.} \cite{murciano_measurement_2023}. Our results agree where they overlap.

\acknowledgements
We are grateful to Yimu Bao, Michael Buchhold, Yaodong Li, and Yijian Zou for useful discussions. This work was supported by UC Berkeley Connect (Z.W.), the Berkeley Physics Undergraduate Research Scholars Program (R.S.), the Gordon and Betty Moore Foundation (S.J.G.), the Department of Energy, Office of Science, Office of High Energy Physics under QuantISED Award DE-SC0019380 (S.J.G.), and in part by the NSF QLCI program through Grant No. OMA-2016245 (E.A. and Z.W.).

\onecolumngrid
\appendix

\section{Free Fermion Simulation}
\label{app:numerics}
Here we summarize some of the technical details required for simulations based on fermionic Gaussian states. On a finite system of size $N$ with periodic boundary conditions, the Hamiltonian reads
\begin{equation}
\label{eq:supp_ham}
  H = -J \sum_{j = 1}^N \qty{ g X_j + Z_j Z_{j+1} } = -iJg \sum_{j = 1}^{N} \gamma_{2j-1} \gamma_{2j} -i J \sum_{j = 1}^{N-1} \gamma_{2j} \gamma_{2j+1} + i J \Pi \gamma_{2N} \gamma_1,
\end{equation}
where we have used the Jordan-Wigner transformation
\begin{equation}
  \gamma_{2j-1} = \qty[ \prod_{i = 1}^{j-1} X_i ] Z_j, \quad \gamma_{2j} = \qty[ \prod_{i = 1}^{j-1} X_i ] Y_j.
\end{equation}
The Majorana fermions $\gamma_j$ satisfy the anticommutation relations $\acomm{\gamma_i}{\gamma_j} = 2 \delta_{ij}$, as well as the identities $X_j = i \gamma_{2j-1} \gamma_{2j}$ and $Z_j Z_{j+1} = i \gamma_{2j} \gamma_{2j+1}$. We have also defined the total parity operator
\begin{equation}
  \Pi = \prod_{j = 1}^N X_j = i^N \prod_{j = 1}^{2N} \gamma_j,
\end{equation}
which appears, with periodic boundary conditions, in the bond connecting sites $j=1$ and $j=N$. Thus, the Majorana representation of the TFIM has antiperiodic boundary conditions in the parity-even ($\Pi = +1$) sector of the Hilbert space, while it has periodic boundary conditions in the parity-odd ($\Pi = -1$) sector. Since the exact ground state of $H$ lies in the parity-even sector \cite{schultz_two-dimensional_1964}, we can freely set $\Pi = +1$ so long as we consider measurements and observables which preserve the parity of the ground state.

The Hamiltonian (\ref{eq:supp_ham}) with $\Pi = 1$ is quadratic, and therefore its ground-state correlations can be efficiently computed \cite{kitaev_unpaired_2001,terhal_classical_2002,knill_fermionic_2001,bravyi_lagrangian_2004}. Let us briefly review the method for a generic quadratic Hamiltonian of the form
\begin{equation}
\label{eq:ham_gen}
  H = \frac{i}{4} \sum_{i,j = 1}^{2N} \gamma_i A_{ij} \gamma_j.
\end{equation}
Here $A$ is a $2N \times 2N$ real antisymmetric matrix, and so can be block-diagonalized into blocks of the form $\varepsilon_{\alpha} (i\sigma^y)$ by a matrix $R \in$ SO~$(2N)$ \cite{zumino_normal_1962}. To do this one can first diagonalize the (fully imaginary) Hermitian matrix $-iA$, whose (real) eigenvalues come in oppositely signed pairs. Note then that the diagonalized $-iA$ is expressed in terms of $2 \times 2$ blocks $\varepsilon_{\alpha} \sigma^z$, which are unitarily related to $\varepsilon_{\alpha} \sigma^y$. Finally we identify $\varepsilon_{\alpha} (i\sigma^y)$ as the real antisymmetric blocks of the transformed matrix $A$, and from this procedure we extract $R$. If we then apply the orthogonal transformation $R$ to the Majoranas (which preserves the anticommutation relations), we obtain
\begin{equation}
  H = \frac{i}{2} \sum_{\alpha = 1}^N \varepsilon_{\alpha} \eta_{2\alpha - 1} \eta_{2\alpha}, \quad \eta_{\alpha} = \sum_{i = 1}^{2N} R_{i \alpha} \gamma_i.
\end{equation}
Choosing each $\varepsilon_{\alpha}$ to be positive, the ground state and ground-state energy are immediately found by demanding $i \eta_{2\alpha-1} \eta_{2\alpha} = -1$. In particular, the two-point correlations in the ground state are
\begin{equation}
  G_{ij} = \expval{i \gamma_i \gamma_j} - i \delta_{ij} = \sum_{\beta, \gamma = 1}^{2N} R_{i \beta} R_{j \gamma} \qty[\expval{i \eta_{\beta} \eta_{\gamma}} - i \delta_{\beta \gamma}] = -\sum_{\alpha = 1}^N (R_{i,2\alpha - 1} R_{j,2\alpha} - R_{i,2\alpha} R_{j,2\alpha-1}),
\end{equation}
where we have used the fact that $\expval{i\eta_{\beta}\eta_{\gamma}}=0$ unless $\beta,\gamma=2\alpha-1,2\alpha$ (in either order). 

The ground state of a quadratic Hamiltonian (\ref{eq:ham_gen}), or more generally a thermal state of any inverse temperature\footnote{To be precise, the set of Gaussian states consist of density matrices of the form $\rho = \frac{1}{\mathcal{Z}} e^{- \beta H}$, where $H$ is of the form (\ref{eq:ham_gen}) and $\mathcal{Z} = \tr e^{-\beta H}$. In this equation, $\beta = \infty$ recovers the ground-state density matrix $\rho = \dyad{\psi}$. Even more generally, $H$ is allowed to have individual single-particle energies $\varepsilon_{\alpha} = \pm \infty$, corresponding to definite fermion parities $\expval{i \eta_{2\alpha - 1} \eta_{2\alpha}} = \mp 1$ amongst other indefinite fermion parities.} $\beta$, is called a Gaussian state \cite{bravyi_lagrangian_2004}. Once the covariance matrix $G_{ij}$ of a Gaussian state has been obtained, all higher-order correlations of the Majoranas are determined via Wick's theorem \cite{terhal_classical_2002}. For example,
\begin{equation}
\label{eq:app_wick}
  i^2\expval{\gamma_i \gamma_j \gamma_k \gamma_{\ell}} = \expval{i \gamma_i \gamma_j} \expval{i \gamma_k \gamma_{\ell}} - \expval{i \gamma_i \gamma_k} \expval{i \gamma_j \gamma_{\ell}} + \expval{i \gamma_i \gamma_{\ell}} \expval{i \gamma_j \gamma_k} = G_{ij} G_{k\ell} - G_{ik} G_{j\ell} + G_{i\ell} G_{jk}.
\end{equation}
In general, a $2n$-point correlation function $i^n \expval{\gamma_{i_1} \ldots \gamma_{i_{2n}}}$ can be computed using the Pfaffian of a submatrix of $G_{ij}$, containing only the rows and columns $i_1$ through $i_{2n}$. Explicitly,
\begin{equation}
\label{eq:app_pfaffian}
	\begin{split}
		i^n \expval{\gamma_{i_1} \ldots \gamma_{i_{2n}}} &= \frac{1}{2^n n!} \sum_{\sigma \in S_{2n}} (-1)^{\sigma} \expval{i \gamma_{i_{\sigma(1)}} \gamma_{i_{\sigma(2)}}} \ldots \expval{i \gamma_{i_{\sigma(2n-1)}} \gamma_{i_{\sigma(2n)}}} \\
		&= \frac{1}{2^n n!} \sum_{\sigma \in S_{2n}} (-1)^{\sigma} G_{i_{\sigma(1)} i_{\sigma(2)}} \ldots G_{i_{\sigma(2n-1)} i_{\sigma(2n)}} \\
		&= \text{Pf}_{i_1, \ldots , i_{2n}} [ G ],
	\end{split}
\end{equation}
where $S_{2n}$ is the permutation group of $2n$ elements, $(-1)^{\sigma} = \pm 1$ is the sign of the permutation $\sigma$, and the indices of Pf denote the subset of rows and coulumns of $G$ appearing in the second line. Such a Pfaffian can be computed efficiently using the algorithm of Ref.~\cite{wimmer2012algorithm}. We also note that Gaussian states necessarily commute with parity, which immediately implies that odd $(2n+1)$-point correlators vanish.

As an application of the Eqs. (\ref{eq:app_wick}) and (\ref{eq:app_pfaffian}), we provide explicit formulas for the correlators $\expval{X_j X_{j+r}} - \expval{X_j} \expval{X_{j+r}}$ and $\expval{Z_j Z_{j+r}}$ employed in the main text. The former correlator is local in the Majorana representation and therefore has a simple representation in terms of the covariance matrix:
\begin{equation}
	\expval{X_j X_{j+r}} - \expval{X_j} \expval{X_{j+r}} = G_{2j-1,2j+2r} G_{2j,2j+2r-1} - G_{2j-1, 2j+2r-1} G_{2j,2j+2r}.
\end{equation}
On the other hand, the latter expression is nonlocal in the Majorana representation, and requires computing a Pfaffian of a $2r \times 2r$ submatrix of $G$:
\begin{equation}
	\expval{Z_j Z_{j+r}} = \expval{i^r \gamma_{2j} \ldots \gamma_{2j+2r-1}} = \text{Pf}_{2j, \ldots , 2j+2r-1}[G].
\end{equation}

Projective measurements of the pairing operators $i \gamma_k \gamma_{\ell}$ preserve the Gaussianity of the ground state \cite{terhal_classical_2002,bravyi_lagrangian_2004}; in particular, measurements of both $X_j = i \gamma_{2j-1} \gamma_{2j}$ and $Z_j Z_{j+1} = i \gamma_{2j} \gamma_{2j+1}$ preserve Gaussianity. Up to a normalization factor, the effect of such a projective measurement on a state $\ket{\psi}$ is
\begin{equation}
	\ket{\psi} \mapsto P^{\pm}_{k \ell} \ket{\psi}, \quad P^{\pm}_{k \ell} = \frac{1 \pm i \gamma_k \gamma_{\ell}}{2} ,
\end{equation}
where the outcomes $i \gamma_k \gamma_{\ell} = \pm 1$ occur with probability $\bra{\psi} P^{\pm}_{k \ell} \ket{\psi}$ respectively, according to the Born rule. Following the measurement, the covariance matrix evolves to
\begin{equation}
	G_{ij} \mapsto G_{ij}' = \frac{\bra{\psi} P^{\pm}_{k \ell} i \gamma_i \gamma_j P^{\pm}_{k \ell} \ket{\psi} }{\bra{\psi} P^{\pm}_{k \ell} \ket{\psi}} ,
\end{equation}
which can be evaluated using Wick's theorem if the initial state $\ket{\psi}$ is Gaussian.

Finally, computation of the entanglement entropy $S(A)$ of a subregion $A$ can be performed efficiently using the covariance matrix. We first note that the reduced density matrix $\rho^A = \tr_{A^c} \dyad{\psi}$ is automatically Gaussian if $\ket{\psi}$ is Gaussian, since all of its correlations can be obtained using Wick's theorem. Its correlation matrix $G^A_{ij}$ is simply a submatrix of $G_{ij}$ with entries from region $A$. We can then infer the spectrum of $\rho^A$ directly from the spectrum of $G^A_{ij}$: block-diagonalizing $G^A_{ij}$ using an orthogonal matrix $R^A_{i \alpha}$,
\begin{equation}
	G^A_{\alpha \beta} = \sum_{i,j \in A} R^A_{i \alpha} R^A_{j \beta} G^A_{ij} = \bigoplus_{\alpha = 1}^{N_A} \begin{pmatrix}
		0 & \lambda_{\alpha} \\
		-\lambda_{\alpha} & 0
	\end{pmatrix} ,
\end{equation}
where $\abs{\lambda_{\alpha}} < 1$. The unique Gaussian reduced density matrix reproducing these correlations is then
\begin{equation}
	\rho^A = \prod_{\alpha = 1}^{N_A} \qty( \frac{1 + i \lambda_{\alpha} \xi_{2\alpha - 1} \xi_{2\alpha}}{2} ), \quad \xi_{\alpha} = \sum_{i \in A} R^A_{i \alpha} \gamma_i ,
\end{equation}
where $N_A$ is the number of sites in region $A$. From this expression we can immediately read off the spectrum of $\rho^A$, and thereby compute the entanglement entropy:
\begin{equation}
	S(A) = - \tr \rho^A \log \rho^A = - \sum_{\alpha = 1}^{N_A} \qty[ \qty( \frac{1 + \lambda_{\alpha}}{2} ) \log \qty( \frac{1 + \lambda_{\alpha}}{2} ) + \qty( \frac{1 - \lambda_{\alpha}}{2} ) \log \qty( \frac{1 - \lambda_{\alpha}}{2} ) ] .
\end{equation}
We can therefore numerically compute $S(A)$ simply by block-diagonalizing $G^A_{ij}$, or equivalently by diagonalizing $i G^A_{ij}$.

\section{Ising Conformal Field Theory}
\label{app:ising_CFT}
In this Appendix we provide a brief account on the relation between the microscopic lattice Hamiltonian (\ref{eq:tfi_ham}) for the TFIM and the continuum action (\ref{eq:ising_CFT}) for the Ising CFT. 
Starting from the Majorana representation (\ref{eq:ham_JW}), we can trivially rewrite the Hamiltonian as
\begin{equation}
	H = - \frac{iJ}{2} \sum_{j = 1}^N \Big\{ \gamma_{2j}(\gamma_{2j+1} - \gamma_{2j-1}) + \gamma_{2j-1}(\gamma_{2j} - \gamma_{2j-2}) + (g-1) (\gamma_{2j-1} \gamma_{2j} - \gamma_{2j} \gamma_{2j-1}) \Big\}.
\end{equation}
In the scaling limit $g \to 1$, the correlation length diverges and the lattice Hamiltonian can be traded for a continuum description. We introduce a lattice spacing $a \to 0$ and a two-component spinor $\hat{\psi}(x=ja) = \frac{1}{\sqrt{2a}} [\gamma_{2j-1}, \gamma_{2j}]^T$, whose components satisfy $\acomm{\hat{\psi}_a(x)}{\hat{\psi}_b(x')} = \frac{1}{a} \delta_{ab} \delta_{j j'} \to \delta_{ab} \delta(x-x')$. Up to irrelevant terms, the Hamiltonian is written in terms of $\hat{\psi}$ as
\begin{equation}
	H = \frac{v}{2} \int \dd{x} \hat{\psi}^T [-i \sigma^x \partial_x + m \sigma^y] \hat{\psi} ,
\end{equation}
where $v = 2Ja$ is the Fermi velocity from the exact solution of the TFIM, and $m = (g-1)/a$ vanishes at the critical point. 

To derive the path integral representation of the above continuum model \cite{shankar_quantum_2017}, we introduce a second copy of the same system, written in terms of a Majorana spinor $\hat{\chi}(x)$. We can then combine $\hat{\psi}$ and $\hat{\chi}$ into a single Dirac spinor, $\hat{D} = \frac{1}{\sqrt{2}} (\hat{\psi} + i \hat{\chi})$, which is an ordinary complex Dirac fermion. We then write the partition function using the usual Grassmann coherent-state path integral, trading the fermion operators $\hat{D}_a$ and $\hat{D}_a^{\dagger}$ for Grassmann numbers $D_a$ and $\bar{D}_a$. Finally, we re-express the ``complex'' Grassmann spinor $D = \frac{1}{\sqrt{2}} (\psi + i \chi)$ in terms of ``real'' Grassmann spinors $\psi$ and $\chi$, which are decoupled from each other, and integrate over $\chi$. Absorbing $v$ into the definition of the imaginary time $\tau$, the result of this computation is the imaginary-time action
\begin{equation}
\label{eq:supp_ising_CFT}
	\mathcal{S}_0[\psi] = \frac{1}{2} \int \dd{\tau} \dd{x} \psi^T [\partial_{\tau} - i \sigma^x \partial_x + m \sigma^y] \psi .
\end{equation}
Following this same procedure for arbitrary correlation functions of $\hat{\psi}$, one finds that each such correlation function is obtained in the path integral representation simply by replacing operators $\hat{\psi}$ with Grassmann numbers $\psi$.

At the critical point $m = 0$, $\mathcal{S}_0[\psi]$ is one of the simplest examples of a CFT \cite{di_francesco_conformal_1997}. The Ising CFT in particular is characterized by two scaling operators $\sigma(\tau, x)$ and $\varepsilon(\tau, x)$, with scaling dimensions $\Delta_{\sigma} = 1/8$ and $\Delta_{\varepsilon} = 1$ respectively. These represent the two relevant perturbations to the Ising critical point, and respectively reproduce the correlations of the operators $Z_j$ and $X_j$ at long distances:
\begin{equation}
\label{eq:supp_ising_correlators}
	\expval{Z_0 Z_r}_{\gs} \sim \expval{\sigma(0) \sigma(x)} = \frac{1}{x^{1/4}}, \quad \expval{X_0 X_r}_{\gs} - \expval{X_0}_{\gs} \expval{X_r}_{\gs} \sim \expval{\varepsilon(0) \varepsilon(x)} = \frac{1}{x^2},
\end{equation}
where we are restricted always to equal-$\tau$ correlations, and we have set $x = ra$. The energy operator\footnote{Here $i : \! \psi_1 \psi_2 \! : \ = i \psi_1 \psi_2 - \expval{i \psi_1 \psi_2} $ denotes normal-ordering. Normal-ordering is \textit{a priori} unnecessary at the fixed point, since the symmetry $\psi \to \sigma^x \psi$ implies $\expval{i \psi_1 \psi_2} = 0$. However, since this symmetry is broken by irrelevant perturbations, we keep normal-ordering to ensure $\expval{\varepsilon(x)} = 0$ and $\varepsilon \to -\varepsilon$ under Kramers-Wannier duality.} $\varepsilon = 2\pi i : \! \psi_1 \psi_2 \! :$ is invariant under the $\mathbb{Z}_2$ Ising symmetry, and can therefore be expressed locally in the fermionic representation. On the other hand, the spin operator $\sigma$ is nonlocal in the fermionic representation. Nevertheless, the correlators (\ref{eq:supp_ising_correlators}) can be obtained both directly within the fermionic representation \cite{bander_quantum-field-theory_1977} or by utilizing bosonization techniques \cite{zuber_quantum_1977,di_francesco_conformal_1997}.

It is well-known that critical one-dimensional systems exhibit logarithmic-scaling entanglement entropy. In particular, it can be shown quite generally that the entanglement entropy of a contiguous subregion $[0:r)$ of length $r$ in the ground state of a one-dimensional CFT is given by \cite{calabrese_entanglement_2009}
\begin{equation}
	S(r) = \frac{c}{3} \log r + b_0 ,
\end{equation}
where $b_0$ is a nonuniveral constant, and $c$ is the so-called \textit{central charge} of the CFT. In the Ising CFT, $c = 1/2$.

\section{Finite-Size Scaling}
\label{app:FSS}
In numerical simulations of finite-sized systems, it is convenient to work with periodic boundary conditions, $j \cong j + N$. Conformal invariance of the low-energy theory then allows for a precise prediction of the behavior of correlation functions $C(r)$ and $G(r)$ as a function of the system size $N$ \cite{di_francesco_conformal_1997}. Specifically, if the continuum model (\ref{eq:supp_ising_CFT}) is taken at the critical point $m = 0$, we expect that correlation functions will transform covariantly under holomorphic mappings $z \mapsto z' = f(z)$ of the complex variable $z = x+i\tau$. If the model is initially defined on the cylinder, such that $x \cong x + L$ with $L = Na$, then we can obtain correlation functions on the cylinder from correlation functions on the infinite plane using the mapping
\begin{equation}
	z' = f(z) = L \tan \qty( \frac{\pi z}{L} ) ,
\end{equation}
which maps the cylinder to the infinite plane. This particular mapping is especially useful for our purposes, since it preserves the $\tau = 0$ line. Since measurements in our models appear as defects along the $\tau = 0$ line of Euclidean spacetime, we can therefore predict the numerically observed effects of measurements on finite-sized systems using analytical calculations in the thermodynamic limit. Specifically, the above mapping suggests that the correlators (\ref{eq:supp_ising_correlators}) on the cylinder are given by\footnote{In CFT, $\sigma$ and $\varepsilon$ are examples of so-called ``primary operators.'' Under a conformal transformation $z \mapsto f(z)$, a generic primary operator $\phi(z,\bar{z})$ transforms inside correlation functions as $\phi(z,\bar{z}) \mapsto \phi'(z',\bar{z}') = [f'(z)]^{-h} [\bar{f}'(\bar{z})]^{-\bar{h}} \phi(z,\bar{z})$, where $(h, \bar{h})$ are the so-called ``conformal dimensions'' of $\phi$. Using $h_{\sigma} = \bar{h}_{\sigma} = 1/16$ and $h_{\varepsilon} = \bar{h}_{\varepsilon} = 1/2$, one immediately obtains the given expressions for the correlations $\expval{\sigma(0) \sigma(x)}_{\gs}^{\text{cyl}}$ and $\expval{\varepsilon(0) \varepsilon(x)}_{\gs}^{\text{cyl}}$ on the cylinder.}
\begin{equation}
	\expval{\sigma(0) \sigma(x)}^{\text{cyl}}_{\gs} = \qty[\frac{L}{\pi} \sin \qty( \frac{\pi x}{L} )]^{-1/4}, \quad \expval{\varepsilon(0) \varepsilon(x)}^{\text{cyl}}_{\gs} = \qty[\frac{L}{\pi} \sin \qty( \frac{\pi x}{L} )]^{-2} .
\end{equation}
We therefore expect the correlators $G(r)$ and $C(r)$, which are \textit{a priori} functions of $r$ and $N$ separately, to be functions of the single variable $s = \frac{N}{\pi} \sin( \frac{\pi r}{N} )$. The infinite-plane behavior is recovered in the limit $N \to \infty$, upon which $s \to r$ for any finite $r$. A similar result holds for the entanglement entropy of a finite system with periodic boundary conditions \cite{calabrese_entanglement_2009}:
\begin{equation}
	S^{\text{cyl}}(r) = \frac{c}{3} \log \qty[ \frac{N}{\pi} \sin \qty( \frac{\pi r}{N} ) ] + b_0' .
\end{equation}
These finite-size expressions allow for excellent scaling collapses of various numerically computed observables across several system sizes, as demonstrated in the main text. 

\section{Continuum Limit of \texorpdfstring{$\hat{M}_{\avg}$}{Mavg} and \texorpdfstring{$\hat{K}_{\avg}$}{Lavg}}
\label{app:cont_limit}
In this Appendix we explain the continuum descriptions of the averaged measurement operators $\hat{M}_{\avg}$ and $\hat{K}_{\avg}$ defined in Eqs.~(\ref{eq:M_avg}) and (\ref{eq:K_avg}) respectively. In particular, we show that these two measurement operators result in defects along the $\tau = 0$ line in Euclidean spacetime; the former of these defects is irrelevant, while the latter contains an exactly marginal perturbation to the Ising CFT. We also show that the irrelevant contributions to the former averaged measurement operator cannot generate marginal terms at higher orders of the perturbative RG -- or, more precisely, that any generated marginal terms are inconsequential to observables in the replica limit.

Starting with $\hat{M}_{\avg}$, the denominator of $n$-replica correlation functions of the form (\ref{eq:G_n}) takes the form of a partition function
\begin{equation}
	\mathcal{Z}_M^{(n)} = \bra{\psi^{\otimes n}_{\gs}} \hat{M}_{\avg} \ket{\psi^{\otimes n}_{\gs}} \propto \expval{ \prod_{j = 1}^N \qty{ 1 + \mu \sum_{r = 1}^{\lfloor n/2 \rfloor} \sum_{1 \leq \alpha_1 < \ldots < \alpha_{2r} \leq n} (i \gamma_{2j-1} \gamma_{2j})^{(\alpha_1)} \ldots (i \gamma_{2j-1} \gamma_{2j})^{(\alpha_{2r})} } }_{\gs}.
\end{equation}
Note that from the definition of $\mathcal{Z}_M^{(n)}$ we have the normalization $\mathcal{Z}^{(n)}_M = 1$ for $p=0$. The notation $(i \gamma_{2j - 1} \gamma_{2j})^{(\alpha)}$ simply means $i \gamma_{2j-1}^{(\alpha)} \gamma_{2j}^{(\alpha)}$, the product of Majorana operators within replica $\alpha$.

We can write $\mathcal{Z}_M^{(n)}$ within the path integral formalism by simply replacing each $\gamma_j$ with the Grassmann field\footnote{The factor of $\sqrt{2}$ arises from our normalization convention for the Majorana operators, $\acomm{\gamma_i}{\gamma_j} = 2\delta_{ij}$, rather than $\acomm{\gamma_i}{\gamma_j} = \delta_{ij}$. It can be obtained by retracing the steps outlined in Appendix \ref{app:ising_CFT} for deriving the path integral representation of the Majorana system.} $\sqrt{2}\psi_j(\tau)$, evaluated at $\tau = 0$:
\begin{equation}
\label{eq:app_Z_M_1}
	\begin{split}
		\mathcal{Z}_M^{(n)} &= \int \prod_{\alpha = 1}^n D\psi^{(\alpha)} \, e^{- \sum_{\alpha = 1}^n \mathcal{S}_0[\psi^{(\alpha)}]} \prod_{j = 1}^N \qty{ 1 + \mu \sum_{r = 1}^{\lfloor n/2 \rfloor} \sum_{1 \leq \alpha_1 < \ldots < \alpha_{2r} \leq n} (2i \psi_{2j-1}(0) \psi_{2j}(0))^{(\alpha_1)} \ldots (2i \psi_{2j-1}(0) \psi_{2j}(0))^{(\alpha_{2r})}  } \\
		&= \int \prod_{\alpha = 1}^n D\psi^{(\alpha)} \, e^{- \sum_{\alpha = 1}^n \mathcal{S}_0[\psi^{(\alpha)}]} \exp{ \mu \sum_{j = 1}^N \sum_{1 \leq \alpha < \beta \leq n} (2i \psi_{2j-1}(0) \psi_{2j}(0))^{(\alpha)}(2i \psi_{2j-1}(0) \psi_{2j}(0))^{(\beta)} + \ldots } ,
	\end{split}
\end{equation}
where the ellipsis denotes four-replica terms and higher. Finally, we take the continuum limit by constructing the continuum Grassmann spinor $\psi(\tau, x=ja) = \frac{1}{\sqrt{a}} [\psi_{2j-1}(\tau), \psi_{2j}(\tau)]^T$. Rewriting $2i \psi_{2j-1} \psi_{2j} = -a \psi^T \sigma^y \psi$, we obtain the result
\begin{equation}
\label{eq:app_Z_M}
	\begin{split}
		\mathcal{Z}^{(n)}_M &= \int\prod_{\alpha = 1}^n D\psi^{(\alpha)} \, e^{- \sum_{\alpha = 1}^n \mathcal{S}_0[\psi^{(\alpha)}]} \exp{ \tilde{\mu} \sum_{1 \leq \alpha < \beta \leq n} \int \dd{x} (\psi^T \sigma^y \psi)^{(\alpha)}(\psi^T \sigma^y \psi)^{(\beta)} + \ldots } \\
		&= \int D \psi \, e^{-\mathcal{S}_0[\psi] - \mathcal{S}^{(n)}_M[\{\psi^{(\alpha)}\}]} ,
	\end{split}
\end{equation}
where $\tilde{\mu} = \mu a$ has dimension $-1$; in the main text we set $a = 1$ for simplicity. In the first equation, the field $\psi$ in the latter exponential must be understood as $\psi(\tau=0,x)$, and in the second line we have defined $\mathcal{S}^{(n)}_M[\{\psi^{(\alpha)}\}]$ as the exponent appearing in braces $\{\cdots\}$ in the first. Considering the quantity in braces as a perturbation to the Ising CFT localized to the $\tau = 0$ line, one immediately finds from dimensional analysis that the parameter $\tilde{\mu}$ is irrelevant.

A similar analysis applies to $\hat{K}_{\avg}$. Performing the same continuum limit as above, we obtain
\begin{equation}
	\begin{split}
		\bra{\psi^{\otimes n}_{\gs}} \hat{K}_{\avg} \ket{\psi^{\otimes n}_{\gs}} &= \mathcal{Z}^{(n)}_K \propto \expval{ \prod_{j = 1}^N \qty{ 1 + \nu \sum_{r = 1}^n \sum_{1 \leq \alpha_1 < \ldots < \alpha_r \leq n} (i \gamma_{2j-1} \gamma_{2j})^{(\alpha_1)} \ldots (i \gamma_{2j-1} \gamma_{2j})^{(\alpha_r)} } }_{\gs} \\
		&= \int \prod_{\alpha = 1}^n D \psi^{(\alpha)} \, e^{-\sum_{\alpha = 1}^n \mathcal{S}_0[\psi]} \prod_{j = 1}^N \qty{ 1 + \nu\sum_{r = 1}^n \sum_{1 \leq \alpha_1 < \ldots < \alpha_r \leq n} (2i \psi_{2j-1} \psi_{2j})^{(\alpha_1)} \ldots (2i \psi_{2j-1} \psi_{2j})^{(\alpha_r)} } \\
		&= \int \prod_{\alpha = 1}^n D \psi^{(\alpha)} \, e^{-\sum_{\alpha = 1}^n \mathcal{S}_0[\psi]} \exp{ - \tilde{\nu} \sum_{\alpha = 1}^n \int \dd{x} (\psi^T \sigma^y \psi)^{(\alpha)} + \ldots } \\
		&= \int \prod_{\alpha = 1}^n D \psi^{(\alpha)} e^{-\sum_{\alpha = 1}^n \mathcal{S}_0[\psi^{(\alpha)}] - \mathcal{S}_K^{(n)}[\{\psi^{(\alpha)} \}] } ,
	\end{split}
\end{equation}
where the ellipsis again denotes higher-order irrelevant terms, including those written explicitly in (\ref{eq:app_Z_M}). As above, in the final line we have defined the perturbation to the action, here $\mathcal{S}_K^{(n)}[\{\psi^{(\alpha)} \}]$, as the exponent appearing in braces in the previous line.

An important question is whether higher orders in the perturbative RG can generate relevant or marginal terms in (\ref{eq:app_Z_M}). It is immediately clear that no relevant terms can be generated: since any Grassmann-even contribution to the action contains at minimum two $\psi$'s with scaling dimension $1/2$, power-counting suggests that there are no relevant perturbations upon restricting to the $\tau = 0$ line. There are, on the other hand, marginal perturbations of the form
\begin{equation}
	\delta \mathcal{S}_{M,2} = -\mu_2 \sum_{\alpha = 1}^n \int \dd{x} (\psi^T \sigma^y \psi)^{(\alpha)} .
\end{equation}
While this term properly respects the replica symmetry, it does not respect the symmetry $\psi^{(\alpha)} \to \sigma^x \psi^{(\alpha)}$ present at the unperturbed critical point, and so it cannot be generated under the perturbative RG.

An important subtlety, however, is that this symmetry is \textit{explicitly broken} by irrelevant perturbations to $\mathcal{S}_0$. If we intend to understand measurements of the microscopic lattice TFIM, rather than measurements of the Ising CFT, then these irrelevant terms must be taken into consideration. These irrelevant perturbations result in a nonzero expectation value $\expval{\psi^T \sigma^y \psi}_{\gs}$. As a result, performing a single step of first-order perturbative RG on $\mathcal{S}_M^{(n)}$ yields
\begin{equation}
    \delta \mathcal{S}_{M,2} = -\tilde{\mu} \sum_{1 \leq \alpha < \beta \leq n} \int \dd{x} \qty[ B (\psi^{T} \sigma^y \psi)^{(\beta)} + (\psi^T \sigma^y \psi)^{(\alpha)} B ] = -\tilde{\mu}B(n-1) \sum_{\alpha = 1}^n \int \dd{x} (\psi^T \sigma^y \psi)^{(\alpha)} ,
\label{eq:SM2}
\end{equation}
where $B$ is a constant originating from the integration of fast modes\footnote{To avoid unnecessary details, we have not explained the perturbative RG explicitly. The constant $B$ is given by the expectation $\expval{\psi_>^T \sigma^y \psi_>}^>_{\gs}$ of the modes with momenta between a shell of width $\dd{\ell}$. The rescaling step is inconsequential, since the term is marginal at tree level.}. We therefore find, at first order in the perturbative RG, that $\mu_2 = \tilde{\mu} B (n-1)$.

While it seems that the RG has generated an exactly marginal term, it is important to note that its prefactor $\mu_2$ is proportional to $n-1$. As a result, it is inconsequential to any correlation function upon taking the replica limit $n \to 1$: performing a perturbative expansion in $\delta S_{M,2}$, each term containing a factor $\mu_2$ will vanish upon taking the replica limit. This observation can be understood simply on physical grounds: since probability conservation requires $\sum_{\vm} \hat{M}_{\vm}^2 = 1$, we must have $\mathcal{Z}^{(1)}_M = \bra{\psi_{\gs}} \ket{\psi_{\gs}} = 1$. As a result, all single-replica contributions to to the action of the form \eqref{eq:SM2} must vanish in the $n \to 1$ limit. The vanishing of $\mu_2$ as $n \to 1$ is therefore completely general, to all orders of the perturbative RG, and this term can be discarded for purposes of computing correlation functions in the replica limit.

\section{Noncommuting Measurements and Observables}
\label{app:noncommuting}
In the main text, our analytical approach relied on assuming that the measurement operator commuted with the observable being investigated. In this Appendix we show how to analyze observables which do not commute with the measurement operators using the same analytical mapping. The details are slightly more technically involved than the approach followed in the main text, but are conceptually similar.

We demonstrate our approach explicitly for $C_{\vm}^2(r) = \expval{Z_0 Z_r}^2_{\vm}$ in the ensemble-average measurement scheme; other cases, such as $C_{\vk}(r)$ in the forced measurement scheme, or $G^Z_{\vm}(r)$ under $ZZ$ measurements (see Appendix~\ref{app:ZZ}), follow similarly. To start, it is useful to note that $\expval{Z_0 Z_r}_{\vm}$ vanishes whenever sites $0$ or $r$ are measured. Using the replica scheme of Eq.~(\ref{eq:replica}), the numerator reads
\begin{equation}
	\begin{split}
		\sum_{\vm} p_{\vm}^n &\expval{Z_0 Z_r}^2_{\vm} = \sum_{\vm} \expval*{\hat{M}_{\vm}^2}_{\gs}^{n-2} \expval*{\hat{M}_{\vm} Z_0 Z_r \hat{M}_{\vm}}_{\gs}^2 \\
		&= (1-p)^{2n} \bra{\psi^{\otimes n}_{\gs}} Z_0^{(0)} Z_0^{(1)} Z_r^{(0)} Z_r^{(1)} \prod_{j \neq 0, r} \qty{ (1-p)^n + p^n \sum_{m_j = \pm 1} \qty( \frac{1 + m_j X_j}{2} )^{\otimes n} } \ket{\psi^{\otimes n}_{\gs}} .
	\end{split}
\end{equation}
To recover the form (\ref{eq:M_avg}) for the averaged measurement operator on the right, we simply multiply and divide by the local terms in braces on the right corresponding to sites $0$ and $r$. We note that these operators have strictly positive spectra, and therefore have well-defined inverses. In particular, 
\begin{equation}
	\begin{split}
		\qty[ (1-p)^n + p^n \sum_{m_j = \pm 1} \qty( \frac{1 + m_j X_j}{2} )^{\otimes n} ]^{-1} &= (1-p)^{-n} \qty[ 1 - \frac{p^n}{(1-p)^n + p^n} \sum_{m_j = \pm 1} \qty( \frac{1 + m_j X_j}{2} )^{\otimes n} ] \\
		&\propto  1 - \omega \sum_{r = 1}^{\lfloor n/2 \rfloor} \sum_{1 \leq \alpha_1 < \ldots < \alpha_{2r} \leq n} X^{(\alpha_1)}_j \ldots X^{(\alpha_{2r})}_j  ,
	\end{split}
\end{equation}
where $\omega$ is a monotonically increasing function of $p$, and we have omitted a $p$-dependent prefactor which does not affect the asymptotic scaling behavior of $\overline{C^2_{\vm}(r)}$. We therefore obtain
\begin{equation}
	\begin{split}
		\sum_{\vm} p_{\vm}^n \expval{Z_0 Z_r}^2_{\vm} &\propto \bra{\psi^{\otimes n}_{\gs}} Z_0^{(0)} Z_0^{(1)} Z_r^{(0)} Z_r^{(1)} \qty[ 1 - \omega \sum_{r = 1}^{\lfloor n/2 \rfloor} \sum_{1 \leq \alpha_1 < \ldots < \alpha_{2r} \leq n} X^{(\alpha_1)}_0 \ldots X^{(\alpha_{2r})}_0 ] \\
		& \hspace{5cm} \times \qty[ 1 - \omega \sum_{r = 1}^{\lfloor n/2 \rfloor} \sum_{1 \leq \alpha_1 < \ldots < \alpha_{2r} \leq n} X^{(\alpha_1)}_r \ldots X^{(\alpha_{2r})}_r ] \hat{M}_{\avg} \ket{\psi^{\otimes n}_{\gs}} .
	\end{split}
\end{equation}
This equation is thus far exact, and is simply a sum of higher-order multi-replica correlation functions with respect to multi-replica ground state coupled by the defect $\hat{M}_{\avg}$. 

At sufficiently long distances, higher-order terms proportional to $\omega$ or $\omega^2$ decay more rapidly than the zeroth order term proportional to $\omega^0$. Since we are anyway primarily concerned with the long-range behavior, we are justified in dropping all of these higher-order terms. Altogether, we find
\begin{equation}
	\overline{C^2_{\vm}(r)} = \lim_{n \to 1} \frac{\sum_{\vm} p_{\vm}^n \expval{Z_0 Z_r}^2_{\vm}}{\sum_{\vm} p_{\vm}^n} \sim \lim_{n \to 1} \frac{\bra{\psi^{\otimes n}_{\gs}} Z^{(0)}_0 Z^{(1)}_0 Z^{(0)}_r Z^{(1)}_r \hat{M}_{\avg} \ket{\psi^{\otimes n}_{\gs}}}{\bra{\psi^{\otimes n}_{\gs}} \hat{M}_{\avg} \ket{\psi^{\otimes n}_{\gs}}} \quad (r \gg 1) .
\end{equation}

\section{Forced Measurement Entanglement Entropy}
\label{app:forced_entanglement}
In Sec.~\ref{subsec:force_EE}, we used the conformal invariance of the Ising CFT to provide a qualitative argument relating the average entanglement entropy $\overline{S_{\vk}(r)}$ of the TFIM following forced projective measurements to the half-system entanglement entropy $S_d(N/2)$ of a dual ground-state defect problem. Here we provide some of the technical details required to complete the argument. 

As in Sec.~\ref{subsec:force_EE}, we are interested in computing the entanglement entropy $S^*(r)$ of the fixed-point model (\ref{eq:defect_FP}), for a subregion of length $r$. This can be computed from the $n \to 1$ limit of the following ratio of two partition functions:
\begin{equation}
\label{eq:app_renyi}
    S^*(r, \epsilon_0) = \lim_{n \to 0} \frac{1}{1-n} \log \qty{ \frac{\mathcal{Z}_n(r,\epsilon_0)}{\mathcal{Z}_1^n} } .
\end{equation}
Here $\mathcal{Z}_1 = \int D\psi \, e^{-\mathcal{S}^*[\psi]}$ is the single-replica partition function, while $\mathcal{Z}_n(r,\epsilon_0)$ is the partition function of an $n$-fold replicated theory subjected to the boundary conditions given in Eq.~(\ref{eq:replica_bc}). Alternatively, we can think of the $n$ replica fields $\psi^{(\alpha)}(\tau,x)$ as a single field defined on an $n$-sheeted Riemann surface, with a branch cut running from $x = -r/2$ to $x = +r/2$ along the $\tau = 0$ axis. Both partition functions contain a defect running along the $\tau = 0$ axis\footnote{To avoid the branch cut lying at exactly the same imaginary time as the defect, we can take the second term in the action (\ref{eq:defect_FP}) to be evaluated at $\tau = 0^-$. By returning to the properly discretized expression for the entanglement entropy and applying the cyclicity of the trace, it is clear that the entanglement cut can be chosen to lie just above the measurement defect.}, as given in (\ref{eq:defect_FP}). In defining (\ref{eq:app_renyi}), we have introduced explicit dependence on a short-distance cutoff $\epsilon_0$ into the definition of both $S^*(r,\epsilon_0)$ and $\mathcal{Z}_n(r,\epsilon_0)$, which widens the branch points at $x = \pm r/2$ into circles of radius $\epsilon_0$. The value of $\epsilon_0$ is fixed but otherwise arbitrary, and is necessary for obtaining a finite expression for the entanglement entropy \cite{holzhey_geometric_1994,calabrese_entanglement_2009}.

In order to exploit conformal invariance of the Ising CFT, it is useful to introduce the complex coordinate $z = x + i \tau$. Then, using the scale invariance of the action (\ref{eq:defect_FP}), we first rescale $z \to 2z / r$. This transformation maps the branch points from $z = \pm r/2$ to $z = \pm 1$, but modifies the short-distance cutoff from $\epsilon_0$ to $2 \epsilon_0 / r$. We therefore have $\mathcal{Z}_n(r,\epsilon_0) = \mathcal{Z}_n(2,2 \epsilon_0 / r)$. Without loss of generality, may therefore set $r = 2$ and compute $\mathcal{Z}_n(2,\epsilon)$, from which we obtain the result for general $r$ by setting $\epsilon = 2 \epsilon_0 / r$.

We next deform the entanglement branch cut from the real line onto the unit semicircle in the upper-half plane, as depicted in Fig.\ref{fig:force_EE}(aii). As discussed in Sec.~\ref{subsec:force_EE}, the precise location of the branch cut on a Riemann surface is immaterial and can be deformed freely so long as its endpoints remain fixed. Alternatively, as a theory of $n$ replica fields, the field redefinitions $\psi^{(\alpha)}(\tau,x) \to \tilde{\psi}^{(\alpha)}(\tau,x)$ of Eq.~(\ref{eq:deform}) can be used to move the branch cut. Explicitly, the boundary condition $\psi^{(\alpha)}(0^-, x) = \psi^{(\alpha+1)}(0^+,x)$ for $\abs{x} < 1$ (i.e., at the bottom boundary of the semicircle $\mathcal{D}$) is equivalent to the continuity condition\footnote{Note that we are being somewhat cavalier about $\psi^{(\alpha)}(\tau,x)$ being a Grassmann field; strictly speaking, the field $\psi^{(\alpha)}(\tau,x)$ is indeterminate, and its continuity is ill-defined. Instead one should consider a properly discretized path integral, where the continuity condition is realized by couplings between lattice points across the $\tau = 0$ boundary.} $\tilde{\psi}^{(\alpha)}(0^-,x) = \tilde{\psi}^{(\alpha)}(0^+,x)$. Similarly, the continuity of $\tilde{\psi}^{(\alpha)}(\tau,x)$ at the top boundary of region $\mathcal{D}$ becomes a matching condition analogous to that of Eq.~(\ref{eq:replica_bc}). We emphasize that this transformation of fields is exactly what is done to deform a branch cut in the analysis of an ordinary complex function defined on a Riemann surface.

Next, we use the conformal transformation
\begin{equation}
\label{eq:app_conf_cylinder}
    z \mapsto z' = f(z) = -i \frac{L}{2\pi} \log z
\end{equation}
which maps the complex plane to a cylinder of circumference $L$, with complex coordinate $z' = x' + i \tau'$ with $x' \cong x' + L$ [see Fig.~\ref{fig:force_EE}(aiii)]. The defect due to measurements, which lies along the $\tau = 0$ line of the original complex plane, maps to \textit{two} timelike defects on the cylinder: the positive real axis maps to the line $x' = 0$, while the negative real axis maps to the line $x' = L/2$. Meanwhile, the branch cut along the unit semicircle in the upper-half plane is mapped to the semicircle from $x' = 0$ to $x' = L/2$ along the $\tau' = 0$ line. 

Under the transformation (\ref{eq:app_conf_cylinder}), a short-distance cutoff of size $\epsilon$ on the complex plane becomes a cutoff of size $L \epsilon / 2\pi$ on the cylinder; explicitly, an infinitesimal rectangle of area $\epsilon^2$ centered on $z = \pm 1$ is mapped to an infinitesimal rectangle of area $(\epsilon L / 2\pi)^2$ on the cylinder. Let $\mathcal{Z}_n^{\text{cyl}}(L,\epsilon)$ denote the partition function of the $n$-fold replicated cylinder, with defects along $x' = 0$ and $x' = L/2$, and with a branch cut between $x' = 0$ and $x' = L/2$ along $\tau' = 0$. Then, the above sequence of mappings has shown the following equivalence of partition functions:
\begin{equation}
    \mathcal{Z}_n(r, \epsilon_0) = \mathcal{Z}_n(2, 2\epsilon_0 / r) = \mathcal{Z}^{\text{cyl}}_n(L, L \epsilon_0 / \pi r) .
\end{equation}

Having formally established the connection between partition functions, we may now make connection with known results from the literature. From previous studies of the entanglement entropy of the TFIM and the Ising CFT in the presence of exactly marginal defects \cite{igloi_entanglement_2009,eisler_entanglement_2010,peschel_exact_2012,brehm_entanglement_2015,roy_entanglement_2022}, it is known that the half-system entanglement entropy with defects at the entangling boundaries exhibits the following form:
\begin{equation}
    S^*_d(L/2, \epsilon) = \lim_{n \to 1} \frac{1}{1-n} \log \qty{ \frac{\mathcal{Z}_n^{\text{cyl}}(L,\epsilon)}{ [\mathcal{Z}_1^{\text{cyl}}]^n } } = \frac{c_{\eff}(\nu)}{3} \log \frac{L}{\epsilon} + b_d(\nu) ,
\end{equation}
where $c^*_{\eff}(\nu)$ is a continuously varying effective central charge which depends on the defect strength $\nu$, while $b_d(\nu)$ is an $L$-independent constant which depends on the precise regularization scheme used. Substituting $\epsilon = L \epsilon_0 / \pi r$, we obtain the following result for the original entanglement entropy (\ref{eq:app_renyi}):
\begin{equation}
    S^*(r, \epsilon_0) = S^*_d(L/2, L \epsilon_0 / \pi r) = \frac{c^*_{\eff}(\nu)}{3} \log \qty( \frac{L}{L \epsilon_0 / \pi r} ) + b_d(\nu) = \frac{c_{\eff}^*(\nu)}{3} \log r + b'(\nu) ,
\end{equation}
where we have absorbed the constant $\log (\pi / \epsilon_0)$ into the definition of $b'(\nu)$. Notably, all dependence on the cylinder circumference $L$ has been eliminated from the final expression; since $L$ can be chosen arbitrarily, this is to be expected. We therefore arrive at the desired result: the entanglement entropy of a subregion of length $r$ in the Ising CFT, in the presence of a marginal defect along the $\tau = 0$ line (such as that arising due to forced projective measurements), scales logarithmically with $r$ with an effective central charge $c^*_{\eff}(\nu)$ that continuously varies with the defect strength.

\section{\texorpdfstring{$Z_jZ_{j+1}$}{ZZ} Measurements}
\label{app:ZZ}
Throughout the main text, we considered measurements of the observable $X_j$ throughout the Ising chain. An alternative scheme is to consider measurements of $Z_j Z_{j+1}$. Since $X_j$ and $Z_j Z_{j+1}$ are related by the Kramers-Wannier duality \cite{kogut_introduction_1979}, it is natural to expect that measurements of these observables have similar effects on the ground state. There are, however, important differences in the behavior of observables.

\subsubsection*{Born Projective \texorpdfstring{$Z_jZ_{j+1}$}{ZZ} Measurements}
We first consider a measurement protocol analogous to that of Sec.~\ref{sec:Born}: for each site $j$, we perform a projective measurement of $Z_j Z_{j+1}$ with probability $p$, and sample the measurement outcome according to the Born rule. Such a protocol is described by a measurement operator $\hat{M}^Z_{\vm}$, which is a product of local measurement operators:
\begin{equation}
    \hat{M}^Z_{\vm} = \prod_{j = 1}^N \hat{M}^Z_{m_j,j}, \quad \hat{M}^Z_{0,j} = \sqrt{1-p}, \quad \hat{M}^Z_{\pm 1, j} = \sqrt{p} \frac{1 \pm Z_j Z_{j+1}}{2} .
\end{equation}
We obtain the state $\ket{\psi^Z_{\vm}}$ with probability $p^Z_{\vm}$, where
\begin{equation}
    \ket{\psi^Z_{\vm}} = \frac{\hat{M}^Z_{\vm} \ket{\psi_{\gs}}}{\sqrt{\expval*{[\hat{M}^Z_{\vm}]^2}_{\gs}}}, \quad p^Z_{\vm} = \expval*{[\hat{M}^Z_{\vm}]^2}_{\gs} .
\end{equation}
The discussion of Sec.~\ref{subsec:ensemble_field_theory} then follows identically, with the caveat that we focus on the correlator $[C^Z_{\vm}(r)]^2 = [\expval{Z_0 Z_r}_{\vm}^Z]^2$ in developing our replica scheme, where $\expval{ \cdot }_{\vm}^Z = \bra{\psi^Z_{\vm}} \cdot \ket{\psi^Z_{\vm}}$. To study $G^Z_{\vm}(r) = \expval{X_0 X_r}_{\vm}^Z - \expval{X_0}_{\vm}^Z \expval{X_r}_{\vm}^Z$, we use the approach of Appendix~\ref{app:noncommuting} to handle noncommuting measurements and observables. Following the same steps, we arrive at an averaged measurement operator $\hat{M}^Z_{\avg}$ which couples the $n$ replicas:
\begin{equation}
    \hat{M}^Z_{\avg} = \sum_{\vm} ([\hat{M}^Z_{\vm}]^2)^{\otimes n} \propto \prod_j \qty{ 1 + \mu \sum_{r = 1}^{\lfloor n/2 \rfloor} \sum_{1 \leq \alpha_1 < \ldots < \alpha_{2r} \leq n} Z^{(\alpha_1)}_j Z^{(\alpha_1)}_{j+1} \ldots Z^{(\alpha_{2r})}_j Z^{(\alpha_{2r})}_{j+1} } .
\end{equation}
The derivation of the continuum limit of $\hat{M}^Z_{\avg}$ then follows nearly identically to that of $\hat{M}_{\avg}$ in Appendix~\ref{app:cont_limit}. The only difference lies in the replacement of the operators $X_j = i \gamma_{2j-1} \gamma_{2j}$ with $Z_j Z_{j+1} = i \gamma_{2j} \gamma_{2j+1}$. The partition function $\mathcal{Z}^{(n)}_{M,Z} = \bra{\psi^{\otimes n}_{\gs}} \hat{M}^Z_{\avg} \ket{\psi^{\otimes n}_{\gs}}$ is then given by the same expression as Eq.~(\ref{eq:app_Z_M_1}), with the replacement\footnote{In the final term, we have integrated by parts and dropped a total derivative, which vanishes under the integral $\int \dd{x}$.}
\begin{equation}
\label{eq:app_X-ZZ_replacement}
    2i \psi_{2j-1} \psi_{2j} \to i \psi_{2j} \psi_{2j+1} = -2i \psi_{2j-1} \psi_{2j} + 2i \psi_{2j} (\psi_{2j+1} - \psi_{2j-1}) = a \psi^T \sigma^y \psi + i a^2 \psi^T \sigma^x \partial_x \psi .
\end{equation}
Altogether, the defect created by $Z_jZ_{j+1}$ measurements sampled according to the Born rule is described by a contribution to the action of the form
\begin{equation}
    \mathcal{S}^{(n)}_{M,Z} = -\mu \sum_{\alpha < \beta} \int \dd{x} \qty[ 
\psi^T \sigma^y \psi + ia \psi^T \sigma^x \partial_x \psi ]^{(\alpha)} \qty[ 
\psi^T \sigma^y \psi + ia \psi^T \sigma^x \partial_x \psi ]^{(\beta)} + \ldots .
\end{equation}
The extra derivative terms $i\psi^T \sigma^x \partial_x \psi$ are further irrelevant compared to the term $\psi^T \sigma^y \psi$. We therefore arrive at the same conclusion as in Sec.~\ref{sec:Born}: a nonzero density of $Z_j Z_{j+1}$ measurements performed on the ground state $\ket{\psi_{\gs}}$ of the TFIM, with outcomes sampled according to the Born rule, do not affect the asymptotic structure of correlation functions or entanglement.

\begin{figure}
    \centering
    \includegraphics{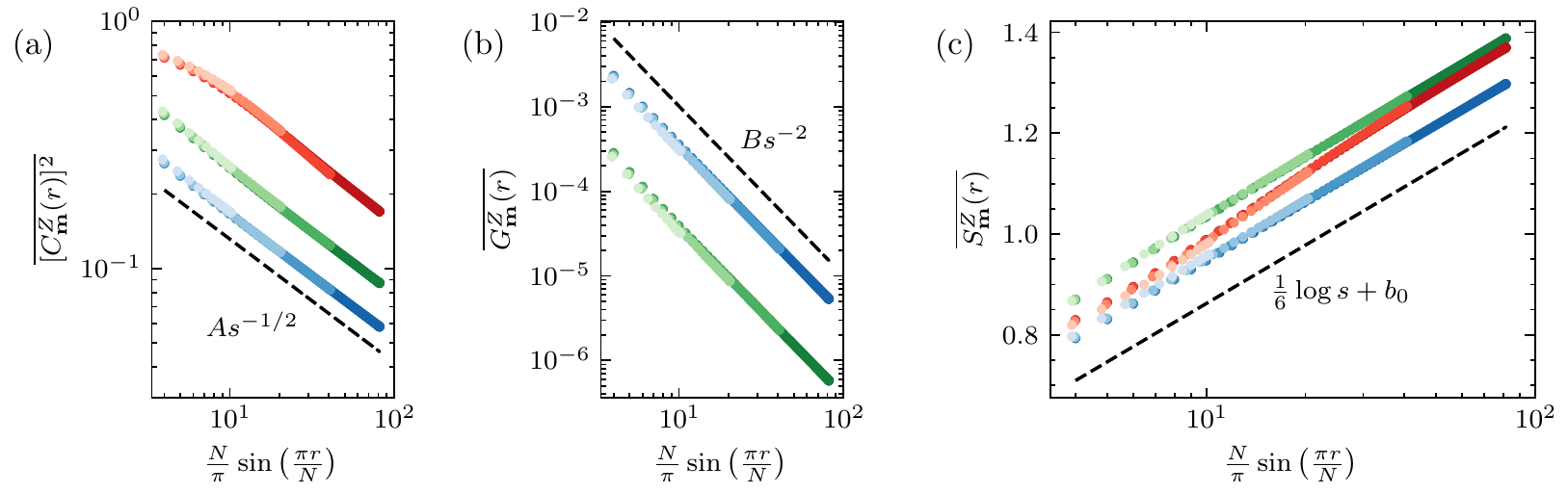}
    \caption{[(a),(b)] Ensemble-averaged correlation functions $\overline{[C^Z_{\vm}(r)]^2}$ and $\overline{G^Z_{\vm}(r)}$ and (c) entanglement entropy $\overline{S^Z_{\vm}(r)}$ following $Z_j Z_{j+1}$ measurements with outcomes sampled according to the Born rule, for measurement probabilities $p = 0.2$ (blue), $0.5$ (green), and $0.8$ (red), and for system sizes $N = 32$, $64$, $128$, and $256$ (light to dark). Data are plotted as a function of $s = \frac{N}{\pi} \sin \qty(\frac{\pi r}{N})$ to achieve scaling collapse of the various system sizes. Similar to the ensemble with $X_j$ measurements [see Sec.~\ref{sec:Born}], both correlation functions retain their power-law scaling with exponents of the unmeasured system at sufficiently long distances, while the entanglement entropy retains its logarithmic scaling with central charge $c = 1/2$ of the unmeasured system.}
    \label{fig:ZZ_born}
\end{figure}

Figure~\ref{fig:ZZ_born} gives the numerically computed correlation functions $\overline{[C^Z_{\vm}(r)]^2}$ and $\overline{G^Z_{\vm}(r)}$ and the entanglement entropy $\overline{S^Z_{\vm}(r)}$, each defined analogously to Eq.~(\ref{eq:obsm}) with the replacement $\ket{\psi_{\vm}} \to \ket{\psi_{\vm}^Z}$. As in the case of $X_j$ measurements, we find excellent scaling collapses by plotting each observable as a function of $s = \frac{N}{\pi} \sin \qty( \frac{\pi r}{N} )$. As expected from the above discussion, we find $\overline{[C^Z_{\vm}(r)]^2} \sim s^{-1/2}$ and $\overline{G^Z_{\vm}(r)} \sim s^{-2}$ at sufficiently large $s$, as well as $\overline{S^Z_{\vm}(r)} \sim \frac{1}{6} \log s + b_4(p)$.

Two differences appear between the numerical results of $Z_j Z_{j+1}$ measurements and $X_j$ measurements. First, whereas the power-law coefficient in Fig.~\ref{fig:born_corr}(a) decreases with increasing $X_j$ measurement probability, the power-law coefficient of Fig.~\ref{fig:ZZ_born}(a) increases with increasing $Z_j Z_{j+1}$ measurement probability. This feature is easily understood on physical grounds: by projecting a large fraction of the ground state onto $X_j = \pm 1$ the short-range ferromagnetic correlations are reduced; in particular, since $\expval{Z_0 Z_r}_{\vm} = 0$ whenever $m_0 = \pm 1$ or $m_r = \pm 1$, $\overline{C_{\vm}^2(r)}$ is bounded above by the probability $(1-p)^2$ that both sites $0$ and $r$ remain unmeasured. In contrast, projecting a large fraction of the ground state onto $Z_j Z_{j+1} = \pm 1$ gives the resulting postmeasurement state $\ket{\psi^Z_{\vm}}$ short-range `spin-glass' structure; in particular, $[C^Z_{\vm}(r)]^2 = +1$ with probability at least $p^r$. Remarkably, even at large $p$ when typical states $\ket{\psi^Z_{\vm}}$ feature such spin-glass structure throughout the majority of the system, at sufficiently long distances the power-law scaling of the ground state is recovered.

Second, whereas the contribution $b_1(p)$ to the entanglement entropy $\overline{S_{\vm}(r)}$ strictly decreases with increasing measurement probability [see Fig.~\ref{fig:born_EE}], the analogous contribution $b_4(p)$ to $\overline{S^Z_{\vm}(r)}$ exhibits non-monotonic behavior [see Fig.~\ref{fig:ZZ_born}(c)]. In contrast to the $X_j$ measurement scheme of Sec.~\ref{sec:Born}, where measurements are localized within subsystem $A = [0:r)$ or its complement and strictly decrease the entanglement entropy, here measurements of $Z_{r-1} Z_r$ or $Z_{-1} Z_0$ are capable of increasing the entanglement entropy between the two subsystems.

\subsubsection*{Forced Projective \texorpdfstring{$Z_jZ_{j+1}$}{ZZ} Measurements}
We can similarly analyze the effects of forced $Z_j Z_{j+1}$ measurements, in which we postselect on the outcome $+1$ for each measurement of $Z_j Z_{j+1}$. Analogously to the discussion of Sec.~\ref{sec:postselection}, we describe this measurement protocol with the measurement operator
\begin{equation}
    \hat{K}^Z_{\vk} = \prod_{j = 1}^N \hat{K}^Z_{k_j, j}, \quad \hat{K}^Z_{0,j} = 1, \quad \hat{K}^Z_{1,j} = \frac{1 + Z_j Z_{j+1}}{2} .
\end{equation}
We then obtain the state $\ket{\psi^Z_{\vk}}$ with probability $p_{\vk}$, where
\begin{equation}
    \ket{\psi^Z_{\vk}} = \frac{\hat{K}^Z_{\vk} \ket{\psi_{\gs}}}{\sqrt{\expval*{\hat{K}^Z_{\vk}}_{\gs}}}, \quad p_{\vk} = p^{\abs{\vk}}(1-p)^{N-\abs{\vk}} ,
\end{equation}
where $\abs{\vk} = \sum_{j = 1}^N k_j$ is the number of measurements performed. The analysis of Sec.~\ref{sec:postselect_replicas} then follows identically, with the replacement (\ref{eq:app_X-ZZ_replacement}) in the averaged measurement operator. The end result is a defect described by the action
\begin{equation}
\label{eq:app_S_K_Z}
    \mathcal{S}^{(n)}_{K,Z} = - \nu \sum_{\alpha = 1}^n \int \dd{x} (\psi^T \sigma^y \psi)^{(\alpha)} + \ldots ,
\end{equation}
where as usual $\psi$ is here evaluated strictly at $\tau = 0$, and the ellipsis again contains irrelevant terms, including the derivative term of Eq.~(\ref{eq:app_X-ZZ_replacement}). Notably, $Z_j Z_{j+1}$ measurements yield an exactly marginal term identical to that of Eq.~(\ref{eq:S_K}), but crucially with the opposite sign. This change of sign does not affect the scaling dimension of $\psi$ and therefore is not expected to affect the asymptotic scaling of $\overline{G_{\vk}^Z(r)}$. On the other hand, it has crucial effects on the behavior of the correlations $C^Z_{\vk}(r) = \expval{Z_0 Z_r}_{\vk}^Z$. As a continuum analog of a two-dimensional classical Ising model with a defect line, the perturbation~(\ref{eq:S_K}) corresponds to weakened bonds along the defect line, and results in weaker ferromagnetic correlations along the defect. In contrast, Eq.~(\ref{eq:app_S_K_Z}) corresponds to \textit{strengthened} bonds along the  defect line, and results in enhanced ferromagnetic correlations. Following the results of Refs.~\cite{bariev_effect_1979,mccoy_two-spin_1980}, we expect $\overline{C_{\vk}^Z(r)} \sim r^{-2 \Delta^Z(p)}$ to again exhibit a continuously varying power law, but with a scaling dimension $\Delta^Z(p)$ which \textit{decreases} with increasing measurement strength. Asymptotically as $p \to 1$, we expect projecting $Z_j Z_{j+1} = +1$ almost everywhere results in near-perfect long-range order, so that $\Delta^Z(p) \to 0$ as $p \to 1$.

\begin{figure}
    \centering
    \includegraphics{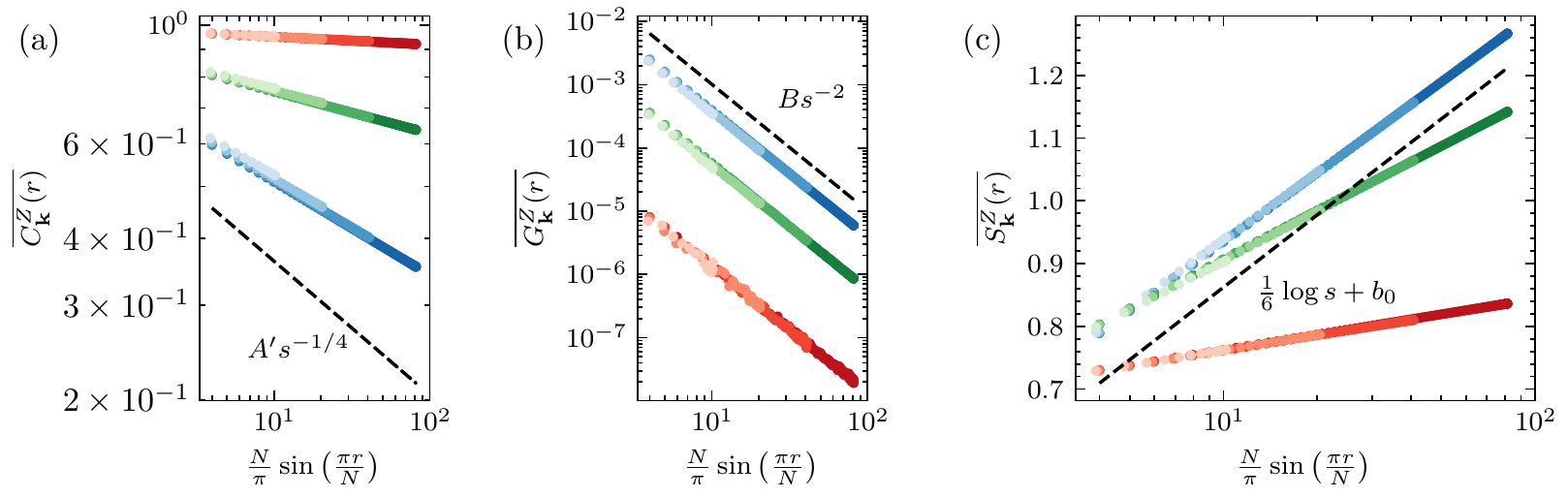}
    \caption{[(a),(b)] Ensemble-averaged correlation functions $\overline{C^Z_{\vk}(r)}$ and $\overline{G^Z_{\vk}(r)}$ and (c) entanglement entropy $\overline{S^Z_{\vk}(r)}$ in the forced-measurement ensemble, for measurement probabilities $p = 0.2$ (blue), $0.5$ (green), and $0.8$ (red), and for system sizes $N = 32$, $64$, $128$, and $256$ (light to dark). Data are plotted as a function of $s = \frac{N}{\pi} \sin \qty( \frac{\pi r}{N} )$ to achieve scaling collapse of the various system sizes. Dotted lines depict the behavior of the unmeasured system. While $\overline{G^Z_{\vk}(r)} \sim s^{-2}$ exhibits the same scaling as the unmeasured system, $\overline{C^Z_{\vk}(r)} \sim s^{-2\Delta^Z(p)}$ exhibits a continuously varying exponent $\Delta^Z(p)$; unlike in the forced $X_j$ measurement scheme, where $\overline{C_{\vk}(r)} \sim r^{-2\Delta(p)}$ featured a monotonically increasing scaling dimension $\Delta(p)$, here the scaling dimension $\Delta^Z(p)$ continuously decreases to zero with increasing measurement probability. The entanglement entropy $\overline{S^Z_{\vk}(r)} \sim \frac{c_{\eff}^Z(p)}{3} \log r + b_5(p)$ again features a continuously decreasing effective central charge $c_{\eff}^Z(p)$, but with a non-monotonic contribution $b_5(p)$.}
    \label{fig:ZZ_force}
\end{figure}

Figure~\ref{fig:ZZ_force} depicts the numerically computed correlation functions $\overline{C^Z_{\vk}(r)}$ and $\overline{G^Z_{\vk}(r)}$ and the entanglement entropy $\overline{S^Z_{\vk}(r)}$, defined analogously to Eq.~(\ref{eq:obsm}) with the replacement $\ket{\psi_{\vm}} \to \ket{\psi^Z_{\vk}}$. As expected, $\overline{G^Z_{\vk}(r)} \sim s^{-2}$ exhibits the same power-law scaling as in the unmeasured system, while $\overline{C^Z_{\vk}(r)} \sim s^{-2 \Delta^Z(p)}$ features a continuously varying power law characterized by a scaling dimension $\Delta^Z(p)$. As $p$ increases, $\Delta^Z(p)$ decreases towards zero, resulting in longer-ranged order parameter correlations. Similar to the entanglement entropy $\overline{S_{\vk}(r)}$ in Sec.~\ref{subsec:force_EE}, the entanglement entropy $\overline{S^Z_{\vk}(r)} \sim \frac{c_{\eff}^Z(p)}{3} \log s + b_5(p)$ exhibits a continuously decreasing effective central charge, which can be understood by mapping to a problem with ordinary timelke impurities as in Sec.~\ref{subsec:force_EE}. 

\section{`No-Click' Measurements}
\label{app_noclick}
Throughout the main text, we have considered two random projective measurement schemes. Focusing on projective measurements provides a closer comparison to typical experimental platforms, while performing measurements randomly throughout space restores translation invariance on average and effectively softens the average strength of the \textit{a priori} completely disentangling projective measurements. One conceptual downside to random measurement schemes, however, is the requirement of replicas in order to perform statistical averages. Random measurement schemes also impose an additional computational overhead due to Monte Carlo sampling, which becomes especially severe in the case of non-self-averaging observables. An alternative deterministic measurement scheme, which retains translation invariance, is to consider the effects of postselected weak measurements which only partially collapse the ground state. Here we consider a particular postselected weak measurement scheme which we call ``no-click'' measurements. A similar measurement scheme was used in the previous work of Ref.~\cite{garratt_measurements_2022}.

\begin{figure}
    \includegraphics{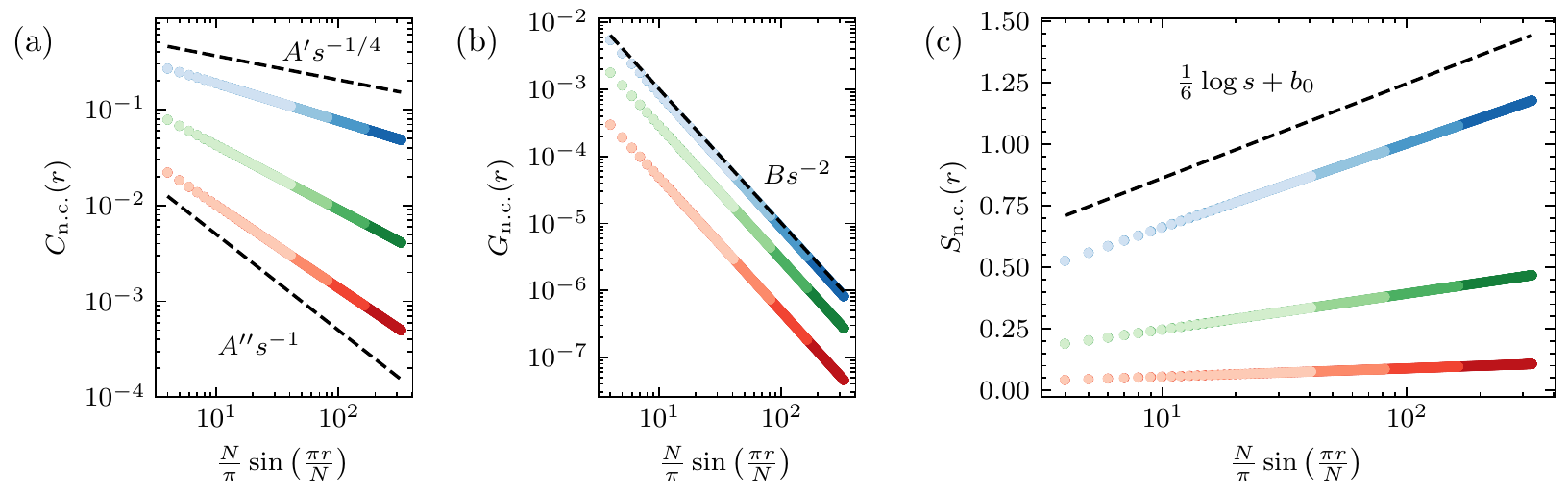}
    \caption{[(a),(b)] No-click correlation functions $C_{\nc}(r)$ and $G_{\nc}(r)$ and (c) entanglement entropy $S_{\nc}(r)$, for measurement strengths $1 - \cos \alpha = 0.1$ (blue), $0.3$ (green), and $0.5$ (red), and for system sizes $N = 128$, $256$, $512$, and $1024$ (light to dark). Data are plotted as a function of $s = \frac{N}{\pi} \sin \qty(\frac{\pi r}{N})$ to achieve scaling collapse of the various system sizes. Dotted lines depict the behavior in the unmeasured system. Similar to the case of forced projective measurements, $G_{\nc}(r) \sim s^{-2}$ exhibits the same power-law scaling as in the unmeasured system, while $C_{\nc}(r) \sim s^{-2 \Delta_{\nc}(\alpha)}$ exhibits a continuously varying power-law exponent and $S_{\nc}(r) \sim \frac{c_{\eff,\nc}(\alpha)}{3} \log s + b_{\nc}(\alpha)$ exhibits a continuously varying effective central charge.}
    \label{fig:noclick}
\end{figure}

To derive the no-click measurement of the observable $X_j$, we imagine introducing an ancillary qubit initialized in the state $\ket{0}$. We then couple this ancillary qubit to the $j$th spin of the system via the unitary
\begin{equation}
\label{eq:app_U_nc}
    U_{\nc} = \exp{-i \alpha \qty( \frac{1 - X_j}{2} ) \otimes \sigma^y } = \qty(\frac{1 + X_j}{2}) + \qty(\frac{1-X_j}{2}) \otimes e^{-i \alpha \sigma^y} ,
\end{equation}
where $\sigma^y$ acts on the ancillary qubit, and $0 \leq \alpha \leq \pi / 2$. Following the evolution by $U_{\nc}$, the state of the system plus ancilla is given by
\begin{equation}
    U_{\nc} \ket{\psi_{\gs}} \otimes \ket{0} = \qty[ 1 + (\cos \alpha -1) \qty( \frac{1 - X_j}{2} ) ] \ket{\psi_{\gs}} \otimes \ket{0} + \sin \alpha \qty( \frac{1 - X_j}{2} ) \ket{\psi_{\gs}} \otimes \ket{1} .
\end{equation}
Finally, we measure the ancilla qubit in the computational basis. A `click' of the measurement apparatus corresponds to the outcome $1$, which projects $X_j$ into the eigenstate $-1$. In the absence of a click, corresponding to the outcome $0$, the amplitude for $X_j = -1$ is only partially suppressed. Postselecting on the latter outcome, the effect of the no-click measurement is given by
\begin{equation}
    \ket{\psi_{\gs}} \mapsto \frac{e^{\lambda X_j}\ket{\psi_{\gs}}}{\expval{e^{2\lambda X_j}}_{\gs}}, 
\end{equation}
where $\lambda$ is a monotonic function of $\alpha$, with $\lambda = \infty$ corresponding to the projective measurement $\alpha = \pi/2$. Performing the same no-click measurement on each qubit, we altogether obtain the state
\begin{equation}
    \ket{\psi_{\nc}} = \frac{\hat{K}_{\nc} \ket{\psi_{\gs}}}{\expval*{\hat{K}_{\nc}^2}_{\gs}}, \quad \hat{K}_{\nc} = \prod_{j = 1}^N e^{\lambda X_j} .
\end{equation}

We are interested in comparing the behavior of observables in the no-click state $\ket{\psi_{\nc}}$ to that of the unmeasured ground state $\ket{\psi_{\gs}}$. For example, the connected energy density correlator is given by
\begin{equation}
    G_{\nc}(r) = \expval{X_0 X_r}_{\nc} - \expval{X_0}_{\nc} \expval{X_r}_{\nc} = \frac{ \expval*{X_0 X_r \hat{K}^2_{\nc}}_{\gs} }{\expval*{\hat{K}^2_{\nc}}_{\gs}} - \frac{ \expval*{X_0 \hat{K}^2_{\nc}}_{\gs} }{\expval*{\hat{K}^2_{\nc}}_{\gs}} \frac{ \expval*{ X_r \hat{K}^2_{\nc}}_{\gs} }{\expval*{\hat{K}^2_{\nc}}_{\gs}} ,
\end{equation}
where we have used $\comm*{X_j}{\hat{M}_{\nc}} = 0$. Order parameter correlations $C_{\nc}(r) = \expval{Z_0 Z_r}_{\nc}$ and entanglement entropy $S_{\nc}(r) = -\tr \rho^A_{\nc} \log \rho^A_{\nc}$ are defined similarly, the former of which can be analyzed within the same framework using the method of Appendix \ref{app:noncommuting}. From this expression, we see that correlations in the postmeasurement state can be obtained from correlations of the following partition function:
\begin{equation}
\label{eq:app_Z_nc}
    \mathcal{Z}_{\nc} = \bra{\psi_{\gs}} \hat{K}^2_{\nc} \ket{\psi_{\gs}} = \int D\psi \, \exp{ -\mathcal{S}_0[\psi] + \tilde{\lambda} \int \dd{x} \psi^T \sigma^y \psi } ,
\end{equation}
where $\tilde{\lambda}$ is a monotonic function of $\lambda$, and in the latter term $\psi$ is evaluated at $\tau = 0$. We immediately see that no-click measurements result in an exactly marginal defect along the $\tau = 0$ line, of exactly the same form as in the forced measurement scheme [see Eq.~(\ref{eq:S_K})]. We therefore expect identical phenomenology for the long-distance correlations: in particular, we expect $G_{\nc}(r) \sim r^{-2}$ to retain the same scaling as in the unmeasured state, while $C_{\nc}(r) \sim r^{-2 \Delta_{\nc}(\alpha)}$ obtains a continuously varying power-law exponent with $\Delta_{\nc}(0) = 1/8$ and $\lim_{\alpha \to \pi / 2} \Delta_{\nc}(\alpha) = 1/2$. We additionally expect the entanglement entropy to exhibit a continuously varying effective central charge, $S_{\nc}(r) \sim \frac{c_{\eff,\nc}(\alpha)}{3} \log r + b_{\nc}(\alpha)$, such that $c_{\eff,\nc}(0) = \pi / 2$ and $c_{\eff,\nc}(\alpha)$ decreases towards zero as $\alpha \to \pi/2$. These qualitative features are verified numerically in Fig.~\ref{fig:noclick}.

\begin{figure}
    \includegraphics{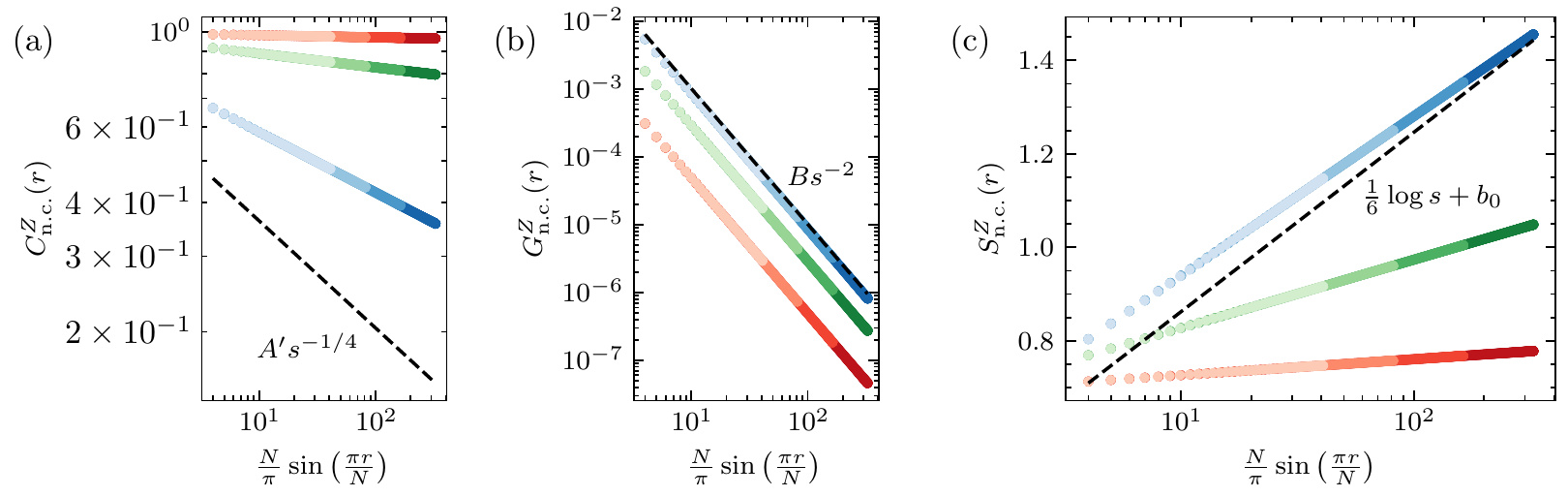}
    \caption{[(a),(b)] No-click correlation functions $C^Z_{\nc}(r)$ and $G^Z_{\nc}(r)$ and (c) entanglement entropy $S^Z_{\nc}(r)$, for measurement strengths $1 - \cos \alpha = 0.1$ (blue), 0.3 (green), and 0.5 (red), and for system sizes $N = 128$, $256$, $512$, and $1024$ (light to dark). Data are plotted as a function of $s = \frac{N}{\pi} \sin \qty( \frac{\pi r}{N} )$ to achieve scaling collapse of the various system sizes. Dotted lines depict the behavior of the unmeasured system. Similar to the case of forced projective $Z_j Z_{j+1}$ measurements, $G^Z_{\nc}(r) \sim s^{-2}$ exhibits the same power-law scaling as the unmeasured system, while $C^Z_{\nc}(r) \sim s^{-2\Delta_{\nc}^Z(\alpha)}$ exhibits a continuously varying power-law exponent and $S^Z_{\nc}(r) \sim \frac{c_{\eff,\nc}^Z}{3} \log s + b^Z_{\nc}(\alpha)$ exhibits a continuously varying effective central charge.}
    \label{fig:noclick_ZZ}
\end{figure}

We may alternatively consider no-click measurements of $Z_j Z_{j+1}$. The unitary $U^Z_{\nc}$ which implements such a measurement is obtained simply by replacing $X_j$ with $Z_j Z_{j+1}$ in Eq.~(\ref{eq:app_U_nc}). The resulting state following $Z_j Z_{j+1}$ no-click measurements for each site $j$ is
\begin{equation}
    \ket{\psi_{\nc}^Z} = \frac{\hat{K}^Z_{\nc} \ket{\psi_{\gs}}}{\expval*{[\hat{K}^Z_{\nc}]^2}_{\gs}}, \quad \hat{K}^Z_{\nc} = \prod_{j = 1}^N e^{\lambda Z_j Z_{j+1}} .
\end{equation}
 Similarly to the discussion of Appendix~\ref{app:ZZ}, replacing $X_j$ by $Z_j Z_{j+1}$ results in the same type of defect as in (\ref{eq:app_Z_nc}) (up to irrelevant terms), but with an altered sign. Explicitly, the partition function used to evaluate correlation functions is given by
 \begin{equation}
     \mathcal{Z}^Z_{\nc} = \bra{\psi_{\gs}} \hat{K}^2_{\nc} \ket{\psi_{\nc}} = \int D\psi \, \exp{ - \mathcal{S}_0[\psi] - \tilde{\lambda}_Z \int \dd{x} \psi^T \sigma^y \psi } ,
 \end{equation}
 where $\lambda_Z$ is a monotonic function of $\lambda$, the latter term in the exponential is again evaluated strictly at $\tau = 0$, and we have neglected irrelevant terms. Identically to the case of forced $Z_j Z_{j+1}$ measurements, we expect $C_{\nc}^Z(r) \sim r^{-2 \Delta_{\nc}^Z(\alpha)}$ to exhibit a continuously \textit{decreasing} power-law exponent $\Delta^Z_{\nc}(\alpha)$ with increasing measurement strength, while $S_{\nc}^Z(r) \sim \frac{c_{\eff,\nc}^Z(\alpha)}{3} \log r + b_{\nc}^Z(\alpha)$ exhibits a continuously decreasing effective central charge $c_{\eff,\nc}^Z(\alpha)$. These features are again verified numerically in Fig.~\ref{fig:noclick_ZZ}.

\twocolumngrid

\bibliographystyle{apsrev4-2}
\bibliography{refs}

\begin{thebibliography}{86}%
\makeatletter
\providecommand \@ifxundefined [1]{%
 \@ifx{#1\undefined}
}%
\providecommand \@ifnum [1]{%
 \ifnum #1\expandafter \@firstoftwo
 \else \expandafter \@secondoftwo
 \fi
}%
\providecommand \@ifx [1]{%
 \ifx #1\expandafter \@firstoftwo
 \else \expandafter \@secondoftwo
 \fi
}%
\providecommand \natexlab [1]{#1}%
\providecommand \enquote  [1]{``#1''}%
\providecommand \bibnamefont  [1]{#1}%
\providecommand \bibfnamefont [1]{#1}%
\providecommand \citenamefont [1]{#1}%
\providecommand \href@noop [0]{\@secondoftwo}%
\providecommand \href [0]{\begingroup \@sanitize@url \@href}%
\providecommand \@href[1]{\@@startlink{#1}\@@href}%
\providecommand \@@href[1]{\endgroup#1\@@endlink}%
\providecommand \@sanitize@url [0]{\catcode `\\12\catcode `\$12\catcode
  `\&12\catcode `\#12\catcode `\^12\catcode `\_12\catcode `\%12\relax}%
\providecommand \@@startlink[1]{}%
\providecommand \@@endlink[0]{}%
\providecommand \url  [0]{\begingroup\@sanitize@url \@url }%
\providecommand \@url [1]{\endgroup\@href {#1}{\urlprefix }}%
\providecommand \urlprefix  [0]{URL }%
\providecommand \Eprint [0]{\href }%
\providecommand \doibase [0]{https://doi.org/}%
\providecommand \selectlanguage [0]{\@gobble}%
\providecommand \bibinfo  [0]{\@secondoftwo}%
\providecommand \bibfield  [0]{\@secondoftwo}%
\providecommand \translation [1]{[#1]}%
\providecommand \BibitemOpen [0]{}%
\providecommand \bibitemStop [0]{}%
\providecommand \bibitemNoStop [0]{.\EOS\space}%
\providecommand \EOS [0]{\spacefactor3000\relax}%
\providecommand \BibitemShut  [1]{\csname bibitem#1\endcsname}%
\let\auto@bib@innerbib\@empty
\bibitem [{\citenamefont {Briegel}\ \emph {et~al.}(2009)\citenamefont
  {Briegel}, \citenamefont {Browne}, \citenamefont {D{\"u}r}, \citenamefont
  {Raussendorf},\ and\ \citenamefont {Van~den
  Nest}}]{briegel_2009_measurement}%
  \BibitemOpen
  \bibfield  {author} {\bibinfo {author} {\bibfnamefont {H.~J.}\ \bibnamefont
  {Briegel}}, \bibinfo {author} {\bibfnamefont {D.~E.}\ \bibnamefont {Browne}},
  \bibinfo {author} {\bibfnamefont {W.}~\bibnamefont {D{\"u}r}}, \bibinfo
  {author} {\bibfnamefont {R.}~\bibnamefont {Raussendorf}},\ and\ \bibinfo
  {author} {\bibfnamefont {M.}~\bibnamefont {Van~den Nest}},\ }\href
  {https://doi.org/10.1038/nphys1157} {\bibfield  {journal} {\bibinfo
  {journal} {Nat. Phys.}\ }\textbf {\bibinfo {volume} {5}},\ \bibinfo {pages}
  {19} (\bibinfo {year} {2009})}\BibitemShut {NoStop}%
\bibitem [{\citenamefont {Li}\ \emph {et~al.}(2021)\citenamefont {Li},
  \citenamefont {Chen}, \citenamefont {Ludwig},\ and\ \citenamefont
  {Fisher}}]{li_2021_conformal}%
  \BibitemOpen
  \bibfield  {author} {\bibinfo {author} {\bibfnamefont {Y.}~\bibnamefont
  {Li}}, \bibinfo {author} {\bibfnamefont {X.}~\bibnamefont {Chen}}, \bibinfo
  {author} {\bibfnamefont {A.~W.~W.}\ \bibnamefont {Ludwig}},\ and\ \bibinfo
  {author} {\bibfnamefont {M.~P.~A.}\ \bibnamefont {Fisher}},\ }\href
  {https://doi.org/10.1103/PhysRevB.104.104305} {\bibfield  {journal} {\bibinfo
   {journal} {Phys. Rev. B}\ }\textbf {\bibinfo {volume} {104}},\ \bibinfo
  {pages} {104305} (\bibinfo {year} {2021})}\BibitemShut {NoStop}%
\bibitem [{\citenamefont {Piroli}\ \emph {et~al.}(2021)\citenamefont {Piroli},
  \citenamefont {Styliaris},\ and\ \citenamefont
  {Cirac}}]{piroli_quantum_2021}%
  \BibitemOpen
  \bibfield  {author} {\bibinfo {author} {\bibfnamefont {L.}~\bibnamefont
  {Piroli}}, \bibinfo {author} {\bibfnamefont {G.}~\bibnamefont {Styliaris}},\
  and\ \bibinfo {author} {\bibfnamefont {J.~I.}\ \bibnamefont {Cirac}},\ }\href
  {https://doi.org/10.1103/PhysRevLett.127.220503} {\bibfield  {journal}
  {\bibinfo  {journal} {Phys. Rev. Lett.}\ }\textbf {\bibinfo {volume} {127}},\
  \bibinfo {pages} {220503} (\bibinfo {year} {2021})}\BibitemShut {NoStop}%
\bibitem [{\citenamefont {Bao}\ \emph {et~al.}(2021{\natexlab{a}})\citenamefont
  {Bao}, \citenamefont {Block},\ and\ \citenamefont
  {Altman}}]{bao_2021_teleportation}%
  \BibitemOpen
  \bibfield  {author} {\bibinfo {author} {\bibfnamefont {Y.}~\bibnamefont
  {Bao}}, \bibinfo {author} {\bibfnamefont {M.}~\bibnamefont {Block}},\ and\
  \bibinfo {author} {\bibfnamefont {E.}~\bibnamefont {Altman}},\ }\href
  {https://arxiv.org/abs/2110.06963} {\bibfield  {journal} {\bibinfo  {journal}
  {arXiv:2110.06963}\ } (\bibinfo {year} {2021}{\natexlab{a}})}\BibitemShut
  {NoStop}%
\bibitem [{\citenamefont {Tantivasadakarn}\ \emph {et~al.}(2021)\citenamefont
  {Tantivasadakarn}, \citenamefont {Thorngren}, \citenamefont {Vishwanath},\
  and\ \citenamefont {Verresen}}]{tantivasadakarn_long_2021}%
  \BibitemOpen
  \bibfield  {author} {\bibinfo {author} {\bibfnamefont {N.}~\bibnamefont
  {Tantivasadakarn}}, \bibinfo {author} {\bibfnamefont {R.}~\bibnamefont
  {Thorngren}}, \bibinfo {author} {\bibfnamefont {A.}~\bibnamefont
  {Vishwanath}},\ and\ \bibinfo {author} {\bibfnamefont {R.}~\bibnamefont
  {Verresen}},\ }\href {https://arxiv.org/abs/2112.01519} {\bibfield  {journal}
  {\bibinfo  {journal} {arXiv:2112.01519}\ } (\bibinfo {year}
  {2021})}\BibitemShut {NoStop}%
\bibitem [{\citenamefont {Verresen}\ \emph {et~al.}(2021)\citenamefont
  {Verresen}, \citenamefont {Tantivasadakarn},\ and\ \citenamefont
  {Vishwanath}}]{verresen_efficiently_2021}%
  \BibitemOpen
  \bibfield  {author} {\bibinfo {author} {\bibfnamefont {R.}~\bibnamefont
  {Verresen}}, \bibinfo {author} {\bibfnamefont {N.}~\bibnamefont
  {Tantivasadakarn}},\ and\ \bibinfo {author} {\bibfnamefont {A.}~\bibnamefont
  {Vishwanath}},\ }\href {https://arxiv.org/abs/2112.03061} {\bibfield
  {journal} {\bibinfo  {journal} {arXiv:2112.03061}\ } (\bibinfo {year}
  {2021})}\BibitemShut {NoStop}%
\bibitem [{\citenamefont {Lin}\ \emph {et~al.}(2023)\citenamefont {Lin},
  \citenamefont {Ye}, \citenamefont {Zou}, \citenamefont {Sang},\ and\
  \citenamefont {Hsieh}}]{lin_2022_probing}%
  \BibitemOpen
  \bibfield  {author} {\bibinfo {author} {\bibfnamefont {C.-J.}\ \bibnamefont
  {Lin}}, \bibinfo {author} {\bibfnamefont {W.}~\bibnamefont {Ye}}, \bibinfo
  {author} {\bibfnamefont {Y.}~\bibnamefont {Zou}}, \bibinfo {author}
  {\bibfnamefont {S.}~\bibnamefont {Sang}},\ and\ \bibinfo {author}
  {\bibfnamefont {T.~H.}\ \bibnamefont {Hsieh}},\ }\href
  {https://doi.org/10.22331/q-2023-02-02-910} {\bibfield  {journal} {\bibinfo
  {journal} {Quantum}\ }\textbf {\bibinfo {volume} {7}},\ \bibinfo {pages}
  {910} (\bibinfo {year} {2023})}\BibitemShut {NoStop}%
\bibitem [{\citenamefont {Lu}\ \emph {et~al.}(2022)\citenamefont {Lu},
  \citenamefont {Lessa}, \citenamefont {Kim},\ and\ \citenamefont
  {Hsieh}}]{lu_measurement_2022}%
  \BibitemOpen
  \bibfield  {author} {\bibinfo {author} {\bibfnamefont {T.-C.}\ \bibnamefont
  {Lu}}, \bibinfo {author} {\bibfnamefont {L.~A.}\ \bibnamefont {Lessa}},
  \bibinfo {author} {\bibfnamefont {I.~H.}\ \bibnamefont {Kim}},\ and\ \bibinfo
  {author} {\bibfnamefont {T.~H.}\ \bibnamefont {Hsieh}},\ }\href
  {https://doi.org/10.1103/PRXQuantum.3.040337} {\bibfield  {journal} {\bibinfo
   {journal} {Phys. Rev. X Quantum}\ }\textbf {\bibinfo {volume} {3}},\
  \bibinfo {pages} {040337} (\bibinfo {year} {2022})}\BibitemShut {NoStop}%
\bibitem [{\citenamefont {Garratt}\ \emph {et~al.}(2023)\citenamefont
  {Garratt}, \citenamefont {Weinstein},\ and\ \citenamefont
  {Altman}}]{garratt_measurements_2022}%
  \BibitemOpen
  \bibfield  {author} {\bibinfo {author} {\bibfnamefont {S.~J.}\ \bibnamefont
  {Garratt}}, \bibinfo {author} {\bibfnamefont {Z.}~\bibnamefont {Weinstein}},\
  and\ \bibinfo {author} {\bibfnamefont {E.}~\bibnamefont {Altman}},\ }\href
  {https://doi.org/10.1103/PhysRevX.13.021026} {\bibfield  {journal} {\bibinfo
  {journal} {Phys. Rev. X}\ }\textbf {\bibinfo {volume} {13}},\ \bibinfo
  {pages} {021026} (\bibinfo {year} {2023})}\BibitemShut {NoStop}%
\bibitem [{\citenamefont {Zhu}\ \emph {et~al.}(2022)\citenamefont {Zhu},
  \citenamefont {Tantivasadakarn}, \citenamefont {Vishwanath}, \citenamefont
  {Trebst},\ and\ \citenamefont {Verresen}}]{zhu_nishimori_2022}%
  \BibitemOpen
  \bibfield  {author} {\bibinfo {author} {\bibfnamefont {G.-Y.}\ \bibnamefont
  {Zhu}}, \bibinfo {author} {\bibfnamefont {N.}~\bibnamefont
  {Tantivasadakarn}}, \bibinfo {author} {\bibfnamefont {A.}~\bibnamefont
  {Vishwanath}}, \bibinfo {author} {\bibfnamefont {S.}~\bibnamefont {Trebst}},\
  and\ \bibinfo {author} {\bibfnamefont {R.}~\bibnamefont {Verresen}},\ }\href
  {https://arxiv.org/abs/2208.11136} {\bibfield  {journal} {\bibinfo  {journal}
  {arXiv:2208.11136}\ } (\bibinfo {year} {2022})}\BibitemShut {NoStop}%
\bibitem [{\citenamefont {Lee}\ \emph {et~al.}(2022{\natexlab{a}})\citenamefont
  {Lee}, \citenamefont {Ji}, \citenamefont {Bi},\ and\ \citenamefont
  {Fisher}}]{lee_2022_decoding}%
  \BibitemOpen
  \bibfield  {author} {\bibinfo {author} {\bibfnamefont {J.~Y.}\ \bibnamefont
  {Lee}}, \bibinfo {author} {\bibfnamefont {W.}~\bibnamefont {Ji}}, \bibinfo
  {author} {\bibfnamefont {Z.}~\bibnamefont {Bi}},\ and\ \bibinfo {author}
  {\bibfnamefont {M.~P.~A.}\ \bibnamefont {Fisher}},\ }\href
  {https://arxiv.org/abs/2208.11699} {\bibfield  {journal} {\bibinfo  {journal}
  {arXiv:2208.11699}\ } (\bibinfo {year} {2022}{\natexlab{a}})}\BibitemShut
  {NoStop}%
\bibitem [{\citenamefont {Tantivasadakarn}\ \emph
  {et~al.}(2022{\natexlab{a}})\citenamefont {Tantivasadakarn}, \citenamefont
  {Verresen},\ and\ \citenamefont
  {Vishwanath}}]{tantivasadakarn_shortest_2022}%
  \BibitemOpen
  \bibfield  {author} {\bibinfo {author} {\bibfnamefont {N.}~\bibnamefont
  {Tantivasadakarn}}, \bibinfo {author} {\bibfnamefont {R.}~\bibnamefont
  {Verresen}},\ and\ \bibinfo {author} {\bibfnamefont {A.}~\bibnamefont
  {Vishwanath}},\ }\href {https://arxiv.org/abs/2209.03964} {\bibfield
  {journal} {\bibinfo  {journal} {arXiv:2209.03964}\ } (\bibinfo {year}
  {2022}{\natexlab{a}})}\BibitemShut {NoStop}%
\bibitem [{\citenamefont {Tantivasadakarn}\ \emph
  {et~al.}(2022{\natexlab{b}})\citenamefont {Tantivasadakarn}, \citenamefont
  {Vishwanath},\ and\ \citenamefont
  {Verresen}}]{tantivasadakarn_hierarchy_2022}%
  \BibitemOpen
  \bibfield  {author} {\bibinfo {author} {\bibfnamefont {N.}~\bibnamefont
  {Tantivasadakarn}}, \bibinfo {author} {\bibfnamefont {A.}~\bibnamefont
  {Vishwanath}},\ and\ \bibinfo {author} {\bibfnamefont {R.}~\bibnamefont
  {Verresen}},\ }\href {https://arxiv.org/abs/2209.06202} {\bibfield  {journal}
  {\bibinfo  {journal} {arXiv:2209.06202}\ } (\bibinfo {year}
  {2022}{\natexlab{b}})}\BibitemShut {NoStop}%
\bibitem [{\citenamefont {Antonini}\ \emph {et~al.}(2022)\citenamefont
  {Antonini}, \citenamefont {Bentsen}, \citenamefont {Cao}, \citenamefont
  {Harper}, \citenamefont {Jian},\ and\ \citenamefont
  {Swingle}}]{antonini_2022_holographic}%
  \BibitemOpen
  \bibfield  {author} {\bibinfo {author} {\bibfnamefont {S.}~\bibnamefont
  {Antonini}}, \bibinfo {author} {\bibfnamefont {G.}~\bibnamefont {Bentsen}},
  \bibinfo {author} {\bibfnamefont {C.}~\bibnamefont {Cao}}, \bibinfo {author}
  {\bibfnamefont {J.}~\bibnamefont {Harper}}, \bibinfo {author} {\bibfnamefont
  {S.-K.}\ \bibnamefont {Jian}},\ and\ \bibinfo {author} {\bibfnamefont
  {B.}~\bibnamefont {Swingle}},\ }\href
  {https://doi.org/10.1007/JHEP12(2022)124} {\bibfield  {journal} {\bibinfo
  {journal} {J. High Energy Phys.}\ }\textbf {\bibinfo {volume} {2022}}\bibinfo
   {number} { (12)},\ \bibinfo {pages} {1}}\BibitemShut {NoStop}%
\bibitem [{\citenamefont {Antonini}\ \emph {et~al.}(2023)\citenamefont
  {Antonini}, \citenamefont {Grado-White}, \citenamefont {Jian},\ and\
  \citenamefont {Swingle}}]{antonini2022holographic}%
  \BibitemOpen
\bibfield  {number} {  }\bibfield  {author} {\bibinfo {author} {\bibfnamefont
  {S.}~\bibnamefont {Antonini}}, \bibinfo {author} {\bibfnamefont
  {B.}~\bibnamefont {Grado-White}}, \bibinfo {author} {\bibfnamefont {S.-K.}\
  \bibnamefont {Jian}},\ and\ \bibinfo {author} {\bibfnamefont
  {B.}~\bibnamefont {Swingle}},\ }\href
  {https://doi.org/10.1007/JHEP02(2023)095} {\bibfield  {journal} {\bibinfo
  {journal} {J. High Energy Phys.}\ }\textbf {\bibinfo {volume} {2023}}\bibinfo
   {number} { (2)},\ \bibinfo {pages} {95}}\BibitemShut {NoStop}%
\bibitem [{\citenamefont {Skinner}\ \emph {et~al.}(2019)\citenamefont
  {Skinner}, \citenamefont {Ruhman},\ and\ \citenamefont
  {Nahum}}]{skinner_measurement-induced_2019}%
  \BibitemOpen
\bibfield  {number} {  }\bibfield  {author} {\bibinfo {author} {\bibfnamefont
  {B.}~\bibnamefont {Skinner}}, \bibinfo {author} {\bibfnamefont
  {J.}~\bibnamefont {Ruhman}},\ and\ \bibinfo {author} {\bibfnamefont
  {A.}~\bibnamefont {Nahum}},\ }\href
  {https://doi.org/10.1103/PhysRevX.9.031009} {\bibfield  {journal} {\bibinfo
  {journal} {Phys. Rev. X}\ }\textbf {\bibinfo {volume} {9}},\ \bibinfo {pages}
  {031009} (\bibinfo {year} {2019})}\BibitemShut {NoStop}%
\bibitem [{\citenamefont {Li}\ \emph {et~al.}(2018)\citenamefont {Li},
  \citenamefont {Chen},\ and\ \citenamefont {Fisher}}]{li_quantum_2018}%
  \BibitemOpen
  \bibfield  {author} {\bibinfo {author} {\bibfnamefont {Y.}~\bibnamefont
  {Li}}, \bibinfo {author} {\bibfnamefont {X.}~\bibnamefont {Chen}},\ and\
  \bibinfo {author} {\bibfnamefont {M.~P.~A.}\ \bibnamefont {Fisher}},\ }\href
  {https://doi.org/10.1103/PhysRevB.98.205136} {\bibfield  {journal} {\bibinfo
  {journal} {Phys. Rev. B}\ }\textbf {\bibinfo {volume} {98}},\ \bibinfo
  {pages} {205136} (\bibinfo {year} {2018})}\BibitemShut {NoStop}%
\bibitem [{\citenamefont {Li}\ \emph {et~al.}(2019)\citenamefont {Li},
  \citenamefont {Chen},\ and\ \citenamefont
  {Fisher}}]{li_measurement-driven_2019}%
  \BibitemOpen
  \bibfield  {author} {\bibinfo {author} {\bibfnamefont {Y.}~\bibnamefont
  {Li}}, \bibinfo {author} {\bibfnamefont {X.}~\bibnamefont {Chen}},\ and\
  \bibinfo {author} {\bibfnamefont {M.~P.~A.}\ \bibnamefont {Fisher}},\ }\href
  {https://doi.org/10.1103/PhysRevB.100.134306} {\bibfield  {journal} {\bibinfo
   {journal} {Phys. Rev. B}\ }\textbf {\bibinfo {volume} {100}},\ \bibinfo
  {pages} {134306} (\bibinfo {year} {2019})}\BibitemShut {NoStop}%
\bibitem [{\citenamefont {Gullans}\ and\ \citenamefont
  {Huse}(2020{\natexlab{a}})}]{gullans_dynamical_2020}%
  \BibitemOpen
  \bibfield  {author} {\bibinfo {author} {\bibfnamefont {M.~J.}\ \bibnamefont
  {Gullans}}\ and\ \bibinfo {author} {\bibfnamefont {D.~A.}\ \bibnamefont
  {Huse}},\ }\href {https://doi.org/10.1103/PhysRevX.10.041020} {\bibfield
  {journal} {\bibinfo  {journal} {Phys. Rev. X}\ }\textbf {\bibinfo {volume}
  {10}},\ \bibinfo {pages} {041020} (\bibinfo {year}
  {2020}{\natexlab{a}})}\BibitemShut {NoStop}%
\bibitem [{\citenamefont {Choi}\ \emph {et~al.}(2020)\citenamefont {Choi},
  \citenamefont {Bao}, \citenamefont {Qi},\ and\ \citenamefont
  {Altman}}]{choi_quantum_2020}%
  \BibitemOpen
  \bibfield  {author} {\bibinfo {author} {\bibfnamefont {S.}~\bibnamefont
  {Choi}}, \bibinfo {author} {\bibfnamefont {Y.}~\bibnamefont {Bao}}, \bibinfo
  {author} {\bibfnamefont {X.-L.}\ \bibnamefont {Qi}},\ and\ \bibinfo {author}
  {\bibfnamefont {E.}~\bibnamefont {Altman}},\ }\href
  {https://doi.org/10.1103/PhysRevLett.125.030505} {\bibfield  {journal}
  {\bibinfo  {journal} {Phys. Rev. Lett.}\ }\textbf {\bibinfo {volume} {125}},\
  \bibinfo {pages} {030505} (\bibinfo {year} {2020})}\BibitemShut {NoStop}%
\bibitem [{\citenamefont {Bao}\ \emph {et~al.}(2020)\citenamefont {Bao},
  \citenamefont {Choi},\ and\ \citenamefont {Altman}}]{bao_theory_2020}%
  \BibitemOpen
  \bibfield  {author} {\bibinfo {author} {\bibfnamefont {Y.}~\bibnamefont
  {Bao}}, \bibinfo {author} {\bibfnamefont {S.}~\bibnamefont {Choi}},\ and\
  \bibinfo {author} {\bibfnamefont {E.}~\bibnamefont {Altman}},\ }\href
  {https://doi.org/10.1103/PhysRevB.101.104301} {\bibfield  {journal} {\bibinfo
   {journal} {Phys. Rev. B}\ }\textbf {\bibinfo {volume} {101}},\ \bibinfo
  {pages} {104301} (\bibinfo {year} {2020})}\BibitemShut {NoStop}%
\bibitem [{\citenamefont {Jian}\ \emph {et~al.}(2020)\citenamefont {Jian},
  \citenamefont {You}, \citenamefont {Vasseur},\ and\ \citenamefont
  {Ludwig}}]{jian_measurement-induced_2020}%
  \BibitemOpen
  \bibfield  {author} {\bibinfo {author} {\bibfnamefont {C.-M.}\ \bibnamefont
  {Jian}}, \bibinfo {author} {\bibfnamefont {Y.-Z.}\ \bibnamefont {You}},
  \bibinfo {author} {\bibfnamefont {R.}~\bibnamefont {Vasseur}},\ and\ \bibinfo
  {author} {\bibfnamefont {A.~W.~W.}\ \bibnamefont {Ludwig}},\ }\href
  {https://doi.org/10.1103/PhysRevB.101.104302} {\bibfield  {journal} {\bibinfo
   {journal} {Phys. Rev. B}\ }\textbf {\bibinfo {volume} {101}},\ \bibinfo
  {pages} {104302} (\bibinfo {year} {2020})}\BibitemShut {NoStop}%
\bibitem [{\citenamefont {Potter}\ and\ \citenamefont
  {Vasseur}(2022)}]{potter_entanglement_2022}%
  \BibitemOpen
  \bibfield  {author} {\bibinfo {author} {\bibfnamefont {A.~C.}\ \bibnamefont
  {Potter}}\ and\ \bibinfo {author} {\bibfnamefont {R.}~\bibnamefont
  {Vasseur}},\ }\bibinfo {title} {Entanglement dynamics in hybrid quantum
  circuits},\ in\ \href {https://doi.org/10.1007/978-3-031-03998-0_9} {\emph
  {\bibinfo {booktitle} {Entanglement in Spin Chains: From Theory to Quantum
  Technology Applications}}}\ (\bibinfo  {publisher} {Springer International
  Publishing},\ \bibinfo {address} {Cham},\ \bibinfo {year} {2022})\ pp.\
  \bibinfo {pages} {211--249}\BibitemShut {NoStop}%
\bibitem [{\citenamefont {Fisher}\ \emph {et~al.}(2023)\citenamefont {Fisher},
  \citenamefont {Khemani}, \citenamefont {Nahum},\ and\ \citenamefont
  {Vijay}}]{fisher_random_2022}%
  \BibitemOpen
  \bibfield  {author} {\bibinfo {author} {\bibfnamefont {M.~P.~A.}\
  \bibnamefont {Fisher}}, \bibinfo {author} {\bibfnamefont {V.}~\bibnamefont
  {Khemani}}, \bibinfo {author} {\bibfnamefont {A.}~\bibnamefont {Nahum}},\
  and\ \bibinfo {author} {\bibfnamefont {S.}~\bibnamefont {Vijay}},\ }\href
  {https://doi.org/10.1146/annurev-conmatphys-031720-030658} {\bibfield
  {journal} {\bibinfo  {journal} {Annu. Rev. Condens. Matter Phys.}\ }\textbf
  {\bibinfo {volume} {14}} (\bibinfo {year} {2023})}\BibitemShut {NoStop}%
\bibitem [{\citenamefont {Sachdev}(2011)}]{sachdev_quantum_2011}%
  \BibitemOpen
  \bibfield  {author} {\bibinfo {author} {\bibfnamefont {S.}~\bibnamefont
  {Sachdev}},\ }\href {https://doi.org/10.1017/CBO9780511973765} {\emph
  {\bibinfo {title} {Quantum Phase Transitions}}},\ \bibinfo {edition} {2nd}\
  ed.\ (\bibinfo  {publisher} {Cambridge University Press},\ \bibinfo {year}
  {2011})\BibitemShut {NoStop}%
\bibitem [{\citenamefont {Belavin}\ \emph {et~al.}(1984)\citenamefont
  {Belavin}, \citenamefont {Polyakov},\ and\ \citenamefont
  {Zamolodchikov}}]{belavin_infinite_1984}%
  \BibitemOpen
  \bibfield  {author} {\bibinfo {author} {\bibfnamefont {A.~A.}\ \bibnamefont
  {Belavin}}, \bibinfo {author} {\bibfnamefont {A.~M.}\ \bibnamefont
  {Polyakov}},\ and\ \bibinfo {author} {\bibfnamefont {A.~B.}\ \bibnamefont
  {Zamolodchikov}},\ }\href {https://doi.org/10.1016/0550-3213(84)90052-X}
  {\bibfield  {journal} {\bibinfo  {journal} {Nucl. Phys. B}\ }\textbf
  {\bibinfo {volume} {241}},\ \bibinfo {pages} {333} (\bibinfo {year}
  {1984})}\BibitemShut {NoStop}%
\bibitem [{\citenamefont {Di~Francesco}\ \emph {et~al.}(1997)\citenamefont
  {Di~Francesco}, \citenamefont {Mathieu},\ and\ \citenamefont
  {Sénéchal}}]{di_francesco_conformal_1997}%
  \BibitemOpen
  \bibfield  {author} {\bibinfo {author} {\bibfnamefont {P.}~\bibnamefont
  {Di~Francesco}}, \bibinfo {author} {\bibfnamefont {P.}~\bibnamefont
  {Mathieu}},\ and\ \bibinfo {author} {\bibfnamefont {D.}~\bibnamefont
  {Sénéchal}},\ }\href
  {https://doi.org/https://doi.org/10.1007/978-1-4612-2256-9} {\emph {\bibinfo
  {title} {Conformal {Field} {Theory}}}}\ (\bibinfo  {publisher} {Springer},\
  \bibinfo {year} {1997})\BibitemShut {NoStop}%
\bibitem [{\citenamefont {Henkel}(1999)}]{henkel1999conformal}%
  \BibitemOpen
  \bibfield  {author} {\bibinfo {author} {\bibfnamefont {M.}~\bibnamefont
  {Henkel}},\ }\href
  {https://doi.org/https://doi.org/10.1007/978-3-662-03937-3} {\emph {\bibinfo
  {title} {Conformal {Invariance} and {Critical} {Phenomena}}}}\ (\bibinfo
  {publisher} {Springer Science \& Business Media},\ \bibinfo {year}
  {1999})\BibitemShut {NoStop}%
\bibitem [{\citenamefont {Cardy}(1984)}]{cardy_conformal_1984}%
  \BibitemOpen
  \bibfield  {author} {\bibinfo {author} {\bibfnamefont {J.}~\bibnamefont
  {Cardy}},\ }\href
  {https://doi.org/https://doi.org/10.1016/0550-3213(84)90241-4} {\bibfield
  {journal} {\bibinfo  {journal} {Nuclear Physics B}\ }\textbf {\bibinfo
  {volume} {240}},\ \bibinfo {pages} {514} (\bibinfo {year}
  {1984})}\BibitemShut {NoStop}%
\bibitem [{\citenamefont {Cardy}(1996)}]{cardy1996scaling}%
  \BibitemOpen
  \bibfield  {author} {\bibinfo {author} {\bibfnamefont {J.}~\bibnamefont
  {Cardy}},\ }\href {https://doi.org/10.1017/CBO9781316036440} {\emph {\bibinfo
  {title} {Scaling and Renormalization in Statistical Physics}}},\ Cambridge
  Lecture Notes in Physics\ (\bibinfo  {publisher} {Cambridge University
  Press},\ \bibinfo {year} {1996})\BibitemShut {NoStop}%
\bibitem [{\citenamefont {Lee}\ \emph {et~al.}(2022{\natexlab{b}})\citenamefont
  {Lee}, \citenamefont {You},\ and\ \citenamefont {Xu}}]{lee_2022_symmetry}%
  \BibitemOpen
  \bibfield  {author} {\bibinfo {author} {\bibfnamefont {J.~Y.}\ \bibnamefont
  {Lee}}, \bibinfo {author} {\bibfnamefont {Y.-Z.}\ \bibnamefont {You}},\ and\
  \bibinfo {author} {\bibfnamefont {C.}~\bibnamefont {Xu}},\ }\href
  {https://arxiv.org/abs/2210.16323} {\bibfield  {journal} {\bibinfo  {journal}
  {arXiv:2210.16323}\ } (\bibinfo {year} {2022}{\natexlab{b}})}\BibitemShut
  {NoStop}%
\bibitem [{\citenamefont {Bao}\ \emph {et~al.}(2023)\citenamefont {Bao},
  \citenamefont {Fan}, \citenamefont {Vishwanath},\ and\ \citenamefont
  {Altman}}]{bao_2023_mixed}%
  \BibitemOpen
  \bibfield  {author} {\bibinfo {author} {\bibfnamefont {Y.}~\bibnamefont
  {Bao}}, \bibinfo {author} {\bibfnamefont {R.}~\bibnamefont {Fan}}, \bibinfo
  {author} {\bibfnamefont {A.}~\bibnamefont {Vishwanath}},\ and\ \bibinfo
  {author} {\bibfnamefont {E.}~\bibnamefont {Altman}},\ }\href
  {https://arxiv.org/abs/2301.05687} {\bibfield  {journal} {\bibinfo  {journal}
  {arXiv:2301.05687}\ } (\bibinfo {year} {2023})}\BibitemShut {NoStop}%
\bibitem [{\citenamefont {Lee}\ \emph {et~al.}(2023)\citenamefont {Lee},
  \citenamefont {Jian},\ and\ \citenamefont {Xu}}]{lee_2023_criticality}%
  \BibitemOpen
  \bibfield  {author} {\bibinfo {author} {\bibfnamefont {J.~Y.}\ \bibnamefont
  {Lee}}, \bibinfo {author} {\bibfnamefont {C.-M.}\ \bibnamefont {Jian}},\ and\
  \bibinfo {author} {\bibfnamefont {C.}~\bibnamefont {Xu}},\ }\href
  {https://arxiv.org/abs/2301.05238} {\bibfield  {journal} {\bibinfo  {journal}
  {arXiv:2301.05238}\ } (\bibinfo {year} {2023})}\BibitemShut {NoStop}%
\bibitem [{\citenamefont {Zou}\ \emph {et~al.}(2023)\citenamefont {Zou},
  \citenamefont {Sang},\ and\ \citenamefont {Hsieh}}]{zou2023channeling}%
  \BibitemOpen
  \bibfield  {author} {\bibinfo {author} {\bibfnamefont {Y.}~\bibnamefont
  {Zou}}, \bibinfo {author} {\bibfnamefont {S.}~\bibnamefont {Sang}},\ and\
  \bibinfo {author} {\bibfnamefont {T.~H.}\ \bibnamefont {Hsieh}},\ }\href
  {https://arxiv.org/abs/2301.07141} {\bibfield  {journal} {\bibinfo  {journal}
  {arXiv:2301.07141}\ } (\bibinfo {year} {2023})}\BibitemShut {NoStop}%
\bibitem [{\citenamefont {Bariev}(1979)}]{bariev_effect_1979}%
  \BibitemOpen
  \bibfield  {author} {\bibinfo {author} {\bibfnamefont {R.}~\bibnamefont
  {Bariev}},\ }\href {http://www.jetp.ras.ru/cgi-bin/dn/e_050_03_0613}
  {\bibfield  {journal} {\bibinfo  {journal} {Sov. Phys. JETP}\ }\textbf
  {\bibinfo {volume} {50}},\ \bibinfo {pages} {613} (\bibinfo {year}
  {1979})}\BibitemShut {NoStop}%
\bibitem [{\citenamefont {McCoy}\ and\ \citenamefont
  {Perk}(1980)}]{mccoy_two-spin_1980}%
  \BibitemOpen
  \bibfield  {author} {\bibinfo {author} {\bibfnamefont {B.~M.}\ \bibnamefont
  {McCoy}}\ and\ \bibinfo {author} {\bibfnamefont {J.~H.~H.}\ \bibnamefont
  {Perk}},\ }\href {https://doi.org/10.1103/PhysRevLett.44.840} {\bibfield
  {journal} {\bibinfo  {journal} {Phys. Rev. Letters}\ }\textbf {\bibinfo
  {volume} {44}},\ \bibinfo {pages} {840} (\bibinfo {year} {1980})}\BibitemShut
  {NoStop}%
\bibitem [{\citenamefont {Igl{\'o}i}\ \emph {et~al.}(1993)\citenamefont
  {Igl{\'o}i}, \citenamefont {Peschel},\ and\ \citenamefont
  {Turban}}]{igloi_1993_inhomogeneous}%
  \BibitemOpen
  \bibfield  {author} {\bibinfo {author} {\bibfnamefont {F.}~\bibnamefont
  {Igl{\'o}i}}, \bibinfo {author} {\bibfnamefont {I.}~\bibnamefont {Peschel}},\
  and\ \bibinfo {author} {\bibfnamefont {L.}~\bibnamefont {Turban}},\ }\href
  {https://doi.org/10.1080/00018739300101544} {\bibfield  {journal} {\bibinfo
  {journal} {Adv. Phys.}\ }\textbf {\bibinfo {volume} {42}},\ \bibinfo {pages}
  {683} (\bibinfo {year} {1993})}\BibitemShut {NoStop}%
\bibitem [{\citenamefont {Oshikawa}\ and\ \citenamefont
  {Affleck}(1997)}]{oshikawa_boundary_1997}%
  \BibitemOpen
  \bibfield  {author} {\bibinfo {author} {\bibfnamefont {M.}~\bibnamefont
  {Oshikawa}}\ and\ \bibinfo {author} {\bibfnamefont {I.}~\bibnamefont
  {Affleck}},\ }\href {https://doi.org/10.1016/S0550-3213(97)00219-8}
  {\bibfield  {journal} {\bibinfo  {journal} {Nucl. Phys. B}\ }\textbf
  {\bibinfo {volume} {495}},\ \bibinfo {pages} {533} (\bibinfo {year}
  {1997})}\BibitemShut {NoStop}%
\bibitem [{\citenamefont {Igl{\'o}i}\ \emph {et~al.}(2009)\citenamefont
  {Igl{\'o}i}, \citenamefont {Szatm{\'a}ri},\ and\ \citenamefont
  {Lin}}]{igloi_entanglement_2009}%
  \BibitemOpen
  \bibfield  {author} {\bibinfo {author} {\bibfnamefont {F.}~\bibnamefont
  {Igl{\'o}i}}, \bibinfo {author} {\bibfnamefont {Z.}~\bibnamefont
  {Szatm{\'a}ri}},\ and\ \bibinfo {author} {\bibfnamefont {Y.-C.}\ \bibnamefont
  {Lin}},\ }\href {https://doi.org/10.1103/PhysRevB.80.024405} {\bibfield
  {journal} {\bibinfo  {journal} {Phys. Rev. B}\ }\textbf {\bibinfo {volume}
  {80}},\ \bibinfo {pages} {024405} (\bibinfo {year} {2009})}\BibitemShut
  {NoStop}%
\bibitem [{\citenamefont {Eisler}\ and\ \citenamefont
  {Peschel}(2010)}]{eisler_entanglement_2010}%
  \BibitemOpen
  \bibfield  {author} {\bibinfo {author} {\bibfnamefont {V.}~\bibnamefont
  {Eisler}}\ and\ \bibinfo {author} {\bibfnamefont {I.}~\bibnamefont
  {Peschel}},\ }\href {https://doi.org/https://doi.org/10.1002/andp.201000055}
  {\bibfield  {journal} {\bibinfo  {journal} {Ann. Phys. (Berl.)}\ }\textbf
  {\bibinfo {volume} {522}},\ \bibinfo {pages} {679} (\bibinfo {year}
  {2010})}\BibitemShut {NoStop}%
\bibitem [{\citenamefont {Gullans}\ and\ \citenamefont
  {Huse}(2020{\natexlab{b}})}]{gullans_2020_scalable}%
  \BibitemOpen
  \bibfield  {author} {\bibinfo {author} {\bibfnamefont {M.~J.}\ \bibnamefont
  {Gullans}}\ and\ \bibinfo {author} {\bibfnamefont {D.~A.}\ \bibnamefont
  {Huse}},\ }\href {https://doi.org/10.1103/PhysRevLett.125.070606} {\bibfield
  {journal} {\bibinfo  {journal} {Phys. Rev. Lett.}\ }\textbf {\bibinfo
  {volume} {125}},\ \bibinfo {pages} {070606} (\bibinfo {year}
  {2020}{\natexlab{b}})}\BibitemShut {NoStop}%
\bibitem [{\citenamefont {Li}\ \emph {et~al.}(2022)\citenamefont {Li},
  \citenamefont {Zou}, \citenamefont {Glorioso}, \citenamefont {Altman},\ and\
  \citenamefont {Fisher}}]{li_2022_cross}%
  \BibitemOpen
  \bibfield  {author} {\bibinfo {author} {\bibfnamefont {Y.}~\bibnamefont
  {Li}}, \bibinfo {author} {\bibfnamefont {Y.}~\bibnamefont {Zou}}, \bibinfo
  {author} {\bibfnamefont {P.}~\bibnamefont {Glorioso}}, \bibinfo {author}
  {\bibfnamefont {E.}~\bibnamefont {Altman}},\ and\ \bibinfo {author}
  {\bibfnamefont {M.~P.~A.}\ \bibnamefont {Fisher}},\ }\href
  {https://arxiv.org/abs/2209.00609} {\bibfield  {journal} {\bibinfo  {journal}
  {arXiv:2209.00609}\ } (\bibinfo {year} {2022})}\BibitemShut {NoStop}%
\bibitem [{\citenamefont {Feng}\ \emph {et~al.}(2022)\citenamefont {Feng},
  \citenamefont {Skinner},\ and\ \citenamefont {Nahum}}]{feng2022measurement}%
  \BibitemOpen
  \bibfield  {author} {\bibinfo {author} {\bibfnamefont {X.}~\bibnamefont
  {Feng}}, \bibinfo {author} {\bibfnamefont {B.}~\bibnamefont {Skinner}},\ and\
  \bibinfo {author} {\bibfnamefont {A.}~\bibnamefont {Nahum}},\ }\href
  {https://arxiv.org/abs/2210.07264} {\bibfield  {journal} {\bibinfo  {journal}
  {arXiv:2210.07264}\ } (\bibinfo {year} {2022})}\BibitemShut {NoStop}%
\bibitem [{\citenamefont {Li}\ and\ \citenamefont
  {Fisher}(2021)}]{li_21_robust}%
  \BibitemOpen
  \bibfield  {author} {\bibinfo {author} {\bibfnamefont {Y.}~\bibnamefont
  {Li}}\ and\ \bibinfo {author} {\bibfnamefont {M.~P.~A.}\ \bibnamefont
  {Fisher}},\ }\href {https://arxiv.org/abs/2108.04274} {\bibfield  {journal}
  {\bibinfo  {journal} {arXiv:2108.04274}\ } (\bibinfo {year}
  {2021})}\BibitemShut {NoStop}%
\bibitem [{\citenamefont {Ippoliti}\ and\ \citenamefont
  {Khemani}(2021)}]{ippoliti_2021_postselection}%
  \BibitemOpen
  \bibfield  {author} {\bibinfo {author} {\bibfnamefont {M.}~\bibnamefont
  {Ippoliti}}\ and\ \bibinfo {author} {\bibfnamefont {V.}~\bibnamefont
  {Khemani}},\ }\href {https://doi.org/10.1103/PhysRevLett.126.060501}
  {\bibfield  {journal} {\bibinfo  {journal} {Phys. Rev. Lett.}\ }\textbf
  {\bibinfo {volume} {126}},\ \bibinfo {pages} {060501} (\bibinfo {year}
  {2021})}\BibitemShut {NoStop}%
\bibitem [{\citenamefont {Nielsen}\ and\ \citenamefont
  {Chuang}(2010)}]{nielsen_quantum_2010}%
  \BibitemOpen
  \bibfield  {author} {\bibinfo {author} {\bibfnamefont {M.~A.}\ \bibnamefont
  {Nielsen}}\ and\ \bibinfo {author} {\bibfnamefont {I.~L.}\ \bibnamefont
  {Chuang}},\ }\href {https://doi.org/10.1017/CBO9780511976667} {\emph
  {\bibinfo {title} {Quantum {Computation} and {Quantum} {Information}}}},\
  \bibinfo {edition} {10th}\ ed.\ (\bibinfo  {publisher} {Cambridge University
  Press},\ \bibinfo {address} {Cambridge ; New York},\ \bibinfo {year}
  {2010})\BibitemShut {NoStop}%
\bibitem [{\citenamefont {Peschel}(2005)}]{peschel_entanglement_2005}%
  \BibitemOpen
  \bibfield  {author} {\bibinfo {author} {\bibfnamefont {I.}~\bibnamefont
  {Peschel}},\ }\href {https://doi.org/10.1088/0305-4470/38/20/002} {\bibfield
  {journal} {\bibinfo  {journal} {J. Phys. A: Math. Gen.}\ }\textbf {\bibinfo
  {volume} {38}},\ \bibinfo {pages} {4327} (\bibinfo {year}
  {2005})}\BibitemShut {NoStop}%
\bibitem [{\citenamefont {Peschel}\ and\ \citenamefont
  {Eisler}(2012)}]{peschel_exact_2012}%
  \BibitemOpen
  \bibfield  {author} {\bibinfo {author} {\bibfnamefont {I.}~\bibnamefont
  {Peschel}}\ and\ \bibinfo {author} {\bibfnamefont {V.}~\bibnamefont
  {Eisler}},\ }\href {https://doi.org/10.1088/1751-8113/45/15/155301}
  {\bibfield  {journal} {\bibinfo  {journal} {J. Phys. A: Math. Theor.}\
  }\textbf {\bibinfo {volume} {45}},\ \bibinfo {pages} {155301} (\bibinfo
  {year} {2012})}\BibitemShut {NoStop}%
\bibitem [{\citenamefont {Brehm}\ and\ \citenamefont
  {Brunner}(2015)}]{brehm_entanglement_2015}%
  \BibitemOpen
  \bibfield  {author} {\bibinfo {author} {\bibfnamefont {E.}~\bibnamefont
  {Brehm}}\ and\ \bibinfo {author} {\bibfnamefont {I.}~\bibnamefont
  {Brunner}},\ }\href {https://doi.org/10.1007/JHEP09(2015)080} {\bibfield
  {journal} {\bibinfo  {journal} {J. High Energy Phys.}\ }\textbf {\bibinfo
  {volume} {2015}}\bibinfo  {number} { (9)},\ \bibinfo {pages}
  {80}}\BibitemShut {NoStop}%
\bibitem [{\citenamefont {Roy}\ and\ \citenamefont
  {Saleur}(2022)}]{roy_entanglement_2022}%
  \BibitemOpen
\bibfield  {number} {  }\bibfield  {author} {\bibinfo {author} {\bibfnamefont
  {A.}~\bibnamefont {Roy}}\ and\ \bibinfo {author} {\bibfnamefont
  {H.}~\bibnamefont {Saleur}},\ }\href
  {https://doi.org/10.1103/PhysRevLett.128.090603} {\bibfield  {journal}
  {\bibinfo  {journal} {Phys. Rev. Lett.}\ }\textbf {\bibinfo {volume} {128}},\
  \bibinfo {pages} {090603} (\bibinfo {year} {2022})}\BibitemShut {NoStop}%
\bibitem [{\citenamefont {Rajabpour}(2015)}]{rajabpour2015post}%
  \BibitemOpen
  \bibfield  {author} {\bibinfo {author} {\bibfnamefont {M.~A.}\ \bibnamefont
  {Rajabpour}},\ }\href
  {https://doi.org/https://doi.org/10.1103/PhysRevB.92.075108} {\bibfield
  {journal} {\bibinfo  {journal} {Phys. Rev. B}\ }\textbf {\bibinfo {volume}
  {92}},\ \bibinfo {pages} {075108} (\bibinfo {year} {2015})}\BibitemShut
  {NoStop}%
\bibitem [{\citenamefont {Rajabpour}(2016)}]{rajabpour2016entanglement}%
  \BibitemOpen
  \bibfield  {author} {\bibinfo {author} {\bibfnamefont {M.~A.}\ \bibnamefont
  {Rajabpour}},\ }\href {https://doi.org/10.1088/1742-5468/2016/06/063109}
  {\bibfield  {journal} {\bibinfo  {journal} {J. Stat. Mech.: Theory Exp.}\
  }\textbf {\bibinfo {volume} {2016}}\bibinfo  {number} { (6)},\ \bibinfo
  {pages} {063109}}\BibitemShut {NoStop}%
\bibitem [{\citenamefont {Chen}\ \emph {et~al.}(2020)\citenamefont {Chen},
  \citenamefont {Li}, \citenamefont {Fisher},\ and\ \citenamefont
  {Lucas}}]{chen_emergent_2020}%
  \BibitemOpen
\bibfield  {number} {  }\bibfield  {author} {\bibinfo {author} {\bibfnamefont
  {X.}~\bibnamefont {Chen}}, \bibinfo {author} {\bibfnamefont {Y.}~\bibnamefont
  {Li}}, \bibinfo {author} {\bibfnamefont {M.~P.~A.}\ \bibnamefont {Fisher}},\
  and\ \bibinfo {author} {\bibfnamefont {A.}~\bibnamefont {Lucas}},\ }\href
  {https://doi.org/10.1103/PhysRevResearch.2.033017} {\bibfield  {journal}
  {\bibinfo  {journal} {Phys. Rev. Research}\ }\textbf {\bibinfo {volume}
  {2}},\ \bibinfo {pages} {033017} (\bibinfo {year} {2020})}\BibitemShut
  {NoStop}%
\bibitem [{\citenamefont {Alberton}\ \emph {et~al.}(2021)\citenamefont
  {Alberton}, \citenamefont {Buchhold},\ and\ \citenamefont
  {Diehl}}]{alberton_entanglement_2021}%
  \BibitemOpen
  \bibfield  {author} {\bibinfo {author} {\bibfnamefont {O.}~\bibnamefont
  {Alberton}}, \bibinfo {author} {\bibfnamefont {M.}~\bibnamefont {Buchhold}},\
  and\ \bibinfo {author} {\bibfnamefont {S.}~\bibnamefont {Diehl}},\ }\href
  {https://doi.org/10.1103/PhysRevLett.126.170602} {\bibfield  {journal}
  {\bibinfo  {journal} {Phys. Rev. Lett.}\ }\textbf {\bibinfo {volume} {126}},\
  \bibinfo {pages} {170602} (\bibinfo {year} {2021})}\BibitemShut {NoStop}%
\bibitem [{\citenamefont {Jian}\ \emph {et~al.}(2022)\citenamefont {Jian},
  \citenamefont {Bauer}, \citenamefont {Keselman},\ and\ \citenamefont
  {Ludwig}}]{jian_criticality_2022}%
  \BibitemOpen
  \bibfield  {author} {\bibinfo {author} {\bibfnamefont {C.-M.}\ \bibnamefont
  {Jian}}, \bibinfo {author} {\bibfnamefont {B.}~\bibnamefont {Bauer}},
  \bibinfo {author} {\bibfnamefont {A.}~\bibnamefont {Keselman}},\ and\
  \bibinfo {author} {\bibfnamefont {A.~W.~W.}\ \bibnamefont {Ludwig}},\ }\href
  {https://doi.org/10.1103/PhysRevB.106.134206} {\bibfield  {journal} {\bibinfo
   {journal} {Phys. Rev. B}\ }\textbf {\bibinfo {volume} {106}},\ \bibinfo
  {pages} {134206} (\bibinfo {year} {2022})}\BibitemShut {NoStop}%
\bibitem [{\citenamefont {Bao}\ \emph {et~al.}(2021{\natexlab{b}})\citenamefont
  {Bao}, \citenamefont {Choi},\ and\ \citenamefont
  {Altman}}]{bao_symmetry_2021}%
  \BibitemOpen
  \bibfield  {author} {\bibinfo {author} {\bibfnamefont {Y.}~\bibnamefont
  {Bao}}, \bibinfo {author} {\bibfnamefont {S.}~\bibnamefont {Choi}},\ and\
  \bibinfo {author} {\bibfnamefont {E.}~\bibnamefont {Altman}},\ }\href
  {https://doi.org/https://doi.org/10.1016/j.aop.2021.168618} {\bibfield
  {journal} {\bibinfo  {journal} {Ann. Phys. (N. Y.)}\ }\textbf {\bibinfo
  {volume} {435}},\ \bibinfo {pages} {168618} (\bibinfo {year}
  {2021}{\natexlab{b}})}\BibitemShut {NoStop}%
\bibitem [{\citenamefont {Turkeshi}\ \emph {et~al.}(2021)\citenamefont
  {Turkeshi}, \citenamefont {Biella}, \citenamefont {Fazio}, \citenamefont
  {Dalmonte},\ and\ \citenamefont
  {Schir\`o}}]{turkeshi_measurement-induced_2021}%
  \BibitemOpen
  \bibfield  {author} {\bibinfo {author} {\bibfnamefont {X.}~\bibnamefont
  {Turkeshi}}, \bibinfo {author} {\bibfnamefont {A.}~\bibnamefont {Biella}},
  \bibinfo {author} {\bibfnamefont {R.}~\bibnamefont {Fazio}}, \bibinfo
  {author} {\bibfnamefont {M.}~\bibnamefont {Dalmonte}},\ and\ \bibinfo
  {author} {\bibfnamefont {M.}~\bibnamefont {Schir\`o}},\ }\href
  {https://doi.org/10.1103/PhysRevB.103.224210} {\bibfield  {journal} {\bibinfo
   {journal} {Phys. Rev. B}\ }\textbf {\bibinfo {volume} {103}},\ \bibinfo
  {pages} {224210} (\bibinfo {year} {2021})}\BibitemShut {NoStop}%
\bibitem [{\citenamefont {Buchhold}\ \emph {et~al.}(2021)\citenamefont
  {Buchhold}, \citenamefont {Minoguchi}, \citenamefont {Altland},\ and\
  \citenamefont {Diehl}}]{buchhold_effective_2021}%
  \BibitemOpen
  \bibfield  {author} {\bibinfo {author} {\bibfnamefont {M.}~\bibnamefont
  {Buchhold}}, \bibinfo {author} {\bibfnamefont {Y.}~\bibnamefont {Minoguchi}},
  \bibinfo {author} {\bibfnamefont {A.}~\bibnamefont {Altland}},\ and\ \bibinfo
  {author} {\bibfnamefont {S.}~\bibnamefont {Diehl}},\ }\href
  {https://doi.org/10.1103/PhysRevX.11.041004} {\bibfield  {journal} {\bibinfo
  {journal} {Phys. Rev. X}\ }\textbf {\bibinfo {volume} {11}},\ \bibinfo
  {pages} {041004} (\bibinfo {year} {2021})}\BibitemShut {NoStop}%
\bibitem [{\citenamefont {Botzung}\ \emph {et~al.}(2021)\citenamefont
  {Botzung}, \citenamefont {Diehl},\ and\ \citenamefont
  {M{\"u}ller}}]{botzung_engineered_2021}%
  \BibitemOpen
  \bibfield  {author} {\bibinfo {author} {\bibfnamefont {T.}~\bibnamefont
  {Botzung}}, \bibinfo {author} {\bibfnamefont {S.}~\bibnamefont {Diehl}},\
  and\ \bibinfo {author} {\bibfnamefont {M.}~\bibnamefont {M{\"u}ller}},\
  }\href {https://doi.org/10.1103/PhysRevB.104.184422} {\bibfield  {journal}
  {\bibinfo  {journal} {Phys. Rev. B}\ }\textbf {\bibinfo {volume} {104}},\
  \bibinfo {pages} {184422} (\bibinfo {year} {2021})}\BibitemShut {NoStop}%
\bibitem [{\citenamefont {Boorman}\ \emph {et~al.}(2022)\citenamefont
  {Boorman}, \citenamefont {Szyniszewski}, \citenamefont {Schomerus},\ and\
  \citenamefont {Romito}}]{boorman_diagnostics_2022}%
  \BibitemOpen
  \bibfield  {author} {\bibinfo {author} {\bibfnamefont {T.}~\bibnamefont
  {Boorman}}, \bibinfo {author} {\bibfnamefont {M.}~\bibnamefont
  {Szyniszewski}}, \bibinfo {author} {\bibfnamefont {H.}~\bibnamefont
  {Schomerus}},\ and\ \bibinfo {author} {\bibfnamefont {A.}~\bibnamefont
  {Romito}},\ }\href {https://doi.org/10.1103/PhysRevB.105.144202} {\bibfield
  {journal} {\bibinfo  {journal} {Phys. Rev. B}\ }\textbf {\bibinfo {volume}
  {105}},\ \bibinfo {pages} {144202} (\bibinfo {year} {2022})}\BibitemShut
  {NoStop}%
\bibitem [{\citenamefont {Turkeshi}\ \emph {et~al.}(2022)\citenamefont
  {Turkeshi}, \citenamefont {Dalmonte}, \citenamefont {Fazio},\ and\
  \citenamefont {Schir\`o}}]{turkeshi_entanglement_2022}%
  \BibitemOpen
  \bibfield  {author} {\bibinfo {author} {\bibfnamefont {X.}~\bibnamefont
  {Turkeshi}}, \bibinfo {author} {\bibfnamefont {M.}~\bibnamefont {Dalmonte}},
  \bibinfo {author} {\bibfnamefont {R.}~\bibnamefont {Fazio}},\ and\ \bibinfo
  {author} {\bibfnamefont {M.}~\bibnamefont {Schir\`o}},\ }\href
  {https://doi.org/10.1103/PhysRevB.105.L241114} {\bibfield  {journal}
  {\bibinfo  {journal} {Phys. Rev. B}\ }\textbf {\bibinfo {volume} {105}},\
  \bibinfo {pages} {L241114} (\bibinfo {year} {2022})}\BibitemShut {NoStop}%
\bibitem [{\citenamefont {Piccitto}\ \emph {et~al.}(2022)\citenamefont
  {Piccitto}, \citenamefont {Russomanno},\ and\ \citenamefont
  {Rossini}}]{piccitto_entanglement_2022}%
  \BibitemOpen
  \bibfield  {author} {\bibinfo {author} {\bibfnamefont {G.}~\bibnamefont
  {Piccitto}}, \bibinfo {author} {\bibfnamefont {A.}~\bibnamefont
  {Russomanno}},\ and\ \bibinfo {author} {\bibfnamefont {D.}~\bibnamefont
  {Rossini}},\ }\href {https://doi.org/10.1103/PhysRevB.105.064305} {\bibfield
  {journal} {\bibinfo  {journal} {Phys. Rev. B}\ }\textbf {\bibinfo {volume}
  {105}},\ \bibinfo {pages} {064305} (\bibinfo {year} {2022})}\BibitemShut
  {NoStop}%
\bibitem [{\citenamefont {Kells}\ \emph {et~al.}(2023)\citenamefont {Kells},
  \citenamefont {Meidan},\ and\ \citenamefont
  {Romito}}]{kells_topological_2023}%
  \BibitemOpen
  \bibfield  {author} {\bibinfo {author} {\bibfnamefont {G.}~\bibnamefont
  {Kells}}, \bibinfo {author} {\bibfnamefont {D.}~\bibnamefont {Meidan}},\ and\
  \bibinfo {author} {\bibfnamefont {A.}~\bibnamefont {Romito}},\ }\href
  {https://doi.org/10.21468/SciPostPhys.14.3.031} {\bibfield  {journal}
  {\bibinfo  {journal} {SciPost Phys.}\ }\textbf {\bibinfo {volume} {14}},\
  \bibinfo {pages} {031} (\bibinfo {year} {2023})}\BibitemShut {NoStop}%
\bibitem [{\citenamefont {Turkeshi}\ and\ \citenamefont
  {Schir\`o}(2023)}]{turkeshi_entanglement_2023}%
  \BibitemOpen
  \bibfield  {author} {\bibinfo {author} {\bibfnamefont {X.}~\bibnamefont
  {Turkeshi}}\ and\ \bibinfo {author} {\bibfnamefont {M.}~\bibnamefont
  {Schir\`o}},\ }\href {https://doi.org/10.1103/PhysRevB.107.L020403}
  {\bibfield  {journal} {\bibinfo  {journal} {Phys. Rev. B}\ }\textbf {\bibinfo
  {volume} {107}},\ \bibinfo {pages} {L020403} (\bibinfo {year}
  {2023})}\BibitemShut {NoStop}%
\bibitem [{\citenamefont {Fradkin}(2013)}]{fradkin_field_2013}%
  \BibitemOpen
  \bibfield  {author} {\bibinfo {author} {\bibfnamefont {E.}~\bibnamefont
  {Fradkin}},\ }\href
  {https://doi.org/https://doi.org/10.1017/CBO9781139015509} {\emph {\bibinfo
  {title} {Field {Theories} of {Condensed} {Matter} {Physics}}}},\ \bibinfo
  {edition} {2nd}\ ed.\ (\bibinfo  {publisher} {Cambridge University Press},\
  \bibinfo {year} {2013})\BibitemShut {NoStop}%
\bibitem [{\citenamefont {Schultz}\ \emph {et~al.}(1964)\citenamefont
  {Schultz}, \citenamefont {Mattis},\ and\ \citenamefont
  {Lieb}}]{schultz_two-dimensional_1964}%
  \BibitemOpen
  \bibfield  {author} {\bibinfo {author} {\bibfnamefont {T.~D.}\ \bibnamefont
  {Schultz}}, \bibinfo {author} {\bibfnamefont {D.~C.}\ \bibnamefont
  {Mattis}},\ and\ \bibinfo {author} {\bibfnamefont {E.~H.}\ \bibnamefont
  {Lieb}},\ }\href {https://doi.org/10.1103/RevModPhys.36.856} {\bibfield
  {journal} {\bibinfo  {journal} {Rev. Mod. Phys.}\ }\textbf {\bibinfo {volume}
  {36}},\ \bibinfo {pages} {856} (\bibinfo {year} {1964})}\BibitemShut
  {NoStop}%
\bibitem [{\citenamefont {Terhal}\ and\ \citenamefont
  {DiVincenzo}(2002)}]{terhal_classical_2002}%
  \BibitemOpen
  \bibfield  {author} {\bibinfo {author} {\bibfnamefont {B.~M.}\ \bibnamefont
  {Terhal}}\ and\ \bibinfo {author} {\bibfnamefont {D.~P.}\ \bibnamefont
  {DiVincenzo}},\ }\href {https://doi.org/10.1103/PhysRevA.65.032325}
  {\bibfield  {journal} {\bibinfo  {journal} {Phys. Rev. A}\ }\textbf {\bibinfo
  {volume} {65}},\ \bibinfo {pages} {032325} (\bibinfo {year}
  {2002})}\BibitemShut {NoStop}%
\bibitem [{\citenamefont {Knill}(2001)}]{knill_fermionic_2001}%
  \BibitemOpen
  \bibfield  {author} {\bibinfo {author} {\bibfnamefont {E.}~\bibnamefont
  {Knill}},\ }\href {https://arxiv.org/abs/quant-ph/0108033} {\bibfield
  {journal} {\bibinfo  {journal} {quant-ph/0108033}\ } (\bibinfo {year}
  {2001})}\BibitemShut {NoStop}%
\bibitem [{\citenamefont {Bravyi}(2004)}]{bravyi_lagrangian_2004}%
  \BibitemOpen
  \bibfield  {author} {\bibinfo {author} {\bibfnamefont {S.}~\bibnamefont
  {Bravyi}},\ }\href {https://arxiv.org/abs/quant-ph/0404180} {\bibfield
  {journal} {\bibinfo  {journal} {quant-ph/0404180}\ } (\bibinfo {year}
  {2004})}\BibitemShut {NoStop}%
\bibitem [{\citenamefont {Calabrese}\ and\ \citenamefont
  {Cardy}(2009)}]{calabrese_entanglement_2009}%
  \BibitemOpen
  \bibfield  {author} {\bibinfo {author} {\bibfnamefont {P.}~\bibnamefont
  {Calabrese}}\ and\ \bibinfo {author} {\bibfnamefont {J.}~\bibnamefont
  {Cardy}},\ }\href {https://doi.org/10.1088/1751-8113/42/50/504005} {\bibfield
   {journal} {\bibinfo  {journal} {J. Phys. A: Math. Theor.}\ }\textbf
  {\bibinfo {volume} {42}},\ \bibinfo {pages} {504005} (\bibinfo {year}
  {2009})}\BibitemShut {NoStop}%
\bibitem [{\citenamefont {Zou}\ \emph {et~al.}(2020)\citenamefont {Zou},
  \citenamefont {Milsted},\ and\ \citenamefont {Vidal}}]{zou_conformal_2020}%
  \BibitemOpen
  \bibfield  {author} {\bibinfo {author} {\bibfnamefont {Y.}~\bibnamefont
  {Zou}}, \bibinfo {author} {\bibfnamefont {A.}~\bibnamefont {Milsted}},\ and\
  \bibinfo {author} {\bibfnamefont {G.}~\bibnamefont {Vidal}},\ }\href
  {https://doi.org/10.1103/PhysRevLett.124.040604} {\bibfield  {journal}
  {\bibinfo  {journal} {Phys. Rev. Lett.}\ }\textbf {\bibinfo {volume} {124}},\
  \bibinfo {pages} {040604} (\bibinfo {year} {2020})}\BibitemShut {NoStop}%
\bibitem [{\citenamefont {Holzhey}\ \emph {et~al.}(1994)\citenamefont
  {Holzhey}, \citenamefont {Larsen},\ and\ \citenamefont
  {Wilczek}}]{holzhey_geometric_1994}%
  \BibitemOpen
  \bibfield  {author} {\bibinfo {author} {\bibfnamefont {C.}~\bibnamefont
  {Holzhey}}, \bibinfo {author} {\bibfnamefont {F.}~\bibnamefont {Larsen}},\
  and\ \bibinfo {author} {\bibfnamefont {F.}~\bibnamefont {Wilczek}},\ }\href
  {https://doi.org/10.1016/0550-3213(94)90402-2} {\bibfield  {journal}
  {\bibinfo  {journal} {Nucl. Phys. B}\ }\textbf {\bibinfo {volume} {424}},\
  \bibinfo {pages} {443} (\bibinfo {year} {1994})}\BibitemShut {NoStop}%
\bibitem [{\citenamefont {Shankar}(2017)}]{shankar_quantum_2017}%
  \BibitemOpen
  \bibfield  {author} {\bibinfo {author} {\bibfnamefont {R.}~\bibnamefont
  {Shankar}},\ }\href {https://doi.org/10.1017/9781139044349} {\emph {\bibinfo
  {title} {Quantum {Field} {Theory} and {Condensed} {Matter}: {An}
  {Introduction}}}}\ (\bibinfo  {publisher} {Cambridge University Press},\
  \bibinfo {year} {2017})\BibitemShut {NoStop}%
\bibitem [{\citenamefont {Nishimori}(2001)}]{nishimori_statistical_2001}%
  \BibitemOpen
  \bibfield  {author} {\bibinfo {author} {\bibfnamefont {H.}~\bibnamefont
  {Nishimori}},\ }\href
  {https://doi.org/10.1093/acprof:oso/9780198509417.001.0001} {\emph {\bibinfo
  {title} {Statistical {Physics} of {Spin} {Glasses} and {Information}
  {Processing}: {An} {Introduction}}}}\ (\bibinfo  {publisher} {Oxford
  University Press},\ \bibinfo {year} {2001})\BibitemShut {NoStop}%
\bibitem [{\citenamefont {Fava}\ \emph {et~al.}(2023)\citenamefont {Fava},
  \citenamefont {Piroli}, \citenamefont {Swann}, \citenamefont {Bernard},\ and\
  \citenamefont {Nahum}}]{fava_nonlinear_2023}%
  \BibitemOpen
  \bibfield  {author} {\bibinfo {author} {\bibfnamefont {M.}~\bibnamefont
  {Fava}}, \bibinfo {author} {\bibfnamefont {L.}~\bibnamefont {Piroli}},
  \bibinfo {author} {\bibfnamefont {T.}~\bibnamefont {Swann}}, \bibinfo
  {author} {\bibfnamefont {D.}~\bibnamefont {Bernard}},\ and\ \bibinfo {author}
  {\bibfnamefont {A.}~\bibnamefont {Nahum}},\ }\href
  {http://arxiv.org/abs/2302.12820} {\bibfield  {journal} {\bibinfo  {journal}
  {arXiv:2302.12820}\ } (\bibinfo {year} {2023})}\BibitemShut {NoStop}%
\bibitem [{\citenamefont {Poboiko}\ \emph {et~al.}(2023)\citenamefont
  {Poboiko}, \citenamefont {P\"{o}pperl}, \citenamefont {Gornyi},\ and\
  \citenamefont {Mirlin}}]{poboiko_theory_2023}%
  \BibitemOpen
  \bibfield  {author} {\bibinfo {author} {\bibfnamefont {I.}~\bibnamefont
  {Poboiko}}, \bibinfo {author} {\bibfnamefont {P.}~\bibnamefont
  {P\"{o}pperl}}, \bibinfo {author} {\bibfnamefont {I.~V.}\ \bibnamefont
  {Gornyi}},\ and\ \bibinfo {author} {\bibfnamefont {A.~D.}\ \bibnamefont
  {Mirlin}},\ }\href {http://arxiv.org/abs/2304.03138} {\bibfield  {journal}
  {\bibinfo  {journal} {arXiv:2304.03138}\ } (\bibinfo {year}
  {2023})}\BibitemShut {NoStop}%
\bibitem [{\citenamefont {Labuhn}\ \emph {et~al.}(2016)\citenamefont {Labuhn},
  \citenamefont {Barredo}, \citenamefont {Ravets}, \citenamefont
  {De~L{\'e}s{\'e}leuc}, \citenamefont {Macr{\`\i}}, \citenamefont {Lahaye},\
  and\ \citenamefont {Browaeys}}]{labuhn_2016_tunable}%
  \BibitemOpen
  \bibfield  {author} {\bibinfo {author} {\bibfnamefont {H.}~\bibnamefont
  {Labuhn}}, \bibinfo {author} {\bibfnamefont {D.}~\bibnamefont {Barredo}},
  \bibinfo {author} {\bibfnamefont {S.}~\bibnamefont {Ravets}}, \bibinfo
  {author} {\bibfnamefont {S.}~\bibnamefont {De~L{\'e}s{\'e}leuc}}, \bibinfo
  {author} {\bibfnamefont {T.}~\bibnamefont {Macr{\`\i}}}, \bibinfo {author}
  {\bibfnamefont {T.}~\bibnamefont {Lahaye}},\ and\ \bibinfo {author}
  {\bibfnamefont {A.}~\bibnamefont {Browaeys}},\ }\href
  {https://doi.org/10.1038/nature18274} {\bibfield  {journal} {\bibinfo
  {journal} {Nature}\ }\textbf {\bibinfo {volume} {534}},\ \bibinfo {pages}
  {667} (\bibinfo {year} {2016})}\BibitemShut {NoStop}%
\bibitem [{\citenamefont {Browaeys}\ and\ \citenamefont
  {Lahaye}(2020)}]{browaeys_2020_many}%
  \BibitemOpen
  \bibfield  {author} {\bibinfo {author} {\bibfnamefont {A.}~\bibnamefont
  {Browaeys}}\ and\ \bibinfo {author} {\bibfnamefont {T.}~\bibnamefont
  {Lahaye}},\ }\href {https://doi.org/10.1038/s41567-019-0733-z} {\bibfield
  {journal} {\bibinfo  {journal} {Nat. Phys.}\ }\textbf {\bibinfo {volume}
  {16}},\ \bibinfo {pages} {132} (\bibinfo {year} {2020})}\BibitemShut
  {NoStop}%
\bibitem [{\citenamefont {Yang}\ \emph {et~al.}(2023)\citenamefont {Yang},
  \citenamefont {Mao},\ and\ \citenamefont {Jian}}]{yang2023entanglement}%
  \BibitemOpen
  \bibfield  {author} {\bibinfo {author} {\bibfnamefont {Z.}~\bibnamefont
  {Yang}}, \bibinfo {author} {\bibfnamefont {D.}~\bibnamefont {Mao}},\ and\
  \bibinfo {author} {\bibfnamefont {C.-M.}\ \bibnamefont {Jian}},\ }\href
  {https://arxiv.org/abs/2301.08255} {\bibfield  {journal} {\bibinfo  {journal}
  {arXiv:2301.08255}\ } (\bibinfo {year} {2023})}\BibitemShut {NoStop}%
\bibitem [{\citenamefont {Murciano}\ \emph {et~al.}(2023)\citenamefont
  {Murciano}, \citenamefont {Sala}, \citenamefont {Liu}, \citenamefont {Mong},\
  and\ \citenamefont {Alicea}}]{murciano_measurement_2023}%
  \BibitemOpen
  \bibfield  {author} {\bibinfo {author} {\bibfnamefont {S.}~\bibnamefont
  {Murciano}}, \bibinfo {author} {\bibfnamefont {P.}~\bibnamefont {Sala}},
  \bibinfo {author} {\bibfnamefont {Y.}~\bibnamefont {Liu}}, \bibinfo {author}
  {\bibfnamefont {R.~S.~K.}\ \bibnamefont {Mong}},\ and\ \bibinfo {author}
  {\bibfnamefont {J.}~\bibnamefont {Alicea}},\ }\href
  {https://arxiv.org/abs/2302.04325} {\bibfield  {journal} {\bibinfo  {journal}
  {arXiv:2302.04325}\ } (\bibinfo {year} {2023})}\BibitemShut {NoStop}%
\bibitem [{\citenamefont {Kitaev}(2001)}]{kitaev_unpaired_2001}%
  \BibitemOpen
  \bibfield  {author} {\bibinfo {author} {\bibfnamefont {A.~Y.}\ \bibnamefont
  {Kitaev}},\ }\href {https://doi.org/10.1070/1063-7869/44/10S/S29} {\bibfield
  {journal} {\bibinfo  {journal} {Phys. Uspekhi}\ }\textbf {\bibinfo {volume}
  {44}},\ \bibinfo {pages} {131} (\bibinfo {year} {2001})}\BibitemShut
  {NoStop}%
\bibitem [{\citenamefont {Zumino}(1962)}]{zumino_normal_1962}%
  \BibitemOpen
  \bibfield  {author} {\bibinfo {author} {\bibfnamefont {B.}~\bibnamefont
  {Zumino}},\ }\href {https://doi.org/10.1063/1.1724294} {\bibfield  {journal}
  {\bibinfo  {journal} {J. Math. Phys.}\ }\textbf {\bibinfo {volume} {3}},\
  \bibinfo {pages} {1055} (\bibinfo {year} {1962})}\BibitemShut {NoStop}%
\bibitem [{\citenamefont {Wimmer}(2012)}]{wimmer2012algorithm}%
  \BibitemOpen
  \bibfield  {author} {\bibinfo {author} {\bibfnamefont {M.}~\bibnamefont
  {Wimmer}},\ }\href {https://doi.org/https://doi.org/10.1145/2331130.2331138}
  {\bibfield  {journal} {\bibinfo  {journal} {ACM Trans. Math. Softw.}\
  }\textbf {\bibinfo {volume} {38}},\ \bibinfo {pages} {1} (\bibinfo {year}
  {2012})}\BibitemShut {NoStop}%
\bibitem [{\citenamefont {Bander}\ and\ \citenamefont
  {Itzykson}(1977)}]{bander_quantum-field-theory_1977}%
  \BibitemOpen
  \bibfield  {author} {\bibinfo {author} {\bibfnamefont {M.}~\bibnamefont
  {Bander}}\ and\ \bibinfo {author} {\bibfnamefont {C.}~\bibnamefont
  {Itzykson}},\ }\href {https://doi.org/10.1103/PhysRevD.15.463} {\bibfield
  {journal} {\bibinfo  {journal} {Phys. Rev. D}\ }\textbf {\bibinfo {volume}
  {15}},\ \bibinfo {pages} {463} (\bibinfo {year} {1977})}\BibitemShut
  {NoStop}%
\bibitem [{\citenamefont {Zuber}\ and\ \citenamefont
  {Itzykson}(1977)}]{zuber_quantum_1977}%
  \BibitemOpen
  \bibfield  {author} {\bibinfo {author} {\bibfnamefont {J.~B.}\ \bibnamefont
  {Zuber}}\ and\ \bibinfo {author} {\bibfnamefont {C.}~\bibnamefont
  {Itzykson}},\ }\href {https://doi.org/10.1103/PhysRevD.15.2875} {\bibfield
  {journal} {\bibinfo  {journal} {Phys. Rev. D}\ }\textbf {\bibinfo {volume}
  {15}},\ \bibinfo {pages} {2875} (\bibinfo {year} {1977})}\BibitemShut
  {NoStop}%
\bibitem [{\citenamefont {Kogut}(1979)}]{kogut_introduction_1979}%
  \BibitemOpen
  \bibfield  {author} {\bibinfo {author} {\bibfnamefont {J.~B.}\ \bibnamefont
  {Kogut}},\ }\href {https://doi.org/10.1103/RevModPhys.51.659} {\bibfield
  {journal} {\bibinfo  {journal} {Rev. Mod. Phys.}\ }\textbf {\bibinfo {volume}
  {51}},\ \bibinfo {pages} {659} (\bibinfo {year} {1979})}\BibitemShut
  {NoStop}%
\end{thebibliography}%

\end{document}